\documentclass[a4paper,11pt]{article}

\usepackage{jheppub,bm,braket,comment}
\usepackage{physics}
\usepackage[T1]{fontenc}
\usepackage{mathrsfs}

\newcommand{\ba}[1]{\begin{align} #1 \end{align}}
\newcommand{\bes}[1]{\begin{equation}\begin{split} #1 \end{split}\end{equation}}
\newcommand{\bsa}[2]{\begin{subequations}\label{#1}\begin{align} #2 \end{align}\end{subequations}}
\newcommand{\kvec}{{\bm k}}
\newcommand{\xvec}{{\bm x}}

\title{\boldmath An analytic evaluation of gravitational particle production of fermions via Stokes phenomenon}

\author[a,b]{Soichiro Hashiba,}
\author[c]{\, Siyang Ling,}
\author[c]{\, and Andrew J. Long}

\affiliation[a]{Department of Physics, Graduate School of Science, \\The University of Tokyo, 7-3-1 Hongo, Tokyo 113-0033, Japan}
\affiliation[b]{Research Center for the Early Universe (RESCEU), Graduate School of Science, \\The University of Tokyo, 7-3-1 Hongo, Tokyo 113-0033, Japan}
\affiliation[c]{Department of Physics and Astronomy, Rice University, Houston, TX 77005, USA}

\preprint{RESCEU-6/22}

\abstract{
The phenomenon of gravitational particle production can take place for quantum fields in curved spacetime.  
The abundance and energy spectrum of gravitationally produced particles is typically calculated by solving the field's mode equations on a time-dependent background metric.  
For purposes of studying dark matter production in an inflationary cosmology, these mode equations are often solved numerically, which is computationally intensive, especially for the rapidly-oscillating high-momentum modes.  
However, these same modes are amenable to analytic evaluation via the Exact Wentzel-Kramers-Brillouin (EWKB) method, where gravitational particle production is a manifestation of the Stokes phenomenon.  
These analytic techniques have been used in the past to study gravitational particle production for spin-0 bosons.  
We extend the earlier work to study gravitational production of spin-1/2 and spin-3/2 fermions.  
We derive an analytic expression for the connection matrix (valid to all orders in perturbations) that relates Bogoliubov coefficients across a Stokes line connecting a merged pair of simple turning points.  
By comparing the analytic approximation with a direct numerical integration of the mode equations, we demonstrate an excellent agreement and highlight the utility of the Stokes phenomenon formalism applied to fermions.  
We discuss the implications for an analytic understanding of catastrophic particle production due to vanishing sound speed, which can occur for a spin-3/2 Rarita-Schwinger field. 
}

\begin{document} 
\maketitle
\flushbottom

\newpage

\section{Introduction}

Gravitational particle production (GPP) corresponds to the creation of particles due to a strong gravitational field~\cite{Parker:1968mv,Sexl:1969ix,Parker:1969au,Zeldovich:1970si,Grib:1970xx,Parker:1971pt,Zeldovich:1971mw,Mamaev:1976tq}.  
It can be understood as the gravitational analog of the Schwinger effect~\cite{Woodhouse:1977,Audretsch:1978qu}.  
Although the phenomenon of GPP has not been observed directly, it is expected to play an important role in black hole evaporation via Hawking radiation~\cite{Hawking:1975}, reheating after inflation~\cite{Ford:1986sy,Hashiba:2018iff,Hashiba:2019mzm}, inflationary quantum fluctuations leading to the generation of primordial curvature perturbations~\cite{Guth:1982ec,Hawking:1982cz,Starobinsky:1982ee,Bardeen:1983qw}, and the creation of cosmological relics such as dark matter~\cite{Kuzmin:1998uv,Kuzmin:1998kk,Chung:1998zb,Chung:1998ua,Kolb:1998ki,Chung:2001cb,Ema:2018ucl,Hashiba:2018tbu,Hashiba:2019mzm,Ema:2019yrd,Li:2019ves}, baryon asymmetry~\cite{Enomoto:2020xlf,Enomoto:2021hfv} and primordial gravitational waves~\cite{Giovannini:1999bh,Tashiro:2003qp,Kunimitsu:2012xx}.  

GPP is studied using the framework of quantum field theory in curved spacetime~\cite{DeWitt:1975ys,Birrell:1982ix,Mukhanov:2007zz,Parker:2009uva}.  
Gravity is treated as a classical background, and in the context of cosmology, the background is taken to be the homogeneous and isotropic Friedmann-(Lema\^itre)-Robertson-Walker (FRW) spacetime~\cite{Friedmann:1922,Friedmann:1924,Lemaitre:1931,Robertson:1935,Robertson:1936a,Robertson:1936b,Walker:1937}.  
This background induces a time-dependent Hamiltonian for the various quantum fields in the theory (as long as they are not conformally coupled to gravity).  
Non-adiabatic evolution of the spectrum populates excited states, which corresponds to particle production.  
To calculate the spectrum and abundance of the gravitationally produced particles, one solves a system of equations of motion for each Fourier mode of the quantum field.  
Imposing Bunch-Davies vacuum initial conditions at early times (during inflation when the relevant modes are deep inside the horizon), one solves for the mode functions' amplitudes at late times (after inflation when modes are inside the horizon and typically non-relativistic), which determines the amount of GPP. 

More accurately, the problem that we seek to solve is gravitational particle production for fermions, which corresponds to the Schr\"odinger equation for a two-level system with a time-dependent and traceless Hamiltonian, $i \dot{\psi}(t) = \Omega(t) \psi(t)$~\cite{Chung:2011ck}.  
At early times the Hamiltonian is asymptotically static and solutions are superpositions of the positive and negative frequency modes, $\alpha^\mathrm{(early)} e^{-i \omega_\mathrm{early} t} + \beta^\mathrm{(early)} e^{i \omega_\mathrm{early} t}$, where $\pm \omega_\mathrm{early}$ are the eigenvalues of $\Omega(-\infty)$.  
Suppose that the Hamiltonian is also static at late times (in practice it is slowly varying), and a general solution takes the form $\alpha^\mathrm{(late)} e^{-i \omega_\mathrm{late} t} + \beta^\mathrm{(late)} e^{i \omega_\mathrm{late} t}$.  
Since the Hermitian Hamiltonian generates unitary time evolution, the coefficients of the positive and negative frequency modes must be related by a unitary transformation: $(\alpha^\mathrm{(late)}, \beta^\mathrm{(late)}) = T (\alpha^\mathrm{(early)}, \beta^\mathrm{(early)})$ where $T \in \mathrm{SU}(2)$ is a special and unitary matrix.  
Even if only the positive frequency mode is present at early times ($\alpha^\mathrm{(early)} = 1$, $\beta^\mathrm{(early)} = 0$), the late-time solution may be a superposition of both positive and negative frequency modes ($\alpha^\mathrm{(late)}, \beta^\mathrm{(late)} \neq 0$) as a result of the nonzero off-diagonal entries in $T$.  
This mode mixing gives rise to GPP.  

We desire to calculate the amplitude $\beta^\mathrm{(late)}$ of the negative frequency mode at late times assuming that only the positive frequency mode is present at early times, since the amount of particle production is proportional to $|\beta^\mathrm{(late)}|^2$.  
This problem has been studied extensively by Voros, Berry, and others~\cite{Berry:1972na,Berry_1982,PMIHES_1988__68__211_0,Berry:1989zz,10.2307/51738,10.2307/51774,lim1991superadiabatic,Berry_1993,Winitzki:2005rw} who derived an approximate relation $|\beta^\mathrm{(late)}|^2 \approx e^{-2 \Psi}$ where $\Psi = i \int_{t_c}^{t_c^\ast} \! \dd{t} \, \omega(t)$, where $\pm \omega(t)$ are the (instantaneous) eigenvalues of the Hamiltonian at time $t$, and where $\omega(t_c) = \omega(t_c^\ast) = 0$ defines its (complex) roots. 
This `master formula' has been used to calculate spin-0, spin-1/2 and spin-1 particle production in an FRW spacetime~\cite{Dabrowski:2014ica,Sou:2021juh}, and it has seen various other applications~\cite{Brezin:1970xf,Froman:2002,Dumlu:2010ua,Kim:2010xm,Kim:2013cka,Hashiba:2020rsi,Taya:2020dco,Dumlu:2020wvd,Hashiba:2021npn,Yamada:2021kqw}.  
The standard derivation of $|\beta^\mathrm{(late)}|^2 \approx e^{-2 \Psi}$ makes use of the adiabatic approximation, and although this approach is typically well justified for the systems of interest, it is worth noting that an exact expression for $|\beta^\mathrm{(late)}|^2$ can be derived using another formalism: the exact WKB method.

The exact WKB (EWKB) method~\cite{PhysRevLett.41.1141,Voros:1983,Aoki:1995,Delabaere1997exact} builds upon the classic (Jeffreys)-Wentzel-Kramers-Brillouin (WKB) method~\cite{Jeffreys:1925,Wentzel:1926aor,Kramers:1926njj,Brillouin:1926blg} for solving a second-order oscillator equation with time-dependent frequency, $\ddot{\phi}(t) + \omega^2(t) \phi(t) = 0$.  
In the EWKB formalism the complex time plane decomposes into disjoint regions where Borel resummation is used to construct pairs of basis functions, corresponding to positive and negative frequency modes, that exactly solve the oscillator equation.  
A global solution is constructed from linear combinations of the basis functions.  
At the boundaries between regions, the coefficients in the linear combination change discontinuously, allowing for mixing of the positive and negative frequency modes; this is the Stokes phenomenon~\cite{Stokes:1864}.  
To solve the problem via the Stokes phenomenon, one calculates a set of connection matrices relating the coefficients in adjacent regions, and the product of these matrices is the matrix $T$ relating the $(\alpha,\beta)$ coefficients in the asymptotically early and late time regions, which gives $\beta^\mathrm{(late)}$.  
One can derive a formal expression for the exact matrix $T$, but in practice the adiabatic approximation is employed to obtain tractable formulas.  

The EWKB method has been used to study GPP in bosonic field theories~\cite{Taya:2020dco} where the field equation takes the form of a second-order oscillator equation, but it has seen limited application for fermionic theories~\cite{Enomoto:2020xlf,Enomoto:2021hfv} where the mode equation takes the form of a two-level Schr\"odinger equation.  
Our central goal is to provide a new derivation of the connection matrix for general theories of free fermions with spin-1/2 or 3/2.
This result appears in Eq.~\eqref{eq:con_mat}, and details of the derivation appear in the appendices.  
Having an analytical result is particularly useful for understanding the spectrum of gravitationally-produced particles at high wavenumber where the standard numerical approaches becomes computationally taxing. 
We also present two toy models so as to illustrate the utility of our connection matrix and validate it against a direct numerical computation. 
The toy models correspond to a spin-1/2 and a spin-3/2 field in an FRW spacetime.  
This is the first time that EWKB has been employed to study spin-3/2 fermions in FRW spacetime.  
Previous work on GPP with spin-3/2 particles~\cite{Giudice:1999yt,Giudice:1999am,Hasegawa:2017hgd,Kolb:2021xfn} (such as supersymmetry's gravitino~\cite{Kallosh:1999jj,Kallosh:2000ve,Bastero-Gil:2000lgf}) used numerical techniques to estimate the spectrum of inflationary GPP, finding that for a range of cosmological models and particle masses, there can be a `catastrophic' production of high-momentum particles~\cite{Hasegawa:2017hgd,Kolb:2021xfn,Kolb:2021nob,Dudas:2021njv,Antoniadis:2021jtg}.  
Thus, as an application of our work we complement previous numerical studies with an analytical understanding of spin-3/2 GPP via the Stokes phenomenon.  

This article is organized as follows.  
In Sec.~\ref{sec:formalism} we study the equations of motion for free spin-1/2 and spin-3/2 fields in an FRW background with arbitrary time dependence, and by studying the Stokes phenomenon we derive an exact expression for the connection matrix $T$ linking the $(\alpha,\beta)$ coefficients at asymptotically early and late times.  
This result yields a `master formula' that allows the Bogoliubov coefficients $\beta_k$ and spectrum $n_k$ to be calculated once a model is specified, and it agrees with the standard result $|\beta^\mathrm{(late)}|^2 \approx e^{-2 \Psi}$ in the adiabatic regime.  
In Sec.~\ref{sec:spin1/2} we apply these general results to study the gravitational production of spin-1/2 particles with a time-dependent effective mass.  
Similarly, in Sec.~\ref{sec:spin3/2} we study the production of spin-3/2 particles with a time-dependent effective sound speed and discuss the implications for catastrophic particle production.  
Finally, in Sec.~\ref{sec:conclusion} we summarize and discuss the applications of our work.  
The article includes also five appendices: diagonalization of a two-level Hermitian Hamiltonian in App.~\ref{app:diagonalize}, construction of the WKB series for spinor fields in App.~\ref{app:wkb_solution}, concrete derivation of the connection matrix for the Landau-Zener model in App.~\ref{app:landau_zener}, detailed derivation of the connection matrix in App.~\ref{app:connection_matrix}, and explicit evaluation of the phase integral in App.~\ref{app:phase_integral}.  
We use natural units and the $(-,+,+)$ convention~\cite{MTW:1973} for gravitational tensors (mostly-minus metric signature). 

\section{Stokes phenomenon for spinor fields}\label{sec:formalism}

We study the mode equations for free Dirac and Rarita-Schwinger fields in an arbitrary FRW spacetime.  
We relate solutions at asymptotically early and late times using the Stokes phenomenon, and we derive expressions for the corresponding connection matrix, Bogoliubov coefficients, and spectrum of gravitationally-produced particles.  

\subsection{Spinor field's mode equation}\label{sec:setup}

We are interested in the solutions of mode equations that take the form 
\begin{subequations}\label{eq:mode_eqn}
\ba{
    i \dv{\eta} \psi(\eta) = \Omega(\eta) \psi(\eta)
    \;,
}
where $\eta$ is conformal time, $\psi(\eta)$ is a 2-dimensional vector of complex mode functions, and $\Omega(\eta)$ is a 2-by-2, time-dependent, Hermitian, and traceless matrix.  
By writing $\psi = (u_A, u_B)$ and $\Omega = ((M,K),(K^\ast,-M))$, such mode equations can also be expressed as 
\ba{
    i \begin{pmatrix} u_A^\prime \\ u_B^\prime \end{pmatrix} = \begin{pmatrix} M(\eta) & K(\eta) \\ K^\ast(\eta) & - M(\eta) \end{pmatrix} \begin{pmatrix} u_A \\ u_B \end{pmatrix}
    \;,
}
\end{subequations}
where primes denote derivatives with respect to $\eta$, $u_A(\eta)$ and $u_B(\eta)$ are complex mode functions, $M(\eta)$ must be real (we also assume $M>0$), and $K(\eta) = \kappa(\eta) e^{i \zeta(\eta)}$ may be complex.
At conformal time $\eta_0$ we impose initial conditions 
\ba{\label{eq:initial_condit}
    \psi(\eta_0) = \psi_0 
    \qquad \text{or equivalently} \qquad 
    u_A(\eta_0) = u_{A,0} 
    \qquad \text{and} \qquad 
    u_B(\eta_0) = u_{B,0} 
}
that respect the normalization condition 
\ba{\label{eq:normalization}
    \psi_0^\dagger \psi_0 = |u_{A,0}|^2 + |u_{B,0}|^2 = 1 
    \;.
}
The complex mode functions -- $\psi_\kvec(\eta)$, $u_{A,\kvec}(\eta)$, and $u_{B,\kvec}(\eta)$ -- as well as the parameters -- $M_\kvec(\eta)$ and $K_\kvec(\eta)$ -- are indexed by a comoving wavevector $\kvec$ or wavenumber $k = |\kvec|$, and these labels are suppressed throughout this section to simplify the notation. 

This general parametrization lets us simultaneously study GPP for both spin-1/2 and spin-3/2 particles on an FRW background.  
As we will see in Secs.~\ref{sec:spin1/2}~and~\ref{sec:spin3/2}, the following choices 
\ba{\label{eq:choices}
    \begin{array}{c||c|c|c}
    & \text{Dirac} & \text{RS}\ (\pm \tfrac{3}{2}) & \text{RS}\ (\pm \tfrac{1}{2}) \\ \hline \hline 
    M(\eta) & a(\eta) m & a(\eta) m & a(\eta) m \\
    K(\eta) & k & k & c_s(\eta) e^{i \zeta(\eta)} k \\
    \end{array}
    \;,
}
yield the mode equations for a spin-1/2 Dirac spinor field, the $\pm 3/2$ helicity modes of a spin-3/2 Rarita-Schwinger field, and the $\pm 1/2$ helicity modes of a Rarita-Schwinger field.
Here $m$ is the particle's mass and $k = |\kvec|$ is the comoving wavenumber.
Also note that the mode equation for a Majorana spinor field is identical to that of a spin-1/2 Dirac spinor field.
The analysis in this section accommodates general analytic functions $K(\eta)$ and $M(\eta)$ (with $M$ real, non-negative), and it reduces to known results for specific models.  

\subsection{Overview}\label{sec:overview}

If the mode equation~\eqref{eq:mode_eqn} were to contain constant parameters $M(\eta) = M_0$ and $K(\eta) = K_0$, then the system could be solved (exactly) by diagonalizing the Hermitian matrix, integrating the two decoupled first-order ordinary differential equations, and obtaining the solutions called \emph{instantaneous positive and negative frequency modes}.
For time-varying $M(\eta)$ and $K(\eta)$, the \emph{WKB method} can be employed to construct a divergent asymptotic \emph{WKB series} that solves Eq.~\eqref{eq:mode_eqn} order-by-order in powers of a formal expansion parameter $\hbar$ that counts powers of the derivatives $M^\prime(\eta)$, $K^\prime(\eta)$ and $K^{*\prime}(\eta)$.
Truncating this asymptotic series at low orders gives the familiar \emph{leading-order WKB solution}, which is an approximate solution to Eq.~\eqref{eq:mode_eqn}. 
Alternatively, applying \emph{Borel resummation} to the divergent WKB series yields an analytic function that solves the mode equation exactly; this approach is called the \emph{exact WKB method} (EWKB method). 
How solutions are decomposed into positive and negative frequency components changes discontinuously at \emph{Stokes lines} in the complex time plane, and for the solution to remain continuous, there must be abrupt mode mixing, which is the \emph{Stokes phenomenon}.  

In the following subsections, we identify the instantaneous positive and negative frequency mode functions, explain how the EWKB method can be used to derive exact solutions of the mode equation, discuss the locality of these solutions and its connection with mode mixing via the Stokes phenomenon, derive the connection matrix to quantify the amount of mode mixing, and discuss the implications for GPP.  
The main results appear in Sec.~\ref{sec:Weber}.  
Additional details are provided in appendices \ref{app:diagonalize}, \ref{app:wkb_solution}, \ref{app:landau_zener}, and \ref{app:connection_matrix}.  

\subsection{Instantaneous positive and negative frequency modes}\label{sec:instantaneous}

The mode equation~\eqref{eq:mode_eqn} has the same form as the Schr\"odinger equation for a two-level system with Hamiltonian $\Omega(\eta)$.  
At any instant in time, the Hermitian matrix $\Omega(\eta)$ can be diagonalized by a unitary transformation with unitary matrix $U(\eta)$.  
We discuss the details of this diagonalization in App.~\ref{app:diagonalize}, and we summarize the key results in this section.  

Let the complex mode functions $\alpha(\eta)$ and $\beta(\eta)$ be defined by
\ba{\label{eq:U_def}
     \mqty(\alpha(\eta) \\ \beta(\eta))
     = U(\eta) \mqty(u_A(\eta) \\ u_B(\eta)) 
     \;,
}
where $U(\eta)$ is a special and unitary matrix, $U(\eta) \in \mathrm{SU}(2)$.  
The initial condition~\eqref{eq:initial_condit} and mode function normalization~\eqref{eq:normalization} become 
\ba{\label{eq:alpha0_beta0}
     \mqty(\alpha(\eta_0) \\ \beta(\eta_0))
     = \mqty(\alpha_0 \\ \beta_0)
     = U(\eta_0) \mqty(u_{A,0} \\ u_{B,0}) 
    \quad \text{where} \quad  
    |\alpha_0|^2 + |\beta_0|^2 = 1
    \;.
}
Note that $0 \leq |\alpha_0| \leq 1$ and $0 \leq |\beta_0| \leq 1$.  
The unitary matrix $U(\eta)$ is chosen to diagonalize the Hermitian matrix $\Omega(\eta)$ as 
\ba{\label{eq:omega_def}
      U \Omega U^{-1} 
      & = \begin{pmatrix} \omega(\eta) & 0 \\ 0 & -\omega(\eta) \end{pmatrix} 
      \qquad \text{with} \qquad 
      \omega(\eta) \equiv \sqrt{\kappa(\eta)^2 + M(\eta)^2} 
      \;,
}
where $\pm \omega(\eta)$ are the eigenvalues of $\Omega(\eta)$.  
This is accomplished by taking 
\ba{\label{eq:U_transform}
    U = \frac{1}{\sqrt{2}} 
    \mqty(\sqrt{1+\frac{M}{\omega}} e^{-i\zeta/2} e^{i\delta} e^{i\Phi} & 
    \sqrt{1-\frac{M}{\omega}} e^{i\zeta/2} e^{i\delta} e^{i\Phi} \\ 
    -\sqrt{1-\frac{M}{\omega}} e^{-i\zeta/2} e^{-i\delta} e^{-i\Phi} & \sqrt{1+\frac{M}{\omega}} e^{i\zeta/2} e^{-i\delta} e^{-i\Phi}) 
}
where 
\ba{\label{eq:Phi_def}
    \Phi(\eta) \equiv \int_{\eta_a}^\eta \! \dd{\eta^\prime} \omega(\eta^\prime) 
    \qquad \text{and} \qquad 
    \delta(\eta) \equiv \int_{\eta_a}^\eta \! \dd{\eta^\prime} \frac{M(\eta^\prime)}{2\omega(\eta^\prime)} \, \zeta^\prime(\eta^\prime) 
    \;. 
}
The lower limit of integration, which we call the \emph{anchor time} $\eta_a$, is arbitrary, and it may differ from the time $\eta_0$ at which the initial conditions are imposed.  
We shall study the phase integral $\Phi(\eta)$ further in later subsections, as its analytic structure in the complex $\eta$ plane is intimately connected with the Stokes phenomenon. 

In terms of the new mode functions, $\alpha(\eta)$ and $\beta(\eta)$, the mode equation~\eqref{eq:mode_eqn} becomes 
\ba{\label{eq:dalpha_dbeta}
    i \mqty( \alpha^\prime \\ \beta^\prime ) = \mqty( 0 & - i \mu(\eta) \, e^{2i \Phi(\eta)} \\ i \mu^\ast(\eta) \, e^{- 2i \Phi(\eta)} & 0 ) \mqty( \alpha \\ \beta )
    \;,
}
where 
\ba{\label{eq:mu_def}
    \mu(\eta) \equiv \frac{\kappa M^\prime - M \kappa^\prime - i \kappa \omega \zeta^\prime}{2 \omega^2} \, e^{2 i \delta}
    \;.
}
Note that $\mu(\eta)$ would vanish if $M(\eta)$ and $K(\eta) = \kappa(\eta) e^{i \zeta(\eta)}$ were static, whereas their nonzero time derivatives lead to a mixing of the $\alpha(\eta)$ and $\beta(\eta)$ mode functions. 

In summary, by a change of variables, solutions of the mode equations \eqref{eq:mode_eqn} along with initial conditions \eqref{eq:initial_condit} can be written as 
\begin{subequations}\label{eq:uA_uB_solution}
\ba{
    \psi(\eta) 
    = \mqty(u_A(\eta) \\ u_B(\eta)) 
    = \alpha(\eta) \, \tilde{\psi}_+(\eta) + \beta(\eta) \, \tilde{\psi}_-(\eta) 
}
with 
\ba{
    \tilde{\psi}_\pm(\eta) \equiv \mqty( \tilde{u}_{A\pm}(\eta) \\ \tilde{u}_{B\pm}(\eta)) \equiv \frac{1}{\sqrt{2}} e^{\mp i(\Phi + \delta)} \mqty(\pm e^{i\zeta / 2} \sqrt{1 \pm \frac{M}{\omega}} \\ e^{-i\zeta / 2} \sqrt{1 \mp \frac{M}{\omega}})
    \;,
}
\end{subequations}
provided that the new complex mode functions, $\alpha(\eta)$ and $\beta(\eta)$, satisfy the mode equations~\eqref{eq:dalpha_dbeta} and the initial conditions~\eqref{eq:alpha0_beta0}.  
We call $\tilde{\psi}_+(\eta)$ and $\tilde{\psi}_-(\eta)$ the \textit{instantaneous positive and negative frequency mode functions}; they are distinguished by the factors $\tilde{\psi}_\pm \propto e^{\mp i \Phi}$.  
They form a unitary basis in the sense that 
\ba{
    \tilde{\mathbb{U}}(\eta) & \equiv \mqty( \tilde{\psi}_+(\eta) & \tilde{\psi}_-(\eta) ) 
    \quad \text{with} \quad 
    \begin{cases}
    \tilde{\mathbb{U}}(\eta)^\dagger \tilde{\mathbb{U}}(\eta) = I \\ 
    \tilde{\mathbb{U}}(\eta) \tilde{\mathbb{U}}(\eta)^\dagger = I \\ 
    \mathrm{det} \, \tilde{\mathbb{U}}(\eta) = 1 
    \end{cases}
    \;,
}
where $I$ is the 2-by-2 identity matrix. 
In the next section we will see that $\tilde{\psi}_\pm(\eta)$ also correspond to the leading-order WKB solution.  

If $\Omega(\eta) = \Omega_0$ were a constant matrix, then $\tilde{\psi}_\pm(\eta)$ would be basis functions spanning the space of solutions, and a general solution~\eqref{eq:uA_uB_solution} would be constructed as a linear combination with constant coefficients $\alpha_0$ and $\beta_0$.  
For varying $\Omega(\eta)$, the instantaneous mode functions are not solutions of the mode equation.  
Nevertheless Eq.~\eqref{eq:uA_uB_solution} plays an important role in the study of GPP where the mode evolution is adiabatic ($\Omega^\prime \approx 0$) at both early and late times.  
Typically we take $\alpha_0 = 1$ and $\beta_0 = 0$ at early times, and we solve for the asymptotically constant values of  $\alpha(\eta)$ and $\beta(\eta)$ at late times.  

\subsection{Exact WKB method}\label{sec:EWKB}

It is difficult to find analytic solutions to the mode equation~\eqref{eq:mode_eqn} for general, time-varying parameters $M(\eta)$ and $K(\eta) = \kappa(\eta) \, e^{i \zeta(\eta)}$. 
However, one can employ perturbation theory~\cite{BenderOrszag:1999} to write the solution as a formal power series, and if the system exhibits a hierarchy of time scales between the rapidly-oscillating and slowly-varying behavior, then the leading terms in the perturbative expansion provide a good approximation to the exact solution. 
A well known way of doing this is the \emph{(Jeffreys)-Wentzel-Kramers-Brillouin (WKB) method}~\cite{Jeffreys:1925,Wentzel:1926aor,Kramers:1926njj,Brillouin:1926blg}.

To implement the WKB method on Eq.~\eqref{eq:mode_eqn}, one first defines a family of equations, 
\ba{\label{eq:mode_eqn_with_hbar}
    \hbar \dv{\eta} \psi(\eta,\hbar) = -i \Omega(\eta) \psi(\eta,\hbar)
    \qquad \text{for} \qquad 
    \psi(\eta,\hbar) \equiv \mqty( u_A(\eta,\hbar) \\ u_B(\eta,\hbar) )
    \;,
}
labeled by a real, positive parameter $\hbar > 0$.  
Each equation should be solved with the same initial condition~\eqref{eq:initial_condit} $\psi(\eta_0,\hbar) = \psi_0$.  
We are interested in solving the equation that has $\hbar = 1$.  
The strategy of the WKB method follows from the observation that a solution is available as $\hbar \to 0$, and the plan is to construct the solution at $\hbar = 1$ as a power series centered on $\hbar = 0$.  
To understand the $\hbar \to 0$ solution, notice that rescaling the time coordinate with $\eta = \eta_0 + \hbar \tilde{\eta}$ removes the factor of $\hbar$ from the left side and gives $\Omega(\eta_0 + \hbar \tilde{\eta})$ on the right side.  
As $\hbar \to 0$ the matrix goes to a constant, $\Omega(\eta_0 + \hbar \tilde{\eta}) \approx \Omega(\eta_0) + \Omega^\prime(\eta_0) \hbar \tilde{\eta}$, and the solution is 
\bes{
    \psi(\eta,\hbar)
    & \simeq 
    \exp[-\frac{i}{\hbar} 
    \Omega(\eta_0)  
    (\eta - \eta_0)]
    \psi_0
    \qquad \text{as $\hbar \to 0$}
    \;,
}
for a time interval $0 \leq \eta - \eta_0 \lesssim |\omega(\eta_0) / \omega^\prime(\eta_0)|$.  
The existence of a calculable solution as $\hbar \to 0$ motivates an ansatz for the general solution as a power series centered on $\hbar = 0$.  

For nonzero $\hbar$, we are searching for solutions of Eq.~\eqref{eq:mode_eqn_with_hbar} that take the form   
\ba{
\label{eq:wkb_solution}
    \psi(\eta,\hbar) \propto \mqty( u_{A,a} \, e^{\frac{i}{\hbar} \int_{\eta_a}^{\eta} \dd{\eta^\prime} S_A(\eta^\prime,\hbar)} \\ u_{B,a} \, e^{\frac{i}{\hbar} \int_{\eta_a}^{\eta} \dd{\eta^\prime} S_B(\eta^\prime,\hbar)}) 
    \quad \text{where} \quad
    S_J(\eta,\hbar) = \sum_{n=0}^{\infty} \hbar^n S_{Jn}(\eta) 
}
for $J = A$ or $B$.  
This expression is known as the \emph{WKB ansatz} or as the \emph{WKB solution}; we will refer to Eq.~\eqref{eq:wkb_solution} as the \emph{WKB series} to emphasize that it is defined as a (formal) series in powers of $\hbar$.  
Notice that the WKB series is defined with respect to an \emph{anchor time} $\eta_a$, which we discuss further in the next subsection.  
The WKB series is constructed by assuming that this $\psi(\eta,\hbar)$ solves the mode equation~\eqref{eq:mode_eqn_with_hbar}, which leads to a set of non-linear equations for the functions $S_J(\eta,\hbar)$ that are solved order-by-order in $\hbar$ to find the series coefficients $S_{Jn}(\eta)$ in terms of $M$, $\kappa$, $\zeta$, and their derivatives.
General solutions are linear combinations of the two  basis functions 
\bes{\label{eq:uJp_uJm_basis}
    \psi_\pm(\eta,\hbar) 
    & \equiv \mqty(u_{A\pm}(\eta,\hbar) \\ u_{B\pm}(\eta,\hbar))
    \equiv \mqty( \mathcal{N}_{A,\pm}(\hbar) \, \tilde{u}_{A\pm}(\eta,\hbar) \, \mathrm{exp}\bigl[ \frac{i}{\hbar} \int_{\eta_a}^\eta \dd{\eta^\prime} \sum_{n=2}^\infty \hbar^n S_{An\pm}(\eta^\prime) \bigr] \\ \mathcal{N}_{B,\pm}(\hbar) \, \tilde{u}_{B\pm}(\eta,\hbar) \, \mathrm{exp}\bigl[ \frac{i}{\hbar} \int_{\eta_a}^\eta \dd{\eta^\prime} \sum_{n=2}^\infty \hbar^n S_{Bn\pm}(\eta^\prime) \bigr] ) 
    \;. 
}
The normalization factors, $\mathcal{N}_{A,\pm}(\hbar)$ and $\mathcal{N}_{B,\pm}(\hbar)$, are independent of time and go to $1$ as $\hbar \to 0$.  
The instantaneous positive/negative frequency mode functions, $\tilde{u}_{A,\pm}(\eta,\hbar)$ and $\tilde{u}_{B,\pm}(\eta,\hbar)$, are given by Eq.~\eqref{eq:uA_uB_solution} with the replacement $\Phi \to \Phi/\hbar$.  
Notice that $\psi_\pm \propto e^{\mp i \Phi(\eta) / \hbar}$, and so these correspond to the \emph{positive and negative frequency mode functions}, respectively.  
See App.~\ref{app:wkb_solution} for a complete derivation of the WKB series~\eqref{eq:normalized_EWKB_soln_for_u}.  

Truncating the WKB series at any finite order in $\hbar$ and setting $\hbar = 1$ yields an approximate solution to the original mode equation~\eqref{eq:mode_eqn}.  
For instance, by dropping $\order{\hbar}$ terms in the exponent ($S_{Jn}$ for $n \geq 2$), we obtain the \emph{leading-order WKB solution}, 
\ba{\label{eq:leading_order_wkb}
    \mqty(u_{A}(\eta) \\ u_{B}(\eta))
	& \simeq \alpha_a \, \tilde{\psi}_+(\eta) + \beta_a \, \tilde{\psi}_-(\eta) 
	\;,
}
where $\alpha_a$ and $\beta_a$ are arbitrary complex constants, and where $\tilde{\psi}_\pm(\eta)$ were defined in Eq.~\eqref{eq:uA_uB_solution}.  
In general, the leading-order WKB solution can always be written as a linear superposition of the instantaneous positive and negative frequency mode functions~\cite{Bjorken:1980ue}.  

If the power series~\eqref{eq:wkb_solution} were a convergent Taylor series at $\hbar = 1$, then Eq.~\eqref{eq:uJp_uJm_basis} would yield the desired basis functions in the space of solutions.  
However, with the exception of special cases such as constant $\Omega(\eta)$, the power series is generally a divergent asymptotic series.  
Nevertheless, it can still be used to construct a convergent series that solves the mode equations even for $\hbar = 1$ by using the mathematical technique of Borel resummation.  
This approach is known as the \textit{exact WKB method} (EWKB method).  
It is based on the pioneering work of Voros~(1983)~\cite{Voros:1983}, and it has been developed by several groups~\cite{Aoki:1995,Delabaere1997exact}. 
In recent year this method has been employed~\cite{Taya:2020dco,Enomoto:2020xlf,Taya:2021dcz,Enomoto:2021hfv} to study solutions of the oscillator equation $\chi^{\prime\prime}(\eta) + \omega^2(\eta) \chi(\eta) = 0$ that appears in bosonic field theories. 
For a clear and concise review of the EWKB method, see e.g.~Sec.~2 of Ref.~\cite{Taya:2020dco} and Appendix~A of Ref.~\cite{Enomoto:2021hfv}.  
In this paper, we adapt the EWKB method to study the mode equations for spinor fields.

The EWKB method uses Borel resummation to transform the divergent asymptotic series~\eqref{eq:wkb_solution} into an analytic function that solves the mode equation even at $\hbar=1$. 
In this context, Borel resummation involves three steps.  
First the basis functions~\eqref{eq:uJp_uJm_basis} defined by the WKB series~\eqref{eq:wkb_solution} are written as~\cite{Takei:2000} 
\ba{\label{eq:uA_uB_EWKB_pre}
    u_{J\pm}(\eta, \hbar) = e^{\mp i \Phi(\eta) / \hbar} \sum_{n=0}^{\infty} \hbar^{n+1/2} \, u_{Jn\pm}(\eta) 
    \;.
}
The series coefficients $u_{Jn\pm}(\eta)$ are related to the $S_{Jn\pm}(\eta)$ by a Taylor series expansion of the exponential.  
Second, taking the \emph{Borel transform} of this series yields 
\ba{\label{eq:Borel_trans}
    (\mathcal{B}u_{J\pm})(\eta, y) = \sum_{n=0}^\infty \frac{u_{Jn\pm}(\eta)}{\Gamma(n+1/2)} \bigl(y \mp i \Phi(\eta)\bigr)^{n - 1/2} 
    \;,
}
which is a function of $y$ instead of $\hbar$.  
Since $\Gamma(n+1/2)$ grows as the factorial of $n$, this power series is convergent as long as $u_{Jn\pm}$ diverges slower than factorial, which is the case for the systems of interest.  
Third, taking the \emph{Laplace transform} of this function yields 
\ba{\label{eq:Laplace_trans}
    (\mathcal{L} \mathcal{B} u_{J\pm})(\eta,\hbar) = \int_{\pm i \Phi(\eta)}^\infty \! \dd{y} \, e^{-y/\hbar} \,  (\mathcal{B}u_{J\pm})(\eta, y) 
    \;,
}
where the path of integration is taken to be parallel to the real $y$ axis. 
Note that a Laplace transformation is, morally speaking, an ``inverse'' Borel transformation.\footnote{If one recalls the definition of the gamma function, $\Gamma(z) = \int_0^\infty \! \dd{x} \, x^{z-1} \, e^{-x}$ for $\mathrm{Re}(z) > 0$, then one can see that the series $A = \sum_{n=0}^\infty a_n = \sum_{n=0}^\infty \int_0^\infty \! \dd{x} \, a_n \, x^{n-1/2} \, e^{-x} / \Gamma(n+1/2)$ and its Borel sum $B = \int_0^\infty \! \dd{x} \, \sum_{n=0}^\infty a_n \, x^{n-1/2} \, e^{-x} / \Gamma(n+1/2)$, differ only in the order of the summation and integration.  When these operations commute, the Laplace transform inverts the Borel transform, and the Borel sum equals the original series.  When they do not commute, the Borel sum defines a new series.}    
This combination of Borel and Laplace transformations yields the \emph{Borel resummation of the WKB series} (or \emph{Borel sum} for short):
\ba{\label{eq:uA_uB_EWKB}
   (\mathcal{L} \mathcal{B} u_{J\pm})(\eta,\hbar) 
   & = \int_{\pm i \Phi(\eta)}^\infty \! \dd{y} \, e^{-y/\hbar} \sum_{n=0}^\infty \frac{u_{Jn\pm}(\eta)}{\Gamma(n+1/2)} \bigl(y \mp i \Phi(\eta)\bigr)^{n - 1/2} 
   \;.
}
Since the WKB series $u_{Jn\pm}(\eta)$ was constructed to solve the mode equation order-by-order in $\hbar$, the Borel resummation of this series is also a solution of the mode equation~\cite{Voros:1983}. 
More precisely, the Borel sum constructed on the anchor time $\eta_a$ is an exact solution of the mode equation in a restricted region of the complex $\eta$ plane that contains $\eta_a$, and this locality is the subject of discussion for the next subsection.  

To summarize, the EWKB method provides a construction of the functions 
\ba{\label{eq:EWKB_basis}
    (\mathcal{L} \mathcal{B} \psi_\pm)(\eta,\hbar) \equiv \mqty((\mathcal{L} \mathcal{B} u_{A\pm})(\eta,\hbar) \\ (\mathcal{L} \mathcal{B} u_{B\pm})(\eta,\hbar))
    \;, 
}
which form a basis in the space of solutions of the mode equation~\eqref{eq:mode_eqn_with_hbar}, and they are called the \emph{EWKB basis solutions}.  
Taking $\hbar = 1$ gives a solution of the original mode equation~\eqref{eq:mode_eqn}.  
To simplify the notation, we will drop the $\mathcal{L} \mathcal{B}$'s in the expressions below; henceforth $u_{J\pm}(\eta,\hbar)$ refers to the Borel sums~\eqref{eq:uA_uB_EWKB} rather than the WKB series~\eqref{eq:wkb_solution}.  
The EWKB basis solutions form a unitary basis at the anchor time $\eta_a$ in the sense that
\bes{\label{eq:unitary_normalization}
    \mathbb{U}(\eta_a,\hbar) \equiv \mqty( \psi_+(\eta_a,\hbar) & \psi_-(\eta_a,\hbar) ) 
    & \quad \text{with} \quad 
    \begin{cases}
    \mathbb{U}(\eta_a,\hbar)^\dagger \mathbb{U}(\eta_a,\hbar) = I \\
    \mathbb{U}(\eta_a,\hbar) \mathbb{U}(\eta_a,\hbar)^\dagger = I \\ 
    \mathrm{det} \, \mathbb{U}(\eta_a,\hbar) = 1 
    \end{cases}
    \;,
}
which defines their normalization.  

\subsection{Stokes regions, connection matrices, and Stokes phenomenon}\label{sec:discussion}

Now we must address an important subtlety regarding the locality of the EWKB basis solutions~\eqref{eq:EWKB_basis}.  
When solving for the WKB series~\eqref{eq:wkb_solution} it was necessary to specify an anchor time $\eta_a$.  
Although the choice of this anchor time is arbitrary, different anchor times lead to different WKB series and different EWKB basis solutions (that may differ by more than just a phase).  
In this subsection, we discuss how the complex-$\eta$ plane decomposes into disjoint Stokes regions, how the positive and negative frequency modes in different regions are related by a linear transformation implemented with a unitary connection matrix, and how the amplitudes of the positive and negative frequency modes change discontinuously when crossing between regions, which is the Stokes phenomenon.  

\begin{figure}[tb]
	\centering
	\includegraphics[width=.70\textwidth]{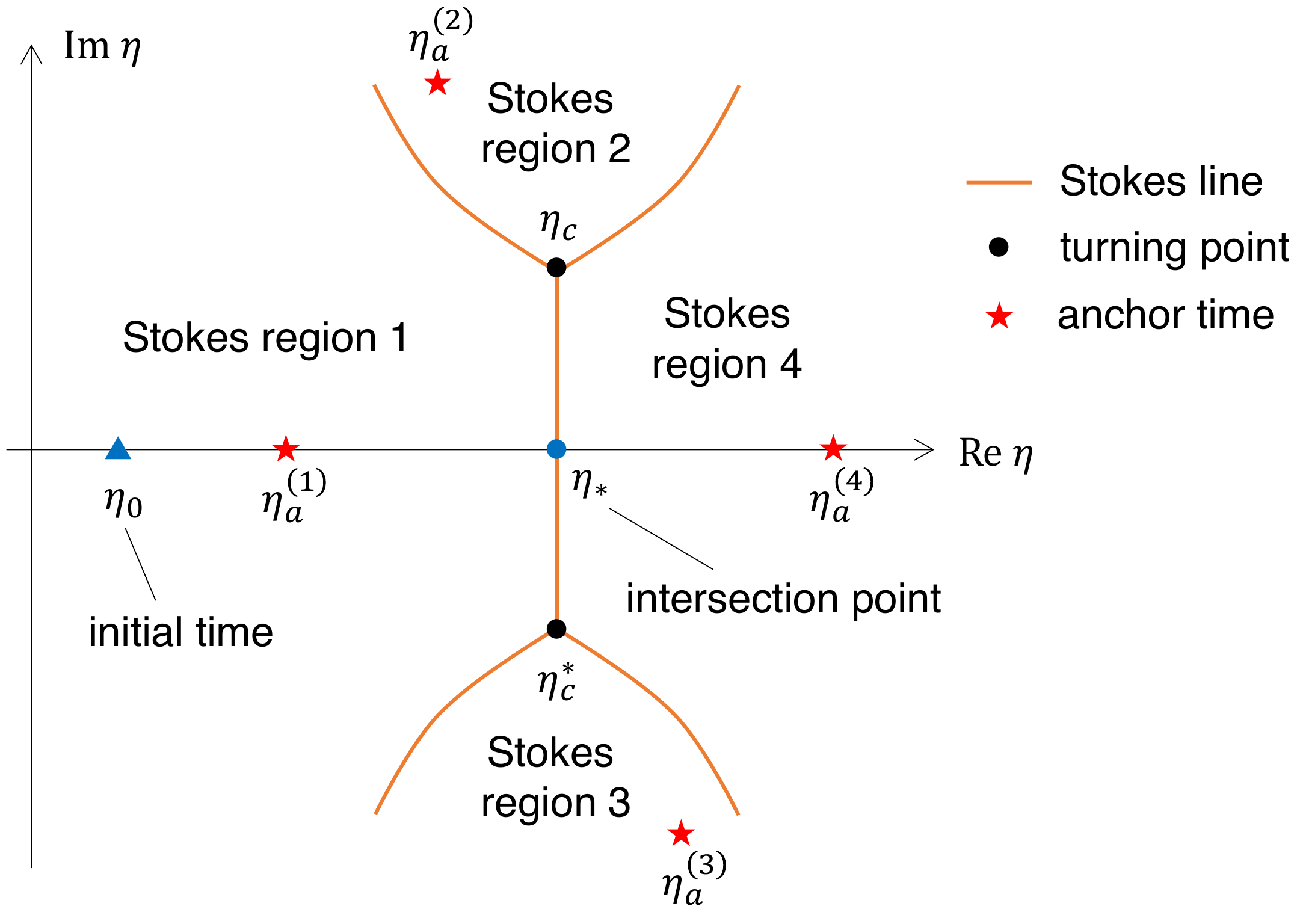}
	\caption{\label{fig:terminology}An illustration showing how the complex plane (conformal time $\eta$) is divided into Stokes regions, bounded by Stokes lines (orange curves), that terminate on turning points (black dots) or extend to infinity.  This structure is totally determined by the roots of $\omega(\eta)$ and its paths of steepest descent/ascent.  In each Stokes region, an EWKB basis solution $\psi_\pm^{(i)}(\eta,\hbar)$ is constructed at an anchor time $\eta_a^{(i)}$ (red stars).  
	}
\end{figure}

The EWKB basis solutions constructed using anchor time $\eta_a$ correspond to the positive and negative frequency modes within a region of the complex-$\eta$ plane containing $\eta_a$.  
(Recall that $\psi_\pm(\eta,\hbar) \propto e^{\mp i \Phi(\eta)/\hbar}$ with $\Phi(\eta) = \int_{\eta_a}^\eta \! \dd{\eta^\prime} \omega(\eta^\prime)$ from Eq.~\eqref{eq:uJp_uJm_basis}.)
This is known as a \emph{Stokes region}, and its boundary is formed by \emph{turning points} and \emph{Stokes lines}; see Fig.~\ref{fig:terminology}. 
A turning point $\eta_c$ is a point in the complex-$\eta$ plane at which $\omega(\eta_c) = 0$.
A Stokes line emanating from turning point $\eta_c$ is a path $\eta(\lambda)$ in the complex-$\eta$ plane along which 
\ba{\label{eq:Stokes1st}
    \Im\left[ i \int_{\eta_c}^{\eta(\lambda)} \dd{\eta'} \, \omega(\eta') \right] = 0 
    \;.
}
Along a Stokes line, the factor $e^{\mp i \Phi(\eta)/\hbar}$ has a constant imaginary part, implying that Stokes lines are also the contours of steepest descent or ascent emanating from a turning point.
Since $\omega(\eta)$ can be expanded as $\omega^2(\eta) = C(\eta - \eta_c) + \mathcal{O}(|\eta - \eta_c|^2)$ around a simple turning point $\eta_c$, Eq.~\eqref{eq:Stokes1st} implies that three Stokes lines emanate from each one simple turning point.
We will discuss why Stokes lines determine the range of EWKB solutions at the end of this section. 

\begin{figure}[tb]
	\centering
	\includegraphics[width=.60\textwidth]{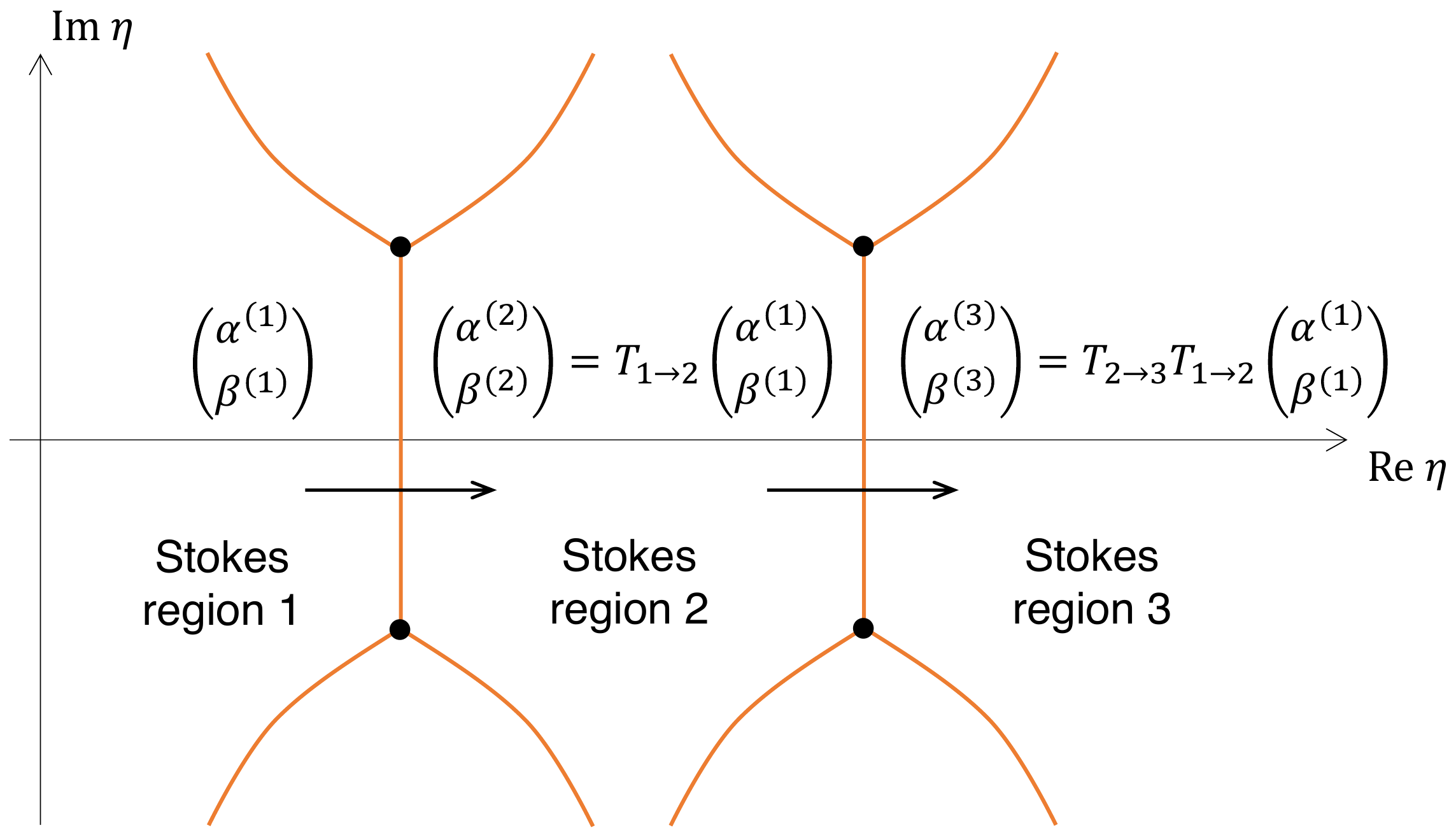}
	\caption{\label{fig:Stokes_concept} 
	An illustration of the Stokes phenomenon. 
    In each Stokes region, a general solution is written as $\alpha^{(i)} \psi_+^{(i)}(\eta,\hbar) + \beta^{(i)} \psi_-^{(i)}(\eta,\hbar)$ where $\alpha^{(i)}$ ($\beta^{(i)}$) multiplies the positive (negative) frequency EWKB basis solution $\psi_+^{(i)}$ ($\psi_-^{(i)}$).  
    The coefficients in adjacent regions are related by a unitary transformation.
    Nonzero off-diagonal entries in the unitary matrix $T_{i\to j}$ imply a mixing of positive and negative frequency modes, which is the Stokes phenomenon.  
    To study cosmological GPP, we take $\alpha^{(0)} = 1$ and $\beta^{(0)} = 0$ in the region containing $\eta \to -\infty$, and the density of gravitationally produced particles is proportional to $|\beta^{(i)}|^2$ in the region containing $\eta \to + \infty$.
	}
\end{figure}

How are local solutions in different Stokes regions related to one another?  
Suppose that a pair of EWKB basis solutions, $\psi_\pm^{(1)}$ and $\psi_\pm^{(2)}$, are constructed at two different real anchor times $\eta_1 < \eta_2$.  
Given a pair of c-number coefficients, $\alpha^{(1)}$ and $\beta^{(1)}$, the function $\psi^{(1)} = \alpha^{(1)} \psi_+^{(1)} + \beta^{(1)} \psi_-^{(1)}$ is a \emph{local solution} in the region containing $\eta_1$. 
The theory of ordinary differential equations tells us that there is a unique \emph{global solution} (namely, valid for all $\eta \in \mathbb{R}$) that has the same initial condition as $\psi^{(1)}$ at $\eta_1$. 
In the region containing $\eta_2$, it must be possible to write the global solution as  $\psi^{(2)} = \alpha^{(2)} \psi_+^{(2)} + \beta^{(2)} \psi_-^{(2)}$ for an appropriate pair of c-number coefficients, $\alpha^{(2)}$ and $\beta^{(2)}$, since the EWKB basis solutions $\psi_\pm^{(2)}$ span the space of all solutions.  
Since the mode equation~\eqref{eq:mode_eqn_with_hbar} is linear, the c-number coefficients must be related by a linear map, 
\ba{
    \mqty(\alpha^{(2)} \\ \beta^{(2)})
    = T_{1\to2} \mqty(\alpha^{(1)} \\ \beta^{(1)})
    \label{eq:connection_matrix}
    \;,
}
where $T_{1\to2}$ is called the \emph{connection matrix}.
See Fig~\ref{fig:Stokes_concept} for a graphical representation.  
If several Stokes lines cross the real $\eta$ axis, the connection matrix is multiplied at each crossing.  

To compute the connection matrix $T_{1\to2}$, we interpret $\psi_\pm^{(1)}$ and $\psi_\pm^{(2)}$ as global solutions consistent with the corresponding local solutions in their respective Stokes region. 
The matching condition
$\alpha^{(1)} \psi_+^{(1)} + \beta^{(1)} \psi_-^{(1)} = \alpha^{(2)} \psi_+^{(2)} + \beta^{(2)} \psi_-^{(2)}$ (now an equality valid globally in $\eta$) is written in matrix form as 
\bes{
    & \mqty(\psi_+^{(1)} & \psi_-^{(1)}) \mqty(\alpha^{(1)} \\ \beta^{(1)}) 
    = \mqty(\psi_+^{(2)} & \psi_-^{(2)}) \mqty(\alpha^{(2)} \\ \beta^{(2)}) 
    \;.
}
Since this relation must hold for arbitrary $\alpha^{(1)}$ and $\beta^{(1)}$, comparing with Eq.~\eqref{eq:connection_matrix} allows the connection matrix to be read off: 
\ba{\label{eq:connection_matrix_formula}
    T_{1\to2}(\hbar) = \mqty(\psi_+^{(2)}(\eta,\hbar) & \psi_-^{(2)}(\eta,\hbar))^{-1} \mqty(\psi_+^{(1)}(\eta,\hbar) & \psi_-^{(1)}(\eta,\hbar)) 
    \;.
}
Notice that although the EWKB basis solutions $\psi_{\pm}^{(1,2)}(\eta,\hbar)$ depend on $\eta$, this time dependence cancels in the connection matrix, which can be evaluated at any $\eta$.  

To derive a familiar expression for $\beta^{(2)}$, first set $\alpha^{(1)} = 1$ and $\beta^{(1)} = 0$.  
Then Eq.~\eqref{eq:connection_matrix} implies 
\bes{
    \beta^{(2)} 
    & = u_{A-}^{(2)\ast}(\eta,\hbar) \,  u_{A+}^{(1)}(\eta,\hbar) + u_{B-}^{(2)\ast}(\eta,\hbar) \,  u_{B+}^{(1)}(\eta,\hbar) 
    \;,
}
where we have used Eq.~\eqref{eq:unitary_normalization}.  
Next evaluate $\eta = \eta_2$ and approximate the EWKB basis solution by the instantaneous positive/negative frequency mode functions \eqref{eq:uA_uB_solution} to obtain 
\bes{
    \beta^{(2)} 
    & \approx - \frac{1}{\sqrt{2}} e^{i\zeta/2} \sqrt{1 - \frac{M}{\omega}} \, u_{A+}^{(1)}(\eta_2,\hbar) + \frac{1}{\sqrt{2}} e^{-i\zeta/2} \sqrt{1 + \frac{M}{\omega}} \, u_{B+}^{(1)}(\eta_2,\hbar) 
    \;.
}
After some algebra, the squared Bogoliubov coefficient can be written as 
\bes{
    |\beta^{(2)}|^2 
    & \approx \frac{1}{2} - \mathrm{Re} \biggl[ \frac{K}{\omega}  u_{B+}^{(1)\ast}(\eta_2,\hbar) u_{A+}^{(1)}(\eta_2,\hbar) \biggr] - \frac{1}{2} \frac{M}{\omega} \Bigl( 1 - 2 |u_{B+}^{(1)}(\eta_2,\hbar)|^2 \Bigr) 
    \;.
}
This is the amplitude of the negative frequency mode at time $\eta_2$ if only the positive frequency mode is present at time $\eta_1$.  
This expression generalizes the standard formula, $|\beta|^2 = (\omega-E)/2\omega$ with $E = k \, \mathrm{Re}[u_+^\ast u_-] + am(1-|u_+|^2)$ in the notation of Ref.~\cite{Peloso:2000hy}, which is familiar from studies of particle production with Dirac spinor fields~\cite{Giudice:1999am,Peloso:2000hy,Adshead:2015kza} (albeit in a different representation of the gamma matrices and with different normalization for the mode functions).  

For the spinor field mode equations that we consider here, one can argue that the connection matrix $T_{1\to2}$ is in $\mathrm{SU}(2)$ if anchor times $\eta_1$ and $\eta_2$ are real.
Notice that the mode equation~\eqref{eq:mode_eqn} takes the same form as the Schr\"odinger equation with Hermitian `Hamiltonian' $\Omega(\eta)$.  
Thus its solutions are generated by unitary time evolution, $\psi(\eta_2) = U(\eta_2,\eta_1) \psi(\eta_1)$.  
Additionally $U$ is special since $\Omega$ is traceless and $\mathrm{det}(e^A) = e^{\mathrm{Tr}(A)}$.  
Evaluating the connection matrix $T_{1\to2}$ using formula \eqref{eq:connection_matrix_formula} at $\eta_2$, we have:
\bes{
    T_{1\to2} 
    &= \mqty(\psi_+^{(2)}(\eta_2) & \psi_-^{(2)}(\eta_2))^{-1} U(\eta_2,\eta_1) \mqty(\psi_+^{(1)}(\eta_1) & \psi_-^{(1)}(\eta_1))
    \;.
}
From the conditions in Eq.~\eqref{eq:unitary_normalization} one can verify that the block matrices built from $\psi_\pm^{(2)}(\eta_2)$ and $\psi_\pm^{(1)}(\eta_1)$ are in $\mathrm{SU}(2)$.  
Since $T_{1\to2}$ is the product of three $\mathrm{SU}(2)$ matrices, it is also in $\mathrm{SU}(2)$. 
The unitary transformation preserves the norm $|\alpha^{(2)}|^2 + |\beta^{(2)}|^2 = |\alpha^{(1)}|^2 + |\beta^{(1)}|^2$, and since we take the initial condition~\eqref{eq:alpha0_beta0} to be $|\alpha_0|^2 + |\beta_0|^2 = 1$, we find $|\alpha^{(i)}|^2 + |\beta^{(i)}|^2 = 1$ for every Stokes region with anchor time $\eta_a^{(i)}$ on the real axis; this is the familiar normalization condition for the Bogoliubov coefficients of fermionic fields.\footnote{For bosonic mode equations, there is a similar argument to show that the connection matrix between EWKB basis solutions is in $SU(1,1)$. Let $\omega^2(\eta)$ be a function that is real for $\eta \in \mathbb{R}$, then the EWKB basis solutions $\chi_\pm$ of $\hbar^2 \chi'' + \omega^2 \chi = 0$ can be normalized such that $\psi_\pm^\dagger(\eta_0) S \psi_\pm(\eta_0) = \pm i$ and $\psi_\pm^\dagger(\eta_0) S \psi_\mp(\eta_0) = 0$ where $S = ((0,-1), (1, 0))$ and $\psi_\pm = (\chi_\pm, \chi_\pm')$.  
Furthermore, $\psi^\dagger S \psi$ is preserved for any solution $\psi = \alpha \psi_+ + \beta \psi_-$, so the expansion coefficients for different EWKB basis solutions must satisfy $|\alpha^{(1)}|^2 - |\beta^{(1)}|^2 = |\alpha^{(2)}|^2 - |\beta^{(2)}|^2$.
The arbitrariness of $\alpha^{(1)}$ and $ \beta^{(1)}$ then implies $T \in U(1,1)$, and an argument similar to the one in the main text further shows that $\mathrm{det}(T) = 1$.  
}

The \emph{Stokes phenomenon} corresponds to a discontinuous change in the coefficients of the positive and negative frequency EWKB basis solutions across a Stokes line.  
It is a consequence of the off-diagonal entries in the connection matrix.  
Consider two adjacent Stokes regions, and consider a pair of anchor times $\eta_1$ and $\eta_2$ that are brought arbitrarily close to the boundary Stokes line.  
Since the EWKB basis solutions change discontinuously from $\psi_\pm^{(1)}$ to $\psi_\pm^{(2)}$ across the Stokes line, in order to ensure continuity of the global solution, the coefficients must also change discontinuously from $\alpha^{(1)}$ and $\beta^{(1)}$ to $\alpha^{(2)}$ and $\beta^{(2)}$.  
In particular a solution with $\beta^{(1)} = 0$ in the first Stokes region may correspond to $\beta^{(2)} \neq 0$ in the second Stokes region.  
This is a manifestation of the Stokes phenomenon.  

Next, we pause to remark upon a potential source of confusion. 
Let $u_{J\pm}^{(1)}(\eta,\hbar)$ and $u_{J\pm}^{(2)}(\eta,\hbar)$ be the WKB series~\eqref{eq:wkb_solution} constructed at anchor times $\eta_1$ and $\eta_2$.  
These functions can be related as 
\ba{\label{eq:relate_u1_u2}
    u_{J\pm}^{(1)}
    \propto e^{\frac{i}{\hbar} \int_{\eta_1}^{\eta} \dd{\eta'} S_{J\pm}}
    = e^{\frac{i}{\hbar} \int_{\eta_1}^{\eta_2} \dd{\eta'} S_{J\pm}} \, e^{\frac{i}{\hbar} \int_{\eta_2}^{\eta} \dd{\eta'} S_{J\pm}}
    \propto e^{\frac{i}{\hbar} \int_{\eta_1}^{\eta_2} \dd{\eta'} S_{J\pm}} \, u_{J\pm}^{(2)}
    \;,
}
implying that $u_{J\pm}^{(2)}$ is proportional to $u_{J\pm}^{(1)}$. 
If this relation carries over to the Borel sums, it would imply $\psi_\pm^{(2)} \propto \psi_\pm^{(1)}$ for the EWKB basis solutions, corresponding to a diagonal connection matrix $T_{1\to2} = \mathrm{diag}(e^{-i\phi}, e^{i\phi})$.  
However, this result is in contradiction with the discussion of Stokes phenomenon above, where it was important that $T_{1\to2}$ can have off-diagonal entries.  
The paradox is resolved by recognizing that Eq.~\eqref{eq:relate_u1_u2} does not carry over to the Borel sums when $\eta_1$ and $\eta_2$ are in different Stokes regions.  
The impediment lies in the Laplace transform integration~\eqref{eq:Laplace_trans}, which runs into a singularity at a Stokes line.  
To be concrete, suppose that anchor times $\eta_1$ and $\eta_2$ are separated by a single Stokes line that runs between $\eta_c$ and $\eta_c^\ast$ and that crosses the real $\eta$ axis at $\eta_\ast$, as shown in Fig.~\ref{fig:terminology}.  
The connection matrix $T_{1\to2}$ relating EWKB basis solutions constructed at $\eta_1$ an $\eta_2$ can be written as~\cite{Peloso:2000hy}
\ba{\label{eq:Tast_def}
    T_{1\to2} = \mqty( e^{i \phi_2} & 0 \\ 0 & e^{-i \phi_2} ) T_\ast \mqty( e^{-i \phi_1} & 0 \\ 0 & e^{i \phi_1} ) 
    \;.
}
The first and third factors account for the analytic continuation of the Borel sums from $\eta_1$ to $\eta_\ast$ approaching from the left, and from $\eta_2$ to $\eta_\ast$ approaching from the right; they are phases given by $e^{-i \phi_1} = u_{J+}^{(1)}(\eta_{\ast L}) / u_{J+}^{(\ast L)}(\eta_{\ast L})$ and $e^{-i \phi_2} = u_{J+}^{(2)}(\eta_2) / u_{J+}^{(\ast R)}(\eta_2)$.  
The 2-by-2 matrix $T_\ast \in \mathrm{SU}(2)$ accounts for discontinuous change in the identification of positive and negative frequency modes at the Stokes line, and it is this factor that has nonzero off-diagonal entries.  
In the next section we use solutions of the Weber equation to calculate $T_\ast$.  

\begin{figure}[tb]
	\centering
	\includegraphics[width=1.0\textwidth]{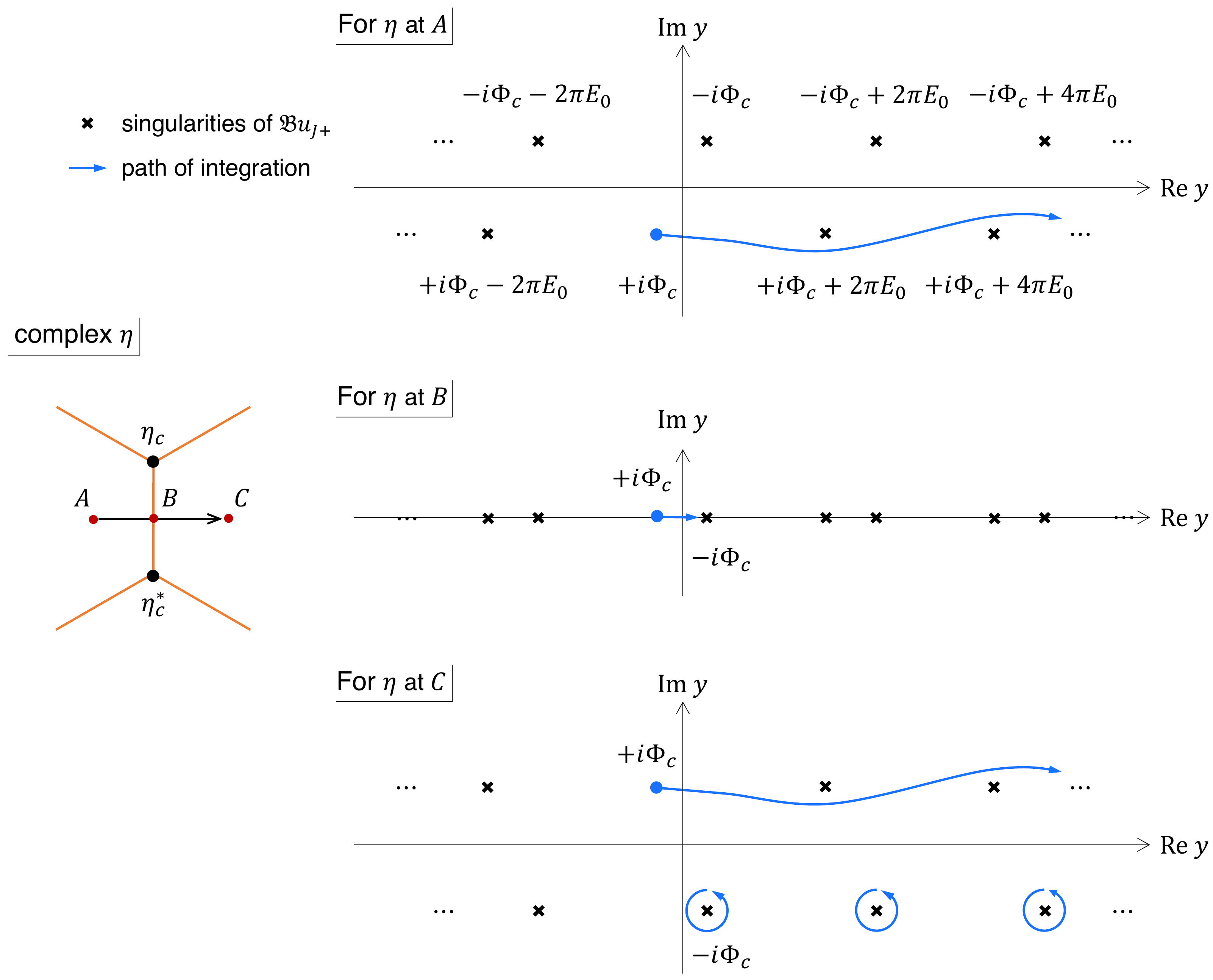}
	\caption{\label{fig:Borel}
	The emergence of a singularity in the Laplace transform upon crossing a Stokes line in the complex time plane.  This figure corresponds to the $u_{J+}$ case.  The Laplace transform~\eqref{eq:uA_uB_EWKB} involves an integral over $y$ from $+i \Phi_c(\eta)$ toward $\mathrm{Re} \, y = \infty$; the blue curve shows a path of integration.  The Borel transformed WKB solutions $\mathcal{B}u_{J+}$ \eqref{eq:Borel_trans} have an infinite number of singularities at $y = \pm i\Phi_c(\eta) + 2\pi n E_0$ ($n \in \mathbb{Z}$), where $E_0$ is real and positive~\cite{Takei:2008}.  For $\eta$ at point $A$ in the first Stokes region of the complex time plane, one queue of these singularities at $y = + i\Phi_c(\eta) + 2\pi n E_0$ lies in the direction of the path of integration, and the path is defined to avoid these poles~\cite{Aoki:1991}.  As $\eta$ is moved to $B$ on the Stokes line, the other queue of singularities at $y = - i\Phi_c(\eta) + 2\pi n E_0$ crosses the path of integration.  As $\eta$ is moved across the Stokes line toward point $C$ in the second Stokes region, the integral picks up the residue from the queue of poles.  Conversely if the integral were initially defined in the second Stokes region, these residues would not be included; this discontinuity is the Stokes phenomenon.  
	}
\end{figure}

Broadly speaking, the significance of Stokes lines can be understood in the following way.  
Suppose that the WKB series is constructed at anchor time $\eta_1$, and the EWKB basis solutions are defined by Borel resummation, which involves a Laplace transform~\eqref{eq:Laplace_trans} with an integral over complex variable $y$.  
The Borel sums can be analytically continued throughout the complex $\eta$ plane, and a Stokes line corresponds to a point at which the analytic structure changes in the complex $y$ plane.  
For concreteness, it is illuminating to consider a system with the Stokes structure shown in Fig.~\ref{fig:terminology}: a single pair of turning points at $\eta_c$ and $\eta_c^\ast$ are connected by a Stokes line that crosses the real $\eta$ axis at $\eta_\ast$.  
As we will see in the next subsection, for such systems the mode equation maps to the Weber equation where the analytic structure is known.  
The Borel transform of the WKB series~\eqref{eq:Borel_trans} has an infinite number of poles at $y = \pm i\Phi(\eta) + 2\pi n E_0$ where $n$ is an integer and $E_0$ is a real and positive constant~\cite{Takei:2008}. 
To calculate $u_{J+}(\eta,\hbar)$ the Laplace transform~\eqref{eq:Laplace_trans} integrates $y$ from $i \Phi(\eta)$ toward $\mathrm{Re} \, y = + \infty$, and the poles at $y = i\Phi(\eta) + 2\pi n E_0$ are avoided by an appropriate choice of the integration path~\cite{Aoki:1991}; see the top panel of Fig.~\ref{fig:Borel}.  
As $\eta$ is analytically continued throughout the complex $\eta$ plane, the poles move around in the complex $y$ plane.  
In particular, all the poles fall along the path of integration when~\cite{Delabaere:1993,Delabaere:1999}\footnote{Whereas $\Phi(\eta) = \int_{\eta_a}^\eta \! \dd{\eta^\prime} \, \omega(\eta^\prime)$ has the anchor time $\eta_a = \eta_1$ as its lower limit of integration, in Eq.~\eqref{eq:Stokes_line2} the Stokes line is defined by a different phase integral $\Phi_c(\eta)$ that has the turning point $\eta_c$ as its lower limit of integration.  This is because EWKB basis solutions at different points in the same Stokes region are related by a phase~\eqref{eq:Tast_def}.  For the arguments in this paragraph, it is convenient to first extract this phase by changing the anchor time to $\eta_\ast$ or $\eta_c$.  Then the poles in the Borel transform are located at $y = \pm i \Phi_c(\eta) + 2\pi n E_0$, and they cross when $\mathrm{Im}[ i \Phi_c(\eta) ] = 0$, which is identical to Eq.~\eqref{eq:Stokes_line2}.  } 
\ba{\label{eq:Stokes_line2}
    \Im\bigl[ i \Phi_c(\eta) \bigr] = 0 
    \qquad \text{where} \qquad 
    \Phi_c(\eta) \equiv \int_{\eta_c}^{\eta} \! \dd{\eta^\prime} \omega(\eta^\prime) 
    \;,
}
and this defines the location of a Stokes line~\eqref{eq:Stokes1st} in the complex $\eta$ plane.  
Analytically continuing $\eta$ across a Stokes line picks up the residues of the poles in the complex $y$ plane.
This is why the EWKB basis solutions constructed at anchor time $\eta_1$ and analytically continued to time $\eta_2$ in another Stokes region will differ (by more than just a phase) from the EWKB basis solutions constructed at anchor time $\eta_2$.  

\subsection{Derivation of the connection matrix via the Weber equation}\label{sec:Weber}

We seek to derive an expression for the connection matrix $T_\ast$ that relates solutions across a Stokes line.  
This task would be intractable if we wanted to study a general time dependence for $\Omega(\eta)$ in the mode equation.  
However, we are considering models of gravitational particle production that lead to a symmetric structure for the Stokes regions with two simple turning points connected by a Stokes line.  
The key insight, which makes it possible to derive the connection matrix, is that the mode equation can be mapped to the Weber equation near such a Stokes line~\cite{Enomoto:2020xlf}. 
Exact solutions to the Weber equation are available and by studying their asymptotic behavior, the connection matrix can be evaluated.  
This section contains the essential points of the derivation, and additional details are provided in appendices \ref{app:landau_zener} and \ref{app:connection_matrix}.  

The mode equation~\eqref{eq:mode_eqn_with_hbar} is a pair of coupled first-order differential equations.  
Taking another derivative yields a pair of decoupled second-order differential equations; see App.~\ref{app:wkb_solution} for details.  
We suppose that the entire complex-$\eta$ plane contains only a single pair of simple turning points, 
$\eta_c$ and $\eta_c^\ast$ with $\mathrm{Im} \, \eta_c > 0$.
They are connected by a Stokes line that divides the real $\eta$ axis into a left Stokes region and a right Stokes region.  
Note that since $\omega(\eta)$ is real along the real $\eta$-axis, the Schwarz reflection principle assures $\omega(\eta^\ast) = [\omega(\eta)]^\ast$ and hence turning points always appear as conjugate pairs.
Under these circumstances, Theorem~2.4 of Ref.~\cite{Aoki:1991} and Theorem~3.1 of Ref.~\cite{10.2977/prims/1195166424} ensures that the mode equation can be recast in the form of the Weber equation, 
\ba{\label{eq:Webereq}
    \left[ \hbar^2 \dv[2]{x} + E(\hbar) + \frac{x^2}{4} \right] w(x,\hbar) = 0 
    \;,
}
around the Stokes line. 
The function $E(\hbar) \equiv \sum_{n = 0}^\infty \hbar^n \, E_n$ is a power series in $\hbar$ that does not depend on $x$.  
The leading coefficient $E_0$ is given by 
\ba{\label{eq:Psi_def}
    \Psi \equiv \pi E_0 \equiv i \Phi_c(\eta_c^\ast) = i \int_{\eta_c}^{\eta_c^\ast} \! \dd{\eta} \, \omega(\eta) 
    \;,
}
which also defines the phase integral $\Psi$.  
Since $\omega(\eta^\ast) = [\omega(\eta)]^\ast$ and $\mathrm{Im} \, \eta_c > 0$, it follows that $E_0$ and $\Psi$ are real and positive. 
The coordinate $x = x(\eta, \hbar) = \sum_{n=0}^{\infty} \hbar^n x_n(\eta)$ is a holomorphic function of conformal time that satisfies $x_0(\eta_c) = 2i\sqrt{\Psi/\pi}$ and $x_0(\eta_c^\ast) = -2i\sqrt{\Psi/\pi}$ and that has the limiting behavior $x_0 \to \pm \infty$ for $\eta \to \pm \infty$ on the real axis.  

Solutions of the Weber equation can be written in terms of the parabolic cylinder functions $D_a(z)$~\cite{abramowitz+stegun}.
Each of the following functions solves the Weber equation~\eqref{eq:Webereq}: 
\ba{\label{eq:Weber_soln}
    w(x,\hbar) \ \propto \quad 
    D_{-\frac12 - i\xi}(z) \, , \quad
    D_{-\frac12 + i\xi}(-iz) \, , \quad 
    D_{-\frac12 - i\xi}(-z) \, , \quad
    D_{-\frac12 + i\xi}(iz) 
    \;,
}
where $\xi(\hbar) = E(\hbar)/\hbar$ and $z(x,\hbar) = e^{-3i\pi/4} x / \sqrt{\hbar}$. 
Any linear combination is also a solution, since the Weber equation is linear.

From the EWKB analysis of Sec.~\ref{sec:discussion}, solutions of the Weber equation can also be written as 
\ba{\label{eq:w_left_right}
    w(x,\hbar) = \begin{cases} \alpha^{(1)} \, w_+^{(1)} + \beta^{(1)} \, w_-^{(1)} & , \ \text{left Stokes region} \\ \alpha^{(2)} \, w_+^{(2)} + \beta^{(2)} \, w_-^{(2)} & , \ \text{right Stokes region} \end{cases}
    \;,
}
where $w_\pm(\eta) \propto e^{\mp i \Phi(\eta)/\hbar}$ are EWKB basis functions constructed from anchor times in the left and right Stokes regions; they are defined in analogy with Eq.~\eqref{eq:uJp_uJm_basis}.  
The left and right regions are related by 
\ba{\label{eq:w_connection_matrix}
    \mqty( \alpha^{(2)} \\ \beta^{(2)} ) = T_\ast \, \mqty( \alpha^{(1)} \\ \beta^{(1)} ) 
    \qquad \text{and} \qquad 
    \mqty( w_+^{(1)} \\ w_-^{(1)} ) = T_\ast^T \, \mqty( w_+^{(2)} \\ w_-^{(2)} )
}
where $T_\ast$ is the connection matrix that we seek to calculate, and $T_\ast^T$ is its transpose.  
Recall from Eq.~\eqref{eq:Tast_def} that $T_\ast$ is the discontinuous contribution to $T_{1\to 2}$ that connects a pair of anchor times that are just to the left and right of the Stokes line.

We are interested in how the basis of parabolic cylinder functions~\eqref{eq:Weber_soln} maps onto the basis of EWKB solutions~\eqref{eq:w_left_right}.  
To identify the matching relations, we investigate the asymptotic behavior at early times ($\eta,x \to -\infty$) and late times ($\eta,x \to +\infty$).  
The parabolic cylinder functions have asymptotic forms given by~\cite{abramowitz+stegun}
\bsa{eq:ParaCyn_asym}{
    D_{-\frac12 - i\xi}(z) 
    & \sim \begin{cases} 
    e^{-\frac{z^2}{4}} \, z^{-\frac12 - i\xi} 
    & , \quad \eta,x \to -\infty \\ 
    e^{-\frac{z^2}{4}} \, z^{-\frac12 - i\xi} + \frac{\sqrt{2\pi}}{\Gamma(\frac12 + i\xi)} e^{\frac{z^2}{4}} \, z^{-\frac12 + i\xi} \, e^{\frac32 i\pi-\pi\xi}
    & , \quad \eta,x \to +\infty 
    \end{cases} \\ 
    D_{-\frac12 + i\xi}(-iz) 
    & \sim \begin{cases} 
    e^{\frac{z^2}{4}} \, z^{-\frac12 + i\xi} \, e^{\frac14 i\pi + \frac{1}{2}\pi\xi} 
    & , \quad \eta,x \to -\infty \\ 
    e^{\frac{z^2}{4}} \, z^{-\frac12 + i\xi} \, e^{-\frac34 i\pi - \frac{3}{2}\pi\xi} + \frac{\sqrt{2\pi}}{\Gamma(\frac12 - i\xi)} e^{-\frac{z^2}{4}} \, z^{-\frac12 - i\xi} e^{-\frac14 i\pi + \frac{1}{2}\pi\xi} 
    & , \quad \eta,x \to +\infty 
    \end{cases} \\ 
    D_{-\frac12 - i\xi}(-z) 
    & \sim \begin{cases} 
    e^{-\frac{z^2}{4}} \, z^{-\frac12 - i\xi} \, e^{\frac12 i\pi-\pi\xi} + \frac{\sqrt{2\pi}}{\Gamma(\frac12 + i\xi)} \, e^{\frac{z^2}{4}} \, z^{-\frac12 + i\xi}
    & , \quad \eta,x \to -\infty \\ 
    e^{-\frac{z^2}{4}} \, z^{-\frac12 - i\xi} \, e^{\frac32 i\pi + \pi\xi} 
    & , \quad \eta,x \to +\infty 
    \end{cases} \\ 
    D_{-\frac12 + i\xi}(iz) 
    & \sim \begin{cases} 
    e^{\frac{z^2}{4}} \, z^{-\frac12 + i\xi} \, e^{-\frac14 i\pi - \frac{1}{2}\pi\xi} + \frac{\sqrt{2\pi}}{\Gamma(\frac12 - i\xi)} e^{-\frac{z^2}{4}} \, z^{-\frac12 - i\xi} \, e^{\frac14 i\pi - \frac{1}{2}\pi\xi}
    & , \quad \eta,x \to -\infty \\ 
    e^{\frac{z^2}{4}} \, z^{-\frac12 + i\xi} \, e^{-\frac14 i\pi - \frac{1}{2}\pi\xi} 
    & , \quad \eta,x \to +\infty 
    \end{cases} 
    \;.
}
Notice that a factor of $e^{-z^2/4} = e^{-i x^2/4\hbar}$ or $e^{z^2/4} = e^{i x^2/4\hbar}$ appears in each expression.  
These factors are the analogs of $e^{\pm i \Phi/\hbar}$, which distinguish positive and negative frequency solutions~\eqref{eq:uA_uB_solution}.  
At early times ($\eta,x \to -\infty$) the correspondence is $-x^2/4 \leftrightarrow \Phi$, since both are increasing functions of time, and $e^{z^2/4} \leftrightarrow e^{-i \Phi/\hbar}$ is the positive frequency mode; whereas, at late times ($\eta,x \to +\infty$) the correspondence is $x^2/4 \leftrightarrow \Phi$, and $e^{-z^2/4} \leftrightarrow e^{-i \Phi/\hbar}$ is the positive frequency mode.  
In this way, the EWKB basis solutions are matched to the parabolic cylinder functions: 
\bsa{eq:w_to_ParaCyn}{
    w_+^{(1)}(x,\hbar) & = \lambda_+^{(1)}(\hbar) \, D_{-\frac{1}{2} + i\xi}(-iz) \\ 
    w_-^{(1)}(x,\hbar) & = \lambda_-^{(1)}(\hbar) \, D_{-\frac{1}{2} - i\xi}(z) \\ 
    w_+^{(2)}(x,\hbar) & = \lambda_+^{(2)}(\hbar) \, D_{-\frac{1}{2} - i\xi}(-z) \\ 
    w_-^{(2)}(x,\hbar) & = \lambda_-^{(2)}(\hbar) D_{-\frac{1}{2} + i\xi}(iz) 
    \;,
}
where the $\lambda$ factors are proportionality constants that do not depend on $x$. To calculate the $\lambda$'s, we can evaluate the functions at $x=z=0$, which gives us:
\bsa{}{
    \lambda_\pm^{(1)}(\hbar) 
    & = \frac{2^{\frac{1}{4} \mp \frac{i\xi}{2}}}{\sqrt{\pi}} \, \Gamma\Bigl(\frac{3}{4} \mp \frac{i\xi}{2}\Bigr) \, w_\pm^{(1)}(0,\hbar) \\
    \lambda_\pm^{(2)}(\hbar) 
    & = \frac{2^{\frac{1}{4} \pm \frac{i\xi}{2}}}{\sqrt{\pi}} \, \Gamma\Bigl(\frac{3}{4} \pm \frac{i\xi}{2}\Bigr) \, w_\pm^{(2)}(0,\hbar) 
    \;.
}
The $w(0,\hbar)$'s are fixed by our chosen normalization. See App.~\ref{app:connection_matrix} for additional details. 

Finally, we are prepared to calculate the connection matrix.  
The four parabolic cylinder functions are not linearly independent, but rather they can be expressed as linear combinations of one another~\cite{abramowitz+stegun}.  
This observation leads to the following relations between the EWKB basis solutions on either side of the Stokes line:\footnote{This matching of asymptotic behaviors of the parabolic cylinder functions has also been used to study preheating of spin-1/2 fermions with an effectively time-dependent mass~\cite{Chung:1999ve,Peloso:2000hy}.}
\ba{\label{eq:w_relation}
    \mqty(w_+^{(1)} \\ w_-^{(1)}) 
    = \mqty(
    \frac{\sqrt{2\pi}}{\Gamma(\frac{1}{2}-i\xi)} \, e^{\frac14 i\pi - \frac12 \pi\xi} \, \frac{\lambda_+^{(1)}}{\lambda_+^{(2)}} & 
    -i e^{-\pi\xi} \, \frac{\lambda_+^{(1)}}{\lambda_-^{(2)}} \\
    i e^{-\pi\xi} \, \frac{\lambda_-^{(1)}}{\lambda_+^{(2)}} & 
    \frac{\sqrt{2\pi}}{\Gamma(\frac{1}{2}+i\xi)} \, e^{-\frac14 i\pi - \frac12 \pi\xi} \, \frac{\lambda_-^{(1)}}{\lambda_-^{(2)}} 
    ) 
    \mqty(w_+^{(2)} \\ w_-^{(2)}) 
    \;.
}
Comparing with Eq.~\eqref{eq:w_connection_matrix} lets us identify the connection matrix $T_\ast$ as 
\ba{\label{eq:T_exact}
    T_\ast = \mqty(
    \frac{\sqrt{2\pi}}{\Gamma(\frac{1}{2}-i\xi)} \, e^{\frac14 i\pi - \frac12 \pi\xi} \, \frac{\lambda_+^{(1)}}{\lambda_+^{(2)}} & 
    i e^{-\pi\xi} \, \frac{\lambda_-^{(1)}}{\lambda_+^{(2)}} \\
    -i e^{-\pi\xi} \, \frac{\lambda_+^{(1)}}{\lambda_-^{(2)}} & 
    \frac{\sqrt{2\pi}}{\Gamma(\frac{1}{2}+i\xi)} \, e^{-\frac14 i\pi - \frac12 \pi\xi} \, \frac{\lambda_-^{(1)}}{\lambda_-^{(2)}} 
    ) 
    \;.
}
Eq.~\eqref{eq:T_exact} is the main result of this section.
Here we have derived an \textit{exact} expression for the connection matrix $T_\ast$ linking the $\alpha,\beta$ coefficients of EWKB basis solutions in adjacent Stokes regions bordering a Stokes line that runs between turning points at $\eta_c$ and $\eta_c^\ast$.  
Although this expression for $T_\ast$ was derived by considering solutions of the Weber equation, it has broader applicability, since we show in appendices~\ref{app:wkb_solution} and \ref{app:connection_matrix} how to recast the mode equations for spin-1/2 and spin-3/2 fields (with a conjugate pair of turning points) into the Weber equation.  

To compute $T_\ast$ for a concrete example, we need to derive $\xi$ and the $\lambda$ ratios in terms of the model variables $M(\eta)$ and $K(\eta)$. 
For a system that remains nearly adiabatic, meaning that $M(\eta)$ and $K(\eta)$ vary slowly such that $\left| \mu(\eta) / \omega(\eta) \right| \ll 1$, it is a good approximation to keep only the lowest orders in $\hbar$.  
In App.~\ref{app:connection_matrix}, we derive the $\lambda$ ratios, show that $\xi = E / \hbar = \Psi / (\pi \hbar) + \order{\hbar^0}$, argue that $\mathrm{Im} \, \xi = - 1/2$, and present $T_*$ to all perturbative orders in $\hbar$. See Eq.~\eqref{eq:connection_matrix_full} for the full result.
Keeping only the leading-order terms, we find that
\ba{\label{eq:con_mat}
    T_\ast \approx \mqty(\sqrt{1 - e^{-2 \Psi / \hbar}} & - i e^{i \vartheta} e^{-\Psi / \hbar} \\ 
     - i e^{-i \vartheta} e^{-\Psi / \hbar} & \sqrt{1 - e^{-2 \Psi / \hbar}}
    ) 
    \;,
}
where $\Psi = i \int_{\eta_c}^{\eta_c^\ast} \! \dd{\eta} \, \omega(\eta)$ was defined in Eq.~\eqref{eq:Psi_def} and where $\vartheta$ is an unimportant phase factor.
This expression is manifestly $T_\ast \in \mathrm{SU}(2)$.  
In particular, if we take $\alpha^{(1)} = 1$ and $\beta^{(1)} = 0$ in the early-time Stokes region, then we find 
\ba{\label{eq:Stokes_beta}
    \beta^{(2)} = - i e^{- \pi \xi} \, \frac{\lambda_+^{(1)}}{\lambda_-^{(2)}} \approx - i e^{-i \vartheta} e^{-\Psi/\hbar}
}
in the late-time Stokes region.  
We will use this relation in the next subsection to evaluate the spectrum of gravitationally produced particles.  

In more general situations, not only one but multiple pairs of turning points may appear.
However, as long as they are well-separated from each other, one can still reduce the mode equation to the Weber equation by focusing on a neighborhood of one Stokes line connecting one of those turning point pairs.

\subsection{Evaluating Bogoliubov coefficients for cosmological GPP}\label{sec:Bogoliubov}

To study cosmological gravitational particle production, we need to solve the mode equation~\eqref{eq:mode_eqn} for each Fourier mode of the gravitationally-produced field.  
In general the Fourier modes are indexed by a comoving wavevector $\kvec$, but since the initial condition and spacetime are spatially isotropic, modes with the same comoving wavenumber $k = |\kvec|$ have the same mode functions, i.e.~$\alpha_\kvec(\eta) = \alpha_k(\eta)$ and $\beta_\kvec(\eta) = \beta_k(\eta)$.  
The initial condition for cosmological GPP is evaluated at an asymptotically early time $\eta_0 = - \infty$, corresponding to the period in our cosmic history known as cosmological inflation.  
Here we assume the \textit{Bunch-Davies} initial condition~\cite{Bunch:1978yq}, which corresponds to setting $\alpha_k^{(0)} = 1$ and $\beta_k^{(0)} = 0$ at $\eta_0 = - \infty$ for the modes with comoving wavenumber $k = |\kvec|$ in the de Sitter spacetime. 

For the systems of interest, the Stokes region in the complex $\eta$ plane containing the asymptotic past $\eta \to - \infty$ will be different from the Stokes region containing the asymptotic future $\eta \to + \infty$, since they will be separated by one or more Stokes lines.  
The observables of interest correspond to the coefficients $\beta_k$ of the modes with wavenumber $k = |\kvec|$ and with negative frequency, i.e.~the mode function is $\propto e^{i\Phi}$, in the region containing the asymptotic future. 
These $\beta_k$'s are the Bogoliubov coefficients linking the Bunch-Davies vacuum state in the asymptotic past to the adiabatic vacuum state in the asymptotic future.  

The Bogoliubov coefficients $\beta_k$ are related to the abundance of particles that result from GPP.  
If $\dd{n}(k,\eta)$ is the (physical) number density of particles per degree of freedom with comoving wavenumber between $k$ and $k + \dd{k}$ at conformal time $\eta$, then~\cite{Parker:2009uva}
\ba{\label{eq:number_density}
    \dd{n}(k,\eta) = n_k(\eta) \, \frac{\dd{k}}{k} 
    \qquad \text{where} \qquad 
    n_k(\eta) = a(\eta)^{-3} \frac{k^3}{2\pi^2} \, |\beta_k|^2 
    \;.
}
These expressions assume that the coeffcients $\alpha_k$ and $\beta_k$ are independent of the wavevector's orientation, and only depend upon its magnitude $k = |\kvec|$, which is the case for the spatially isotropic cosmologies that we study.
Note that $\dd{n}$ is the number density per degree of freedom. To account for all degrees of freedom, one should sum up the contribution from the positive/negative helicity states and the particle/antiparticle states, which means multiplying $n_k$ by 4 for a Dirac model or 2 for a Majorana model.

The Bogoliubov coefficients $\alpha_k$ and $\beta_k$ for the region containing the asymptotic future are derived from the corresponding coefficients in the asymptotic past (Bunch-Davies initial condition) by applying a sequence of connection rules \eqref{eq:connection_matrix} at each intervening Stokes line.  
For the example spacetimes that we study in the next sections, there will only be one relevant Stokes line connecting a single pair of turning points at complex times $\eta_c$ and $\eta_c^\ast$. 
For such systems, we can apply the connection rule \eqref{eq:con_mat} and evaluate the Bogoliubov coefficient as in Eq.~\eqref{eq:Stokes_beta}.
Then the squared Bogoliubov coefficient, which appears in the number density~\eqref{eq:number_density}, evaluates to~\cite{Voros:1983,10.2307/51774}\footnote{In Ref.~\cite{Chung:1998bt} a similar formula (up to a factor of $\pi^2/9 \simeq 1.1$) is derived by approximating $\alpha_k \approx 1$ and applying the steepest descent method to integrate Eq.~\eqref{eq:dalpha_dbeta}. See also Ref.~\cite{Enomoto:2020xlf} for the comparison between the Stokes phenomenon method and the steepest descent method.}
\begin{align}\label{eq:formula}
	|\beta_k|^2 
	\approx e^{-2\Psi_k} = \exp\left( -2i \int_{\eta_c}^{\eta_c^\ast} \! \dd{\eta} \, \omega_k(\eta) \right)
	\;,
\end{align}
where we have set $\hbar = 1$.  
Since $\Psi_k > 0$, this result is consistent with the Pauli exclusion principle $|\beta_k|^2 < 1$. 
From this expression, we observe that a larger $\mathrm{Im}\,\eta_c$ leads to a larger phase integral and a smaller $|\beta_k|^2$, corresponding to less GPP.  
In the following sections we employ Eq.~\eqref{eq:formula} to calculate fermionic GPP for a couple of example models. 

\section{Spin-1/2 Dirac field}\label{sec:spin1/2}

In this section, we first present the well-known mode equations for a Dirac spinor field in an FRW spacetime.  
Then we use the general results on Stokes phenomenon from Sec.~\ref{sec:formalism} to study gravitational particle production due to the field's effectively time-dependent mass.  

\subsection{Mode equations on an FRW background}\label{sec:spin1/2_FRW}

The quantum excitations of a Dirac spinor field correspond to spin-1/2 particles and antiparticles.
A Dirac spinor field $\chi(x)$ with a mass $m$ in a curved spacetime is described by the action~\cite{Parker:2009uva,Freedman:2012zz,Herring:2020cah}
\begin{align}\label{eq:dirac_action_curved}
	S 
	= \int \! \dd[4]{x} \sqrt{-g} \, \biggl( \frac{i}{2} \bar{\chi} \underline{\gamma}^\mu (\nabla_\mu \chi) - \frac{i}{2} (\nabla_\mu \bar{\chi}) \underline{\gamma}^\mu \chi - m\bar{\chi}\chi \biggr) 
\end{align}
where $\bar{\chi} \equiv \chi^\dag \gamma^0$ is the Dirac conjugate field.  
The metric determinant $g(x)$, the local gamma matrix $\underline{\gamma}^\mu(x)$, and the covariant derivative $\nabla_\mu$ depend on the metric $g_{\mu\nu}(x)$.  

In an FRW spacetime, $\dd{s}^2 = a^2(\eta)[\dd{\eta}^2 - \dd{x}^2 - \dd{y}^2 - \dd{z}^2]$ with scale factor $a(\eta)$ at conformal time $\eta$, the field equation takes the form 
\begin{align}\label{eq:Dirac_eom}
	\bigl( i \gamma^\mu \partial_\mu - am \bigr) \bigl( a^{3/2} \chi \bigr) = 0
	\;,
\end{align}
where $\gamma^\mu$ (without the underline) is the usual gamma matrix in Minkowski space and $\partial_\mu = \partial/\partial x^\mu$.  
Notice that the nonzero mass $m \neq 0$ is a source of conformal symmetry breaking.  
Solutions of this equation can be written as~\cite{Chung:2011ck} 
\begin{align}\label{eq:fermi_mode}
	\chi(\xvec,\eta) = a(\eta)^{-3/2} \int \! \! \frac{\dd[3]{k}}{(2\pi)^{3/2}} \sum_{s = \pm} \biggl( a_{\kvec, s} \, U_{\kvec, s}(\eta) \, e^{i \kvec \cdot \xvec} + b_{\kvec, s}^\dag \, V_{\kvec, s}(\eta) \, e^{-i\kvec \cdot \xvec} \biggr)
	\;,
\end{align}
where $a_{\kvec,s}$ and $b_{\kvec,s}$ are ladder operators for particle states and antiparticle states, respectively.  
The time-dependent spinor wavefunctions can be further decomposed as~\cite{Chung:2011ck} 
\bsa{eq:decomp_2spinor}{
	U_{\kvec, s}(\eta) 
	& = \begin{pmatrix}
		u_{A, k} (\eta) \\
		s \, u_{B, k} (\eta)
    \end{pmatrix}
    \otimes h_{\hat{\kvec}, s} \\ 
	V_{\kvec, s}(\eta) 
	& = -i \gamma^2 \, U_{\kvec, s}^\ast (\eta), 
}
where $k = |\kvec|$ is the comoving wavenumber, $\hat{\kvec} \equiv \kvec/k$ is the unit vector, and $h_{\hat{\kvec}, s}$ is the two-component eigenspinor of the helicity operator.  
The c-number mode functions, $u_{A,k}(\eta)$ and $u_{B,k}(\eta)$, are constrained to satisfy the normalization condition $|u_{A,k}(\eta)|^2 + |u_{B,k}(\eta)|^2 = 1$, which is a consequence of the quantization conditions on the field and ladder operators.  
Using this ansatz, the Dirac equation becomes~\cite{Chung:2011ck} 
\ba{\label{eq:Dirac_eq1}
    i \begin{pmatrix}
		u_{A, k}^\prime \\
		u_{B, k}^\prime
	\end{pmatrix}
	= \begin{pmatrix}
		a(\eta) m & k \\
		k & -a(\eta) m
	\end{pmatrix}
	\begin{pmatrix}
		u_{A, k} \\
		u_{B, k}
	\end{pmatrix}
	\;,
}
which is the mode equation of the spin-$1/2$ Dirac spinor field.
We see that Eq.~\eqref{eq:Dirac_eq1} has the same form as the general mode equation that we studied earlier \eqref{eq:mode_eqn}, and it corresponds to setting $M(\eta) = a(\eta) m$ and $K(\eta) = k$.  
The modes evolve in response to an effectively time-dependent mass $a(\eta) m$, which is the source of non-adiabatic particle production.

\subsection{GPP from an effectively time-dependent mass}\label{eq:spin1/2_production}

In this subsection, we provide a concrete example of gravitational particle production for spin-$1/2$ particles in an FRW spacetime.  
In order to assess the reliability of the analytic Stokes phenomenon calculation, we evaluate the number density of gravitationally produced particles using this formalism, and we compare that with the density obtained by numerically solving the Dirac field's mode equation~\eqref{eq:Dirac_eq1}.  

We study GPP in an FRW spacetime with scale factor 
\begin{align}\label{eq:a_for_spin1/2}
    a(\eta) = \frac{a_0}{2} \biggl( 1 + \tanh{\frac{\eta}{\tau}} \biggr) 
    \;,
\end{align}
where $0 < a_0$ and $0 < \tau$.  
Notice that $a(\eta)$ smoothly evolves from $a(-\infty) = 0$ to $a(+\infty) = a_0$, where the transition region is localized to a time window $- \tau \lesssim \eta \lesssim \tau$.  
This spacetime does not correspond to a realistic cosmology.  
Nevertheless, the localization of nonadiabaticity to a finite time window is a feature that it shares with most inflationary cosmologies wherein the transition from the quasi-de Sitter phase to a matter- or radiation-dominated phase after inflation is charaterized by non-adiabatic mode evolution~\cite{Chung:1998zb}.  
We use this toy model to study how fermion GPP depends on the duration of the non-adiabatic period, and we expect a similar qualitative behavior for realistic inflationary cosmologies.  

Drawing upon the analysis of Sec.~\ref{sec:instantaneous}, solutions of \eqref{eq:Dirac_eq1} can be written as~\eqref{eq:uA_uB_solution} 
\bes{\label{eq:Dirac_soln_1}
    & \mqty(u_{A,k}(\eta) \\ u_{B,k}(\eta))
	= \alpha_k(\eta) \, \tilde{\psi}_{+,k}(\eta) + \beta_k(\eta) \, \tilde{\psi}_{-,k}(\eta) \\ 
	& \quad \text{where} \quad 
	\tilde{\psi}_{\pm,k}(\eta) = \frac{1}{\sqrt{2}} e^{\mp i\Phi_k}
	\begin{pmatrix}
		\pm\sqrt{1 \pm \frac{am}{\omega_k}} \\
		\sqrt{1 \mp \frac{am}{\omega_k}}
	\end{pmatrix}
	\;,
}
where we have used $K(\eta) = k$, which is a real constant; we have defined  $\omega_k(\eta) = \sqrt{k^2 + a(\eta)^2 m^2}$, which is the effective time-dependent angular frequency; and we have written the phase integral~\eqref{eq:Phi_def} as $\Phi_k(\eta) \equiv \int_{\eta_a}^\eta \! \dd{\eta^\prime} \omega_k(\eta^\prime)$, which depends on the arbitrary anchor time $\eta_a$.  
This relation trades the rapidly-varying mode functions $u_{A,k}(\eta)$ and $u_{B,k}(\eta)$ for the slowly-varying mode functions $\alpha_k(\eta)$ and $\beta_k(\eta)$ that are required to obey Eq.~\eqref{eq:dalpha_dbeta}.  
Given initial condition $\alpha_k(\eta_0) = \alpha_{k,0}$ and $\beta_k(\eta_0) = \beta_{k,0}$, where $\abs{\alpha_{k,0}}^2 + \abs{\beta_{k,0}}^2 = 1$, the solution will satisfy $\abs{\alpha_k(\eta)}^2 + \abs{\beta_k(\eta)}^2 = 1$ at all $\eta\in\mathbb{R}$. 
To study cosmological GPP, we choose the initial condition $\alpha_k(-\infty)=1$ and $\beta_k(-\infty)=0$. 
As the conformal time $\eta$ evolves from $-\infty$ to $+\infty$, $\abs{\beta_k(\eta)}^2$ increases from $0$ and converges to a positive constant; this positive constant is what we need to determine the particle number density.

On the other hand, the EWKB analysis of Secs.~\ref{sec:EWKB}~and~\ref{sec:discussion} provided an alternative expression for solutions of the mode equation.  
In the $i^\mathrm{th}$ Stokes region, solutions of \eqref{eq:Dirac_eq1} can be written as a linear combination of the local EWKB basis solutions~\eqref{eq:uJp_uJm_basis}
\bes{
\label{eq:Dirac_soln_2}
    \mqty(u_{A,k}^{(i)}(\eta) \\ u_{B,k}^{(i)}(\eta)) 
    & = \alpha_k^{(i)} \, \psi_{+,k}^{(i)}(\eta) + \beta_k^{(i)} \, \psi_{-,k}^{(i)}(\eta) 
    \;,
}
where the complex coefficients, $\alpha^{(i)}_k$ and $\beta^{(i)}_k$, satisfy $|\alpha^{(i)}|^2 + |\beta^{(i)}|^2 = 1$. 
The coefficients in different Stokes regions are related by an $\mathrm{SU}(2)$ connection matrix, which can be determined via the Stokes phenomenon calculation described in Sec.~\ref{sec:discussion} and App.~\ref{app:connection_matrix}. 
The Stokes regions containing $\eta \to -\infty$ and $\eta \to +\infty$ are called the early and late regions, respectively.  
The initial condition corresponds to $\alpha^{(\mathrm{early})} = 1$ and $\beta^{(\mathrm{early})} = 0$ in the early region.  
The quantity of interest for cosmological GPP is $\beta^{(\mathrm{late})}$, which can be evaluated using the connection matrix between the early and late regions.  

We have described two schemes for evaluating exact solutions of the Dirac field's mode equation~\eqref{eq:Dirac_eq1}. 
When $a(\eta)$ is slowly-varying, $|\mu(\eta)| \ll \omega(\eta)$ in Eq.~\eqref{eq:dalpha_dbeta}, $\alpha_k(\eta)$ and $\beta_k(\eta)$ are approximately constant, and the leading-order WKB solution~\eqref{eq:leading_order_wkb} is a good approximation to the exact EWKB basis solution, 
\ba{
    \psi_{\pm,k}^{(i)}(\eta) \approx \tilde{\psi}_{\pm,k}(\eta)
    \;.
}
At such times, we can match the two schemes by equating Eqs.~\eqref{eq:Dirac_soln_1}~and~\eqref{eq:Dirac_soln_2} to find $|\alpha_k(\eta)| \simeq |\alpha^{(i)}_k|$ and $|\beta_k(\eta)| \simeq |\beta^{(i)}_k|$. 
Going back to our choice \eqref{eq:a_for_spin1/2} for $a(\eta)$, it is clear that $a(\eta)$ is slow-varying at both early ($\eta < - \tau$) and late ($\eta > \tau$) times. 
Matching the two schemes at late times tells us
\ba{
    |\beta^{(\mathrm{late})}_k|^2 
    = \lim_{\eta \to \infty} \abs{\beta_k(\eta)}^2 
    \;.
}
This relationship provides the first of two approaches that we will use to evaluate $|\beta^{(\mathrm{late})}_k|^2$.  
The mode equations~\eqref{eq:dalpha_dbeta} for $\alpha_k(\eta)$ and $\beta_k(\eta)$ can be solved using numerical techniques; the initial condition is imposed at a finite, early time $\eta \ll - \tau$.  
In practice, the numerical value of $|\beta_k(\eta)|^2$ converges to a constant fairly quickly once $\eta \gtrsim \tau$, and that constant is taken to be our numerical estimate of $|\beta^{(\mathrm{late})}_k|^2$.  

\begin{figure}[t]
	\centering
	\includegraphics[width=.6\textwidth]{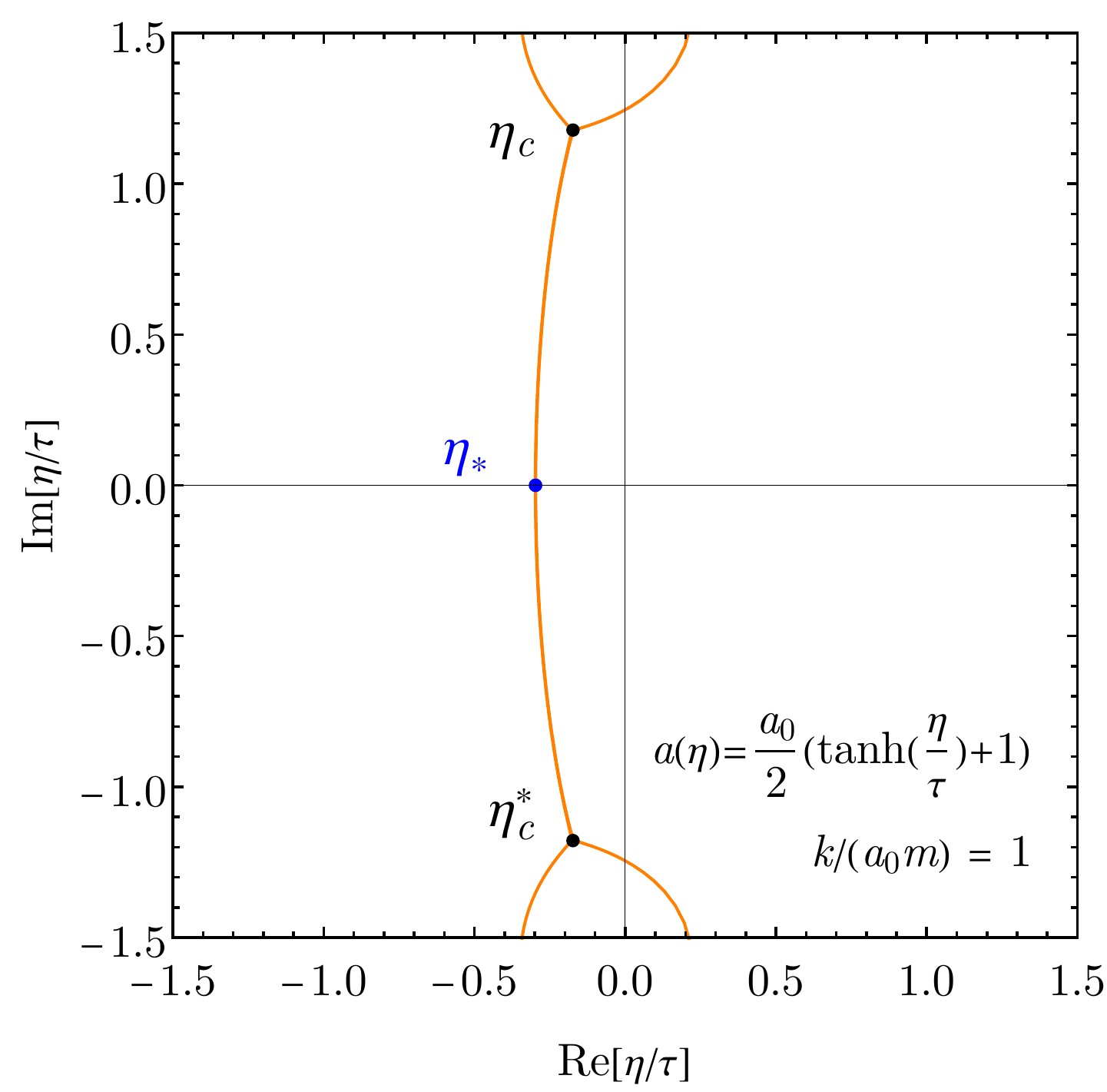}
	\caption{\label{fig:dirac_stokes} 
	An illustration of the complex-$\eta$ plane showing turning points $\eta_c$ and $\eta_c^\ast$ (black dots), crossing point $\eta_*$ (blue dot) and Stokes lines (orange curves) close to the real axis.  Additional turning points are found at larger values of $|\mathrm{Im}[\eta/\tau]|$.  The early Stokes region containing $\eta \to -\infty$ and the late Stokes region containing $\eta \to +\infty$ share a common border at the central Stokes line shown here.  The real axis intersects this Stokes line at the blue dot.  
	}
\end{figure}

As a second approach, we evaluate $|\beta^{(\mathrm{late})}_k|^2$ using the analytic Stokes phenomenon analysis.  
The structure of the Stokes regions is controlled by the turning points and Stokes lines, which are shown in Fig.~\ref{fig:dirac_stokes}.  
The turning points are solutions of $0 = \omega_k^2(\eta) = k^2 + a_0^2 m^2 (1 + \tanh \eta/\tau)^2 / 4$ in the complex-$\eta$ plane.  
Along the real-$\eta$ axis, the early region containing $\eta \to -\infty$ and late region containing $\eta \to +\infty$ are separated by a single Stokes line that runs between a pair of turning points at 
\ba{
    \eta_c = \tau \ \,\mathrm{arctanh}\biggl( -1 + \frac{2 i k}{a_0 m} \biggr)
}
and $\eta_c^\ast$.  
In the early region the initial condition is implemented by taking $\alpha_k^{(\mathrm{early})} = 1$ and $\beta_k^{(\mathrm{early})} = 0$.  
A connection matrix $T_k$ relates the coefficients in the early and late regions across the single Stokes line: 
\ba{
    \mqty(\alpha_k^{(\mathrm{late})} \\ \beta_k^{(\mathrm{late})}) 
    = T_k \mqty(\alpha_k^{(\mathrm{early})} \\ \beta_k^{(\mathrm{early})}) 
    \approx \mqty(\sqrt{1 - e^{-2\Psi_k}} \\ e^{-\Psi_k} )
    \;,
}
where we have dropped irrelevant phases.
A formal expression for the exact connection matrix appears in Eq.~\eqref{eq:T_exact}, and here we have written the approximate expression from Eq.~\eqref{eq:con_mat}, where the phase integral $\Psi_k$ is defined by Eq.~\eqref{eq:Psi_def}, and it evaluates to 
\ba{\label{eq:dirac_phase_integral}
    \Psi_k = i \int_{\eta_c}^{\eta_c^*} \! \dd{\eta} \, \omega_k(\eta)
    = \frac{\pi}{2} \tau \Bigl( \sqrt{k^2 + a_0^2 m^2} + k - a_0 m \Bigr)
    \;;
}
see App.~\ref{app:spin_1/2_phase_integral} for a derivation of $\Psi_k$. 
Finally, the analytic Stokes phenomenon analysis yields the result~\eqref{eq:formula} 
\ba{\label{eq:Dirac_betasq_Stokes}
    |\beta_k^{(\mathrm{late})}|^2 
    \approx e^{-2 \Psi_k}
    = \mathrm{exp}\Bigl[ - \pi \tau \Bigl( \sqrt{k^2 + a_0^2 m^2} + k - a_0 m \Bigr) \Bigr]
}
for the squared coefficient of the negative frequency mode in the late region.  
The approximations above are reliable when the mode evolution remains nearly adiabatic, which corresponds to $|\mu_k(\eta)/\omega_k(\eta)| \ll 1$ for real time $\eta$.  
The non-adiabaticity $\mu_k(\eta)$ is maximal for $\eta = \mathrm{Re} \, \eta_c = (\tau/4) \log[ k^2 / (k^2 + a_0^2 m^2) ]$, when it evaluates to 
\ba{
    | \mu_k(\mathrm{Re} \, \eta_c) | \, \tau  
    = \frac{a_0 m}{8 k} \left( 1 + \frac{k}{\sqrt{k^2 + a_0^2 m^2}} \right) \, \left( \mathrm{sech} \log \sqrt{\frac{k}{\sqrt{k^2 + a_0^2 m^2}}} \right)^2 
    \;.
}
For modes with high wavenumber, this ratio decreases as $k^{-1}$, meaning that the evolution of these modes remains more nearly adiabatic, and the approximate expressions above should be reliable.  
For low-wavenumber modes, this ratio goes to a constant $1/2$, and we expect the adiabatic approximation to be fairly reliable.  

\begin{figure}[t]
	\centering
	\includegraphics[width=.49\textwidth]{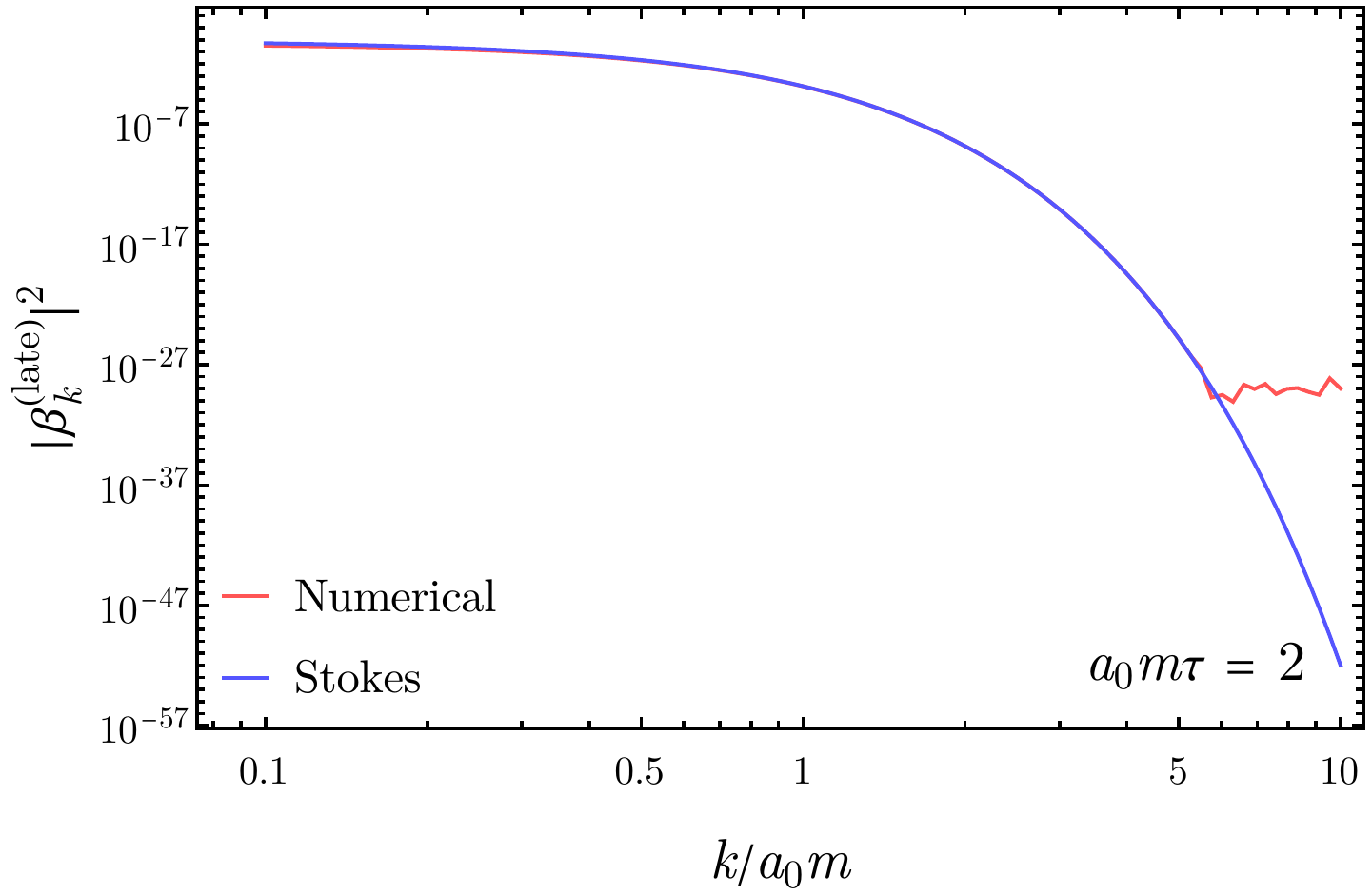}	
	\includegraphics[width=.49\textwidth]{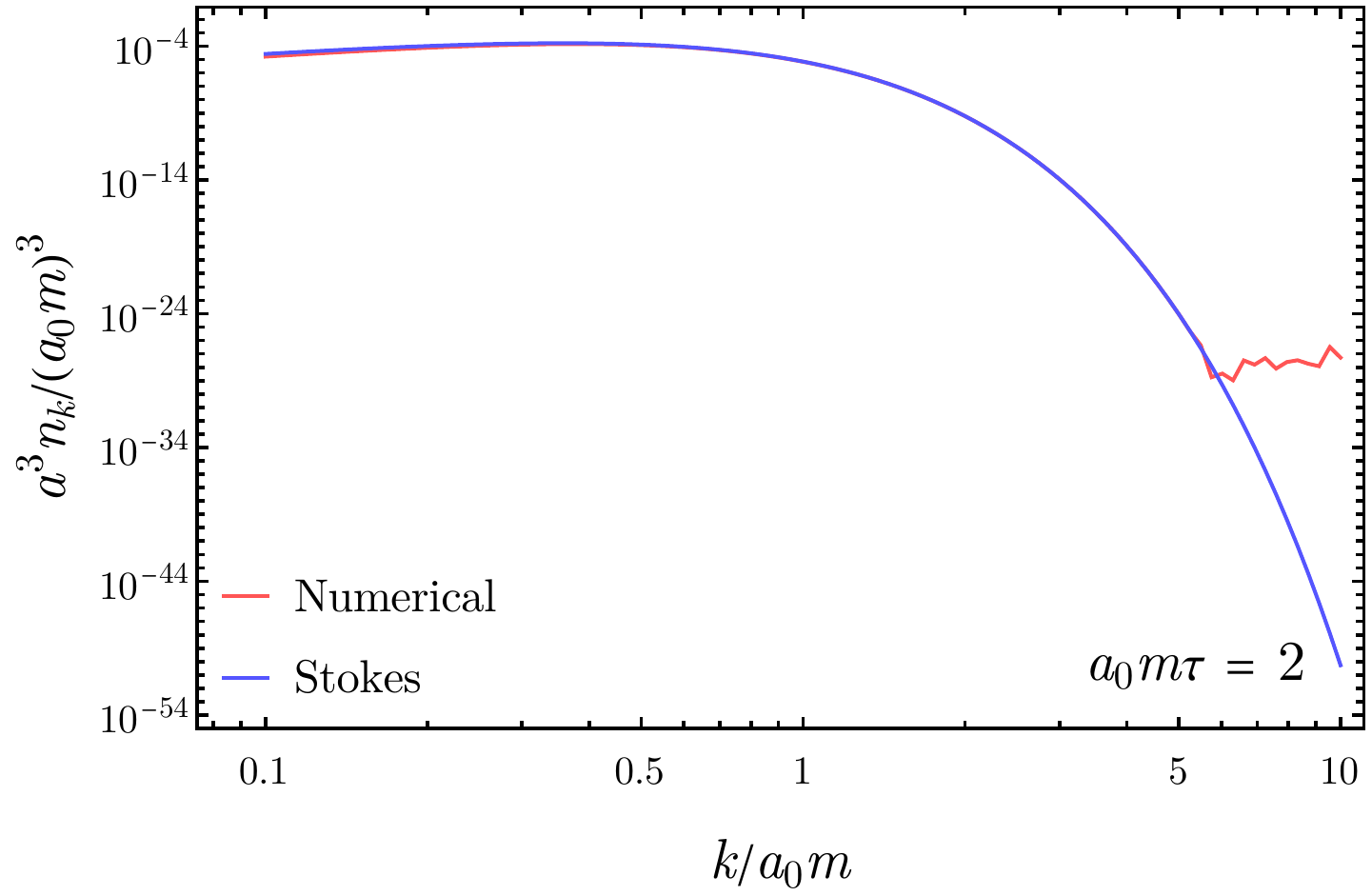}
	\caption{\label{fig:dirac_spectrum} 
	A comparison of GPP calculations via direct numerical integration of the mode equations (red) and analytical analysis with Stokes phenomenon (blue).  
	The left panel shows the squared Bogoliubov coefficient $|\beta_k^\mathrm{(late)}|^2$, the right panel shows the comoving particle number density (per d.o.f) spectrum $a^3 n_k / (a_0 m)^3 = (k/a_0 m)^3 |\beta_k^\mathrm{(late)}|^2 / 2\pi^2$, and the horizontal axis the comoving wavenumber $k/a_0 m$.  
	We show a model with $a_0 m \tau = 2$; other choices of the particle's mass $m$ or duration of non-adiabaticity $\tau$ lead to qualitatively similar results.  
	Both the numerical solution and the Stokes solution agree well over a wide range of wavenumbers.  
	At low $k$ the solutions differ by a factor of $\approx 2$ as the adiabatic approximation becomes less reliable, and at high $k$ the numerical solution behaves poorly due to issues with numerical precision. 
	}
\end{figure}

Having calculated $|\beta_k^\mathrm{(late)}|^2$ by these two approaches, the results are presented and compared in Fig.~\ref{fig:dirac_spectrum}.  
We show the Bogoliubov coefficients for a range of wavenumbers while fixing the model parameters $m$ and $\tau$.  
This figure highlights the excellent agreement between the two approaches over a wide range of wavenumbers, including $k/a_0 m \approx 1$ where the spectrum $n_k$ is peaked.  
Toward larger wavenumbers up to $k/a_0 m \approx 5$ the agreement improves, which is expected since the adiabatic approximation is increasingly reliable at large $k$.  
At very high wavenumbers above $k/a_0 m \approx 5$, the numerical result deviates from the analytical one, since the calculation suffers from numerical precision issues.  
This comparison emphasizes the importance of having analytical results for understanding the correct scaling behavior in asymptotic regimes.  
At small wavenumbers, where the adiabatic approximation is less reliable, the analytical Stokes result deviates from the numerical one by a factor of approximately $2$.  
However the total number of gravitationally produced particles is controlled by the peak of the spectrum, and for this observable the two approaches agree to better than an order $1$ factor.  

The analytic analysis with Stokes phenomenon has an important advantage over the numerical approach.  
Namely, it yields an analytic approximation formula~\eqref{eq:Dirac_betasq_Stokes} that is reliable at high $k$.
This analytic expression reveals that the asymptotic behavior toward large wavenumber is an exponential decrease and the exponential `decay' rate can be read off of Eq.~\eqref{eq:Dirac_betasq_Stokes} as $|\beta_k^\mathrm{(late)}|^2 \sim e^{-2 \pi k \tau}$.  
On the other hand, the numerical calculation is challenging at large wavenumber since the mode functions oscillate rapidly, and small time steps are required to resolve the oscillations, which extends the program's run time.
Moreover, at high-$k$ the magnitude of $\abs{\beta_k}$ becomes so small that numerical procedures may run into precision issues, leading to inaccurate results.
In Fig.~\ref{fig:dirac_spectrum}, the $k / a_0 m > 5$ tail of the numerical result is a jagged curve that behaves differently from the rest of the curve; this signifies that the numerical result is unreliable in that range, and illustrates a shortcoming of the numerical method.

\section{Spin-3/2 Rarita-Schwinger field}\label{sec:spin3/2}

This section parallels our approach in Sec.~\ref{sec:spin1/2}.  
We present the Rarita-Schwinger field's mode equations, which are different for its $\pm 1/2$ and $\pm 3/2$ helicity components.  
We employ the Stokes phenomenon to study gravitational particle production due to its effectively time-dependent sound speed in an FRW spacetime.  

\subsection{Mode equations on an FRW background}\label{sec:spin3/2_FRW}

 The quantum excitations of a Rarita-Schwinger field correspond to spin-$3/2$ particles and antiparticles~\cite{Rarita:1941mf}.  
A Rarita-Schwinger field $\chi_\mu$ with mass $m$ in a curved spacetime is described by the action~\cite{Freedman:2012zz} 
\begin{align}\label{eq:RS_action}
    S = \int \! \dd[4]{x} \, \sqrt{-g} \, \biggl(\frac{i}{2} \bar{\chi}_\mu \underline{\gamma}^{\mu\rho\nu} (\nabla_\rho \chi_\nu) - \frac{i}{2} (\nabla_\rho \bar{\chi}_\mu) \underline{\gamma}^{\mu\rho\nu} \chi_\nu + 2m \bar{\chi}_\mu \underline{\Sigma}^{\mu\nu} \chi_\nu \biggr)
    \;,
\end{align}
where $\bar{\chi}_\mu \equiv \chi_\mu^\dag \gamma^0$ is the Dirac conjugate field.  
The local tensors $\underline{\gamma}^{\mu\rho\nu}(x) = \frac12 \underline{\gamma}^{[\mu} \underline{\gamma}^{\nu]}$ and $\underline{\Sigma}^{\mu\nu}(x) = \underline{\gamma}^{[\mu} \underline{\gamma}^\rho \underline{\gamma}^{\nu]}$ are constructed from the local gamma matrix $\underline{\gamma}^\mu(x)$, where the square brackets denote anti-symmetrization.  
In an FRW spacetime with scale factor $a(\eta)$ the field equation becomes~\cite{Kolb:2021xfn} 
\begin{align}
    \Bigl( i \gamma^{\mu\rho\nu} \partial_\rho - 2i a H \gamma^{[\mu} \eta^{\nu]0} + 2am \Sigma^{\mu\nu} \Bigr) \bigl( a^{1/2} \chi_\nu \bigr) = 0
    \;,
\end{align}
where $H(\eta) = a^\prime(\eta) / a(\eta)^2$ is the Hubble parameter, where $\eta^{\mu\nu}$ is the inverse Minkowski metric, and where $\gamma^{\mu\rho\nu}$, $\gamma^\mu$, and $\Sigma^{\mu\nu}$ are the usual spinor matrices in Minkowski space.  
Similarly to the Dirac field, each Fourier mode of a Rarita-Schwinger field can be decomposed onto helicity eigenspinors.  
For the modes with helicity $\pm 3/2$, the mode functions obey an equation that's identical to the Dirac case \eqref{eq:Dirac_eq1}, and consequently all the results from Sec.~\ref{sec:spin1/2} carry over to the $\pm 3/2$ helicity modes of a Rarita-Schwinger field~\cite{Kallosh:1999jj,Giudice:1999yt,Giudice:1999am,Kallosh:2000ve,Bastero-Gil:2000lgf,Kolb:2021xfn}.     
However, the (longitudinal) modes with helicity $\pm 1/2$ instead obey the mode equation~\cite{Hasegawa:2017hgd,Kolb:2021xfn} 
\begin{align}\label{eq:EoM3/2}
	i \begin{pmatrix} 
    u_{A, k}^\prime \\
    u_{B, k}^\prime
	\end{pmatrix} 
	= \begin{pmatrix} 
    am & k c_s e^{i\zeta} \\
    k c_s e^{-i\zeta} & -am
	\end{pmatrix}
	\begin{pmatrix}
    u_{A, k} \\
    u_{B, k}
	\end{pmatrix}
	\;,
\end{align}
where $c_s(\eta) e^{i \zeta(\eta)}$ is an effectively time-dependent c-number sound speed that differs from $c_s = 1$ due to the cosmological expansion.  
The magnitude and phase of the complex sound speed are~\cite{Hasegawa:2017hgd,Kolb:2021xfn} 
\bes{\label{eq:cs_def}
    c_s(\eta) & = \frac{|-R/3 - H^2 + 3 m^2|}{3 (H^2 + m^2)} \\ 
    e^{i \zeta(\eta)} & = \frac{-R/3 - H^2 + 3m^2}{|-R/3 - H^2 + 3m^2|} \biggl( \frac{m^2 - H^2}{m^2 + H^2} + i \frac{2mH}{m^2 + H^2} \biggr)
    \;,
}
where $R(\eta) = - 6 a^{\prime\prime}/a^3$ is the Ricci scalar curvature.  
In a Minkowski spacetime with $H = 0$ and $R = 0$, the sound speed reduces to $c_s(\eta) = 1$.  
In an FRW cosmology near the end of inflation, $c_s(\eta)$ can vanish.  
Note that the mode equation \eqref{eq:EoM3/2} has the same form as Eq.~\eqref{eq:mode_eqn} that we studied in Sec.~\ref{sec:formalism}, and it corresponds to setting $M(\eta) = a(\eta) m$ and $K(\eta) = c_s(\eta) k e^{i \zeta(\eta)}$.

\subsection{GPP from an effectively time-dependent sound speed}\label{eq:spin3/2_production}

In this subsection, we study the gravitational production of spin-$3/2$ particles with helicity $\pm 1/2$ (longitudinally-polarized) due to the effectively time-dependent sound speed \eqref{eq:cs_def}.  
As in Sec.~\ref{eq:spin1/2_production}, we shall compare the abundance of gravitationally-produced particles obtained by directly solving the mode equations \eqref{eq:EoM3/2} and by an analytic Stokes phenomenon analysis.

\begin{figure}[t]
	\centering
	\includegraphics[width=.49\textwidth]{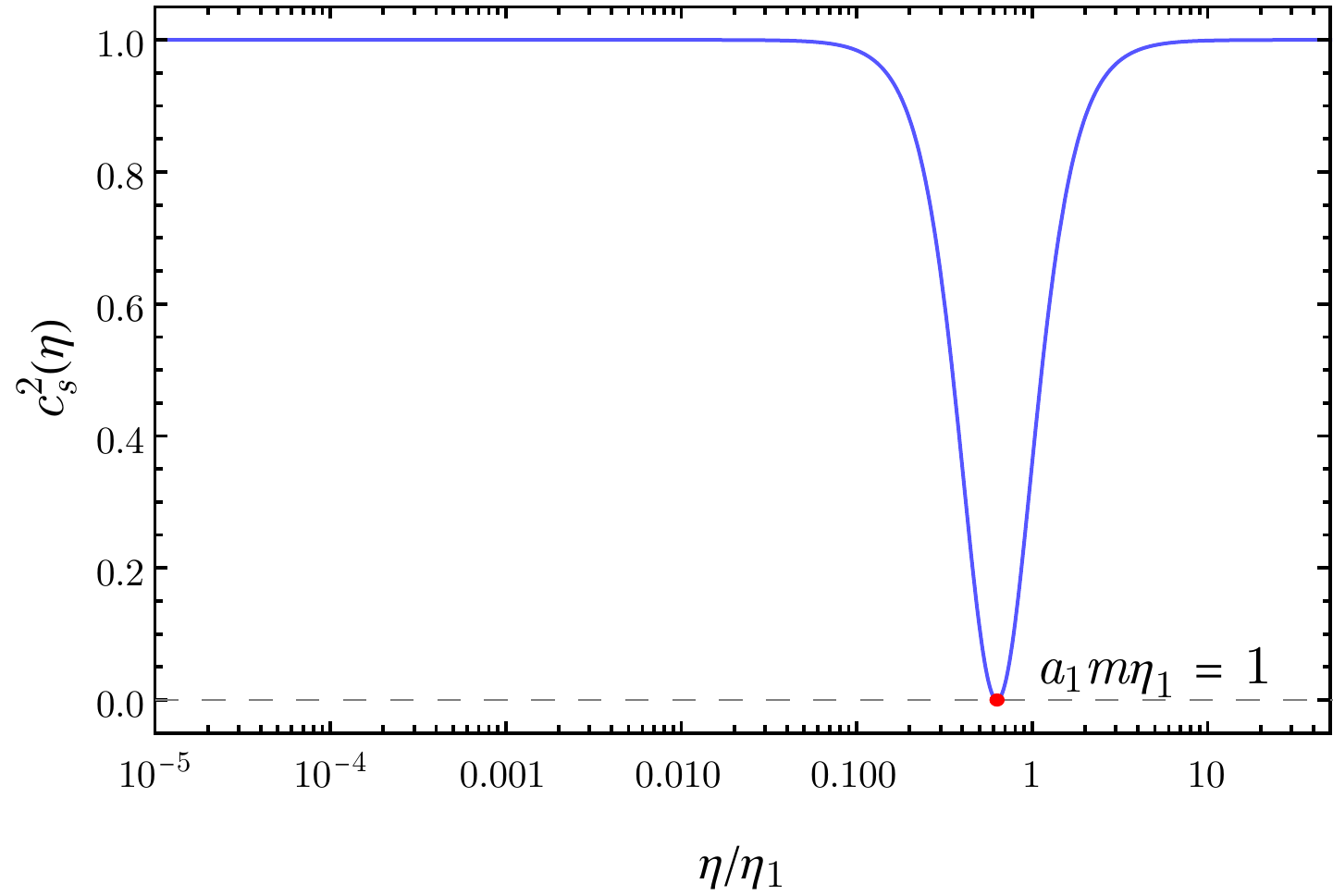}
	\includegraphics[width=.49\textwidth]{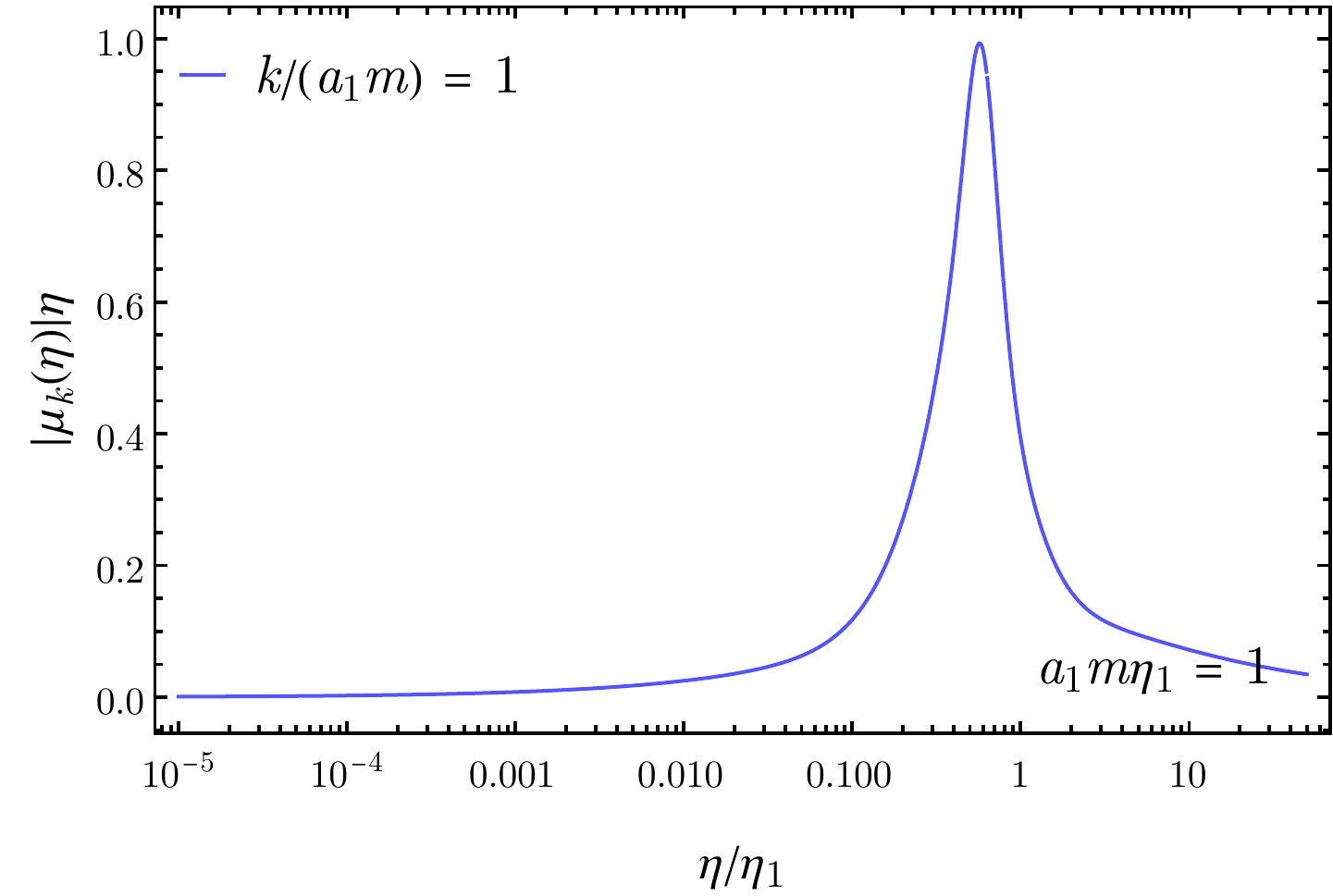}
	\caption{\label{fig:rs_non_adiabaticity}The left and right figure respectively show the squared sound speed $c_s^2$ and the non-adiabaticity $|\mu_k(\eta)| \, \eta$ as functions of conformal time $\eta$.  The sound speed asymptotes to $1$ at both early time and late time; at $\eta = \eta_v$, the sound speed drops to $0$, as indicated by the red dot. Vanishing sound speed is the condition for catastrophic production of particles to happen. The right figure shows that the non-adiabaticity $|\mu_k(\eta)| \, \eta$ is most enhanced around $\eta_v$, reaching a maximum value of $3k \eta_v/2 \simeq 0.945 (k/a_1 m) (a_1 m \eta_1)^{1/3}$. }
\end{figure}

For this comparison, we assume the spacetime to be FRW with scale factor 
\begin{align}\label{eq:spin3/2_a}
    a(\eta) = a_1 \sqrt{ \eta / \eta_1} 
\end{align}
where $0 < a_1$, $0 < \eta_1$, and $0 < \eta$.  
The squared sound speed \eqref{eq:cs_def} evaluates to 
\begin{align}
    c_s^2(\eta) = \left( \frac{1 - 4 a_1^2 m^2 \eta^3 / \eta_1}{1 + 4 a_1^2 m^2 \eta^3 / \eta_1} \right)^2
    \;,
\end{align}
and Fig.~\ref{fig:rs_non_adiabaticity} illustrates its time dependence.  
Notice that $c_s^2(\eta)$ remains $\approx 1$ except for a single dip at 
\ba{\label{eq:etav_def}
    \eta_v \equiv \frac{\eta_1^{1/3}}{(2 a_1 m)^{2/3}}
    \qquad \text{where} \qquad 
    c_s(\eta_v) = 0
    \;.
}
Previous studies have shown that a momentarily vanishing sound speed is a sufficient condition for a `catastrophic' production of particles~\cite{Hasegawa:2017hgd,Kolb:2021xfn}.

The spacetime in Eq.~\eqref{eq:spin3/2_a} corresponds to a kination-dominated universe; i.e. the cosmological medium has equation of state $w=1$.
Although such a spacetime is inconsistent with the cosmic history of our universe, it provides a realistic time-dependent sound speed $c_s$, and thus constitutes a reasonable toy model for which we can study GPP analytically.
In realistic inflationary cosmologies, the spacetime evolves from quasi-de-Sitter at early times into FRW at late times.  
Consequently, the sound speed $c_s$ asymptotes to $1$ at both early and late times, where $c_s$ can be calculated from Eq.~\eqref{eq:cs_def} and is illustrated in Ref.~\cite{Kolb:2021xfn}.  
In order to study the effect of a time-dependent sound speed in an inflationary cosmology via Stokes phenomenon, we must find an analytic function $a(\eta)$ giving a sound speed $c_s(\eta)$ that asymptotes to $1$ at early and late times,
but deviates from $1$ in the transition phase.
Simply defining $a(\eta)$ as a piecewise function that joins a de-Sitter universe with a FRW universe will not do,
since that would break the analyticity of $a(\eta)$.
Instead, we choose $a(\eta)$ to be a power law $a \sim \eta^p$; our choice in \eqref{eq:spin3/2_a} then corresponds to $p=1/2$, which yields $c_s \to 1$ as $\eta \to 0$ and $\eta \to \infty$, fulfilling our requirements. In fact, $p = 1/2$ is the only choice that yields the correct asymptotic behavior for $c_s$ as $\eta \to 0$; for example, a radiation dominated universe with $a \sim \eta$ will give $c_s \to 1/9$ as $\eta \to 0$.

As discussed already in Sec.~\ref{sec:spin1/2}, the results of Sec.~\ref{sec:formalism} provide two approaches for solving Eq.~\eqref{eq:EoM3/2}. 
Following the first approach, solutions of the mode equations are written as linear combinations of the instantaneous positive and negative frequency modes
\bes{\label{eq:chiAB_mix_3/2}
    & \mqty(u_{A,k}(\eta) \\ u_{B,k}(\eta))
	= \alpha_k(\eta) \, \tilde{\psi}_{+,k}(\eta) + \beta_k(\eta) \, \tilde{\psi}_{-,k}(\eta) \\ 
	& \quad \text{where} \quad 
	\tilde{\psi}_{\pm,k}(\eta) = \frac{1}{\sqrt{2}} e^{\mp i(\Phi_k + \delta_k)}
	\begin{pmatrix}
		\pm e^{i\zeta/2} \sqrt{1 \pm \frac{am}{\omega_k}} \\
		e^{-i\zeta/2} \sqrt{1 \mp \frac{am}{\omega_k}}
	\end{pmatrix}
	\;,
}
where $\omega_k = \sqrt{c_s^2 k^2 + a^2 m^2}$ is the effective frequency, where $\zeta$ is the phase of the sound speed given in Eq.~\eqref{eq:cs_def}, and where 
\ba{
    \Phi_k(\eta) \equiv \int_{\eta_a}^\eta \! \dd{\eta^\prime} \, \omega_k(\eta^\prime) 
    \;,\quad
    \delta_k(\eta) \equiv \int_{\eta_a}^\eta \dd{\eta^\prime} \, \frac{a(\eta') m}{2\omega_k(\eta')}\zeta^\prime(\eta')
    \label{eq:rs_EWKB_soln}
}
are the phase integrals defined in Eq.~\eqref{eq:Phi_def} at anchor time $\eta_a$. 
The time-dependent coefficients, $\alpha_k(\eta)$ and $\beta_k(\eta)$, can be determined by solving Eq.~\eqref{eq:dalpha_dbeta} numerically along with the initial conditions $\alpha_k(0) = 1$ and $\beta_k(0) = 0$, and we are interested in calculating $\beta_k(+\infty)$.  
Following the second approach, solutions are written as a linear combination of EWKB basis solutions:
\ba{
    \mqty(u_{A,k}^{(i)}(\eta) \\ u_{B,k}^{(i)}(\eta))
	= \alpha_k^{(i)} \, {\psi}_{+,k}^{(i)}(\eta) + \beta_k^{(i)} \, {\psi}_{-,k}^{(i)}(\eta) 
	\;,
}
where $\alpha_k^{(i)}$ and $\beta_k^{(i)}$ are complex constants, and where $\psi_{\pm,k}^{(i)}$ denote the EWKB basis solutions in the $i^\mathrm{th}$ Stokes region. 
We set $\alpha_k^\mathrm{(early)} = 1$ and $\beta_k^\mathrm{(early)} = 0$ in the region containing $\eta = 0$, and we are interested in calculating $\beta_k^\mathrm{(late)}$ in the region containing $\eta \to +\infty$

\begin{figure}[t]
	\centering
	\includegraphics[width=.6\textwidth]{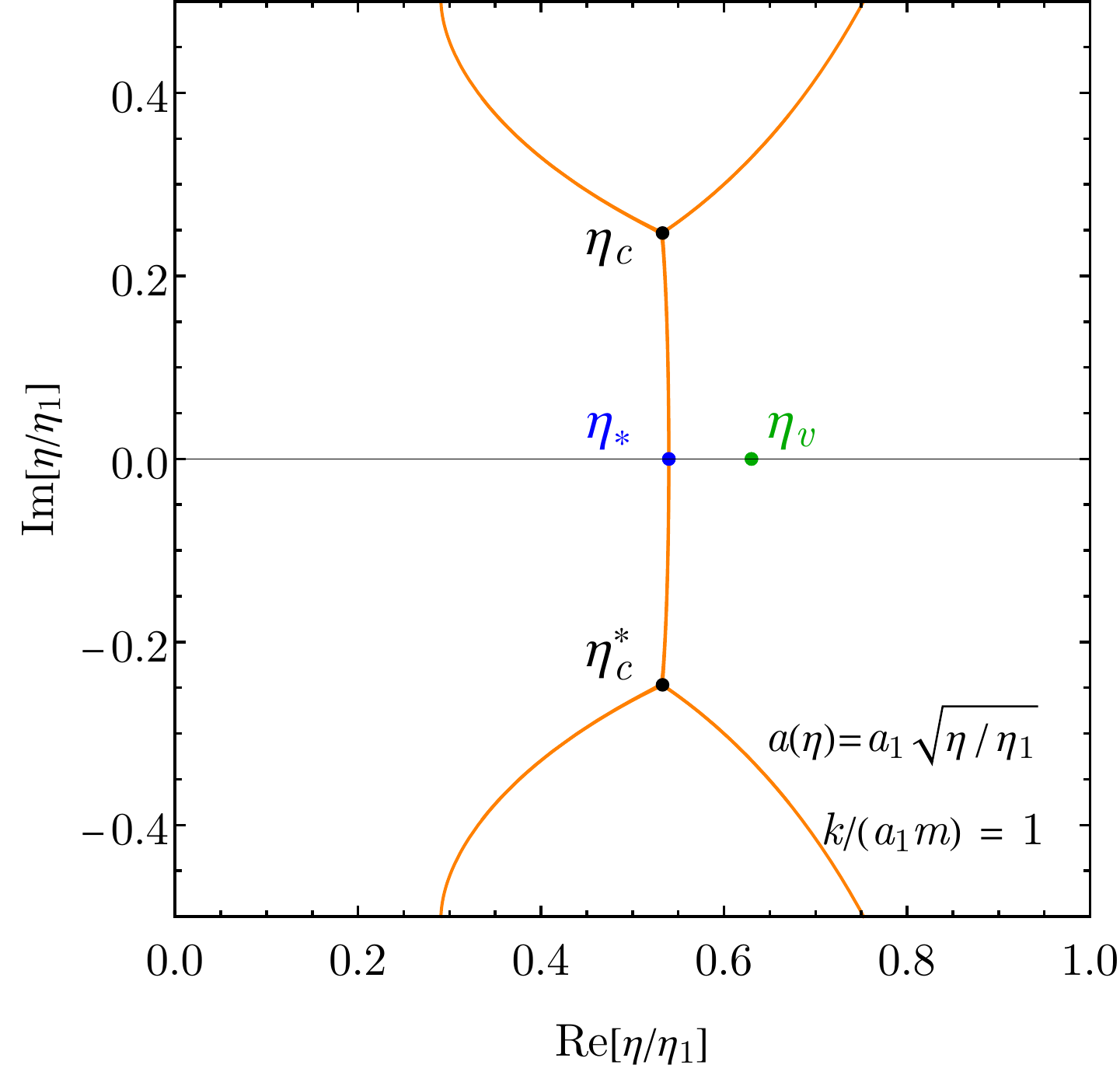}
	\caption{\label{fig:rs_stokes} 
	An illustration of the complex-$\eta$ plane showing turning points $\eta_c$ and $\eta_c^\ast$ (black dots) and Stokes lines (orange curves) close to the real axis.  
	The early Stokes region containing $\eta \to 0$ and the late Stokes region containing $\eta \to +\infty$ share a common border at the central Stokes line shown here.  The real axis intersects this Stokes line at the blue dot.  The green dot is the point of vanishing sound speed $c_s$. As $k \to +\infty$, the black and blue dots approach the green dot. 
	}
\end{figure}

We calculate $\beta_k^\mathrm{(late)}$ using the analytic results of the Stokes phenomenon analysis in Sec.~\ref{sec:formalism}.  
The turning points are the complex roots of the angular frequency, $\omega_k(\eta)=0$, in the half plane $\Re[\eta] > 0$. 
The squared frequency is 
\ba{\label{eq:omegak_spin32}
    \omega_k^2(\eta) = \frac{1}{\eta_v^2} \frac{\frac{\eta}{\eta_v} \bigl( 1 + \frac{\eta^3}{\eta_v^3} \bigr)^2 + 4 \bigl( k^2 \eta_v^2 \bigr) \bigl( 1 - \frac{\eta^3}{\eta_v^3} \bigr)^2}{4 \bigl( 1 + \frac{\eta^3}{\eta_v^3} \bigr)^2} 
}
where $\eta_v = (2 a_1 m)^{-2/3} \eta_1^{1/3}$ is the time at which $c_s^2(\eta_v) = 0$.  
Solving for $\eta_c$ exactly is not possible, as it would require finding the roots of a seventh-order polynomial.  
However, for sufficiently large $k$, there are only two turning points near the real-$\eta$ axis: $\eta_c$ and $\eta_c^*$ (we use convention $\Im[\eta_c] > 0$), and they are connected by a Stokes line. 
See Fig.~\ref{fig:rs_stokes} for an illustration for the Stokes structure. 
Additional turning points are found at larger values of $|\mathrm{Im}[\eta/\tau]|$, which are not shown in the figure, and which furnish a negligible correction to the connection matrix.  
The formalism in Sec.~\ref{sec:discussion} then tells us how to find $\beta_k^{(\mathrm{late})}$ by analytically evaluating a connection matrix.

For large values of $k$, approximate analytic expressions are available for the turning points, $\eta_c$ and $\eta_c^\ast$, and for the phase integral $\Psi_k$. 
As $k \to +\infty$, the second term in Eq.~\eqref{eq:omegak_spin32} dominates, and the turning points approach $\eta_c, \eta_c^* \to \eta_v$. 
Thus, two roots of $\omega_k(\eta)$ can be expanded as a series in powers of $1/k$:
\begin{align}
    \eta_c = \eta_v + i\frac{1}{3k} + \order{k^{-2}} 
    \;.
\end{align}
Also, $\omega_k(\eta)$ can be written as a series around $\eta_v$: 
\begin{align}
    \omega_k(\eta_v + \Delta \eta) 
    = a(\eta_v) m + \omega_k'(\eta_v)\Delta\eta + \order{\Delta\eta^2} 
    \;.
\end{align}
Combining these series gives us the phase integral~\eqref{eq:Psi_def} up to order $1/k$:
\begin{align}\label{eq:3/2_psi}
    \Psi_k 
    = i\int_{\eta_c}^{\eta_c^*} \! \dd{\eta} \, \omega_k(\eta) 
    = \frac{\pi a(\eta_v) m}{6 k} + \order{k^{-2}}
    = \frac{\pi}{12 k \eta_v} + \order{k^{-2}} 
    \;;
\end{align}
see App.~\ref{app:spin_3/2_phase_integral} for details of the derivation.  
Finally, the squared Bogoliubov coefficient is calculated using Eq.~\eqref{eq:formula} in the adiabatic approximation, which gives 
\begin{align}
    \label{eq:3/2_beta_k_stokes}
    |\beta_k^{(\mathrm{late})}|^2 \approx \exp(- \frac{\pi}{6 k \eta_v})
    \;.
\end{align}
Since $\eta_v > 0$ we find $0 < |\beta_k^{(\mathrm{late})}| < 1$ as required for Fermi statistics.  

\begin{figure}[t]
	\centering
	\includegraphics[width=.49\textwidth]{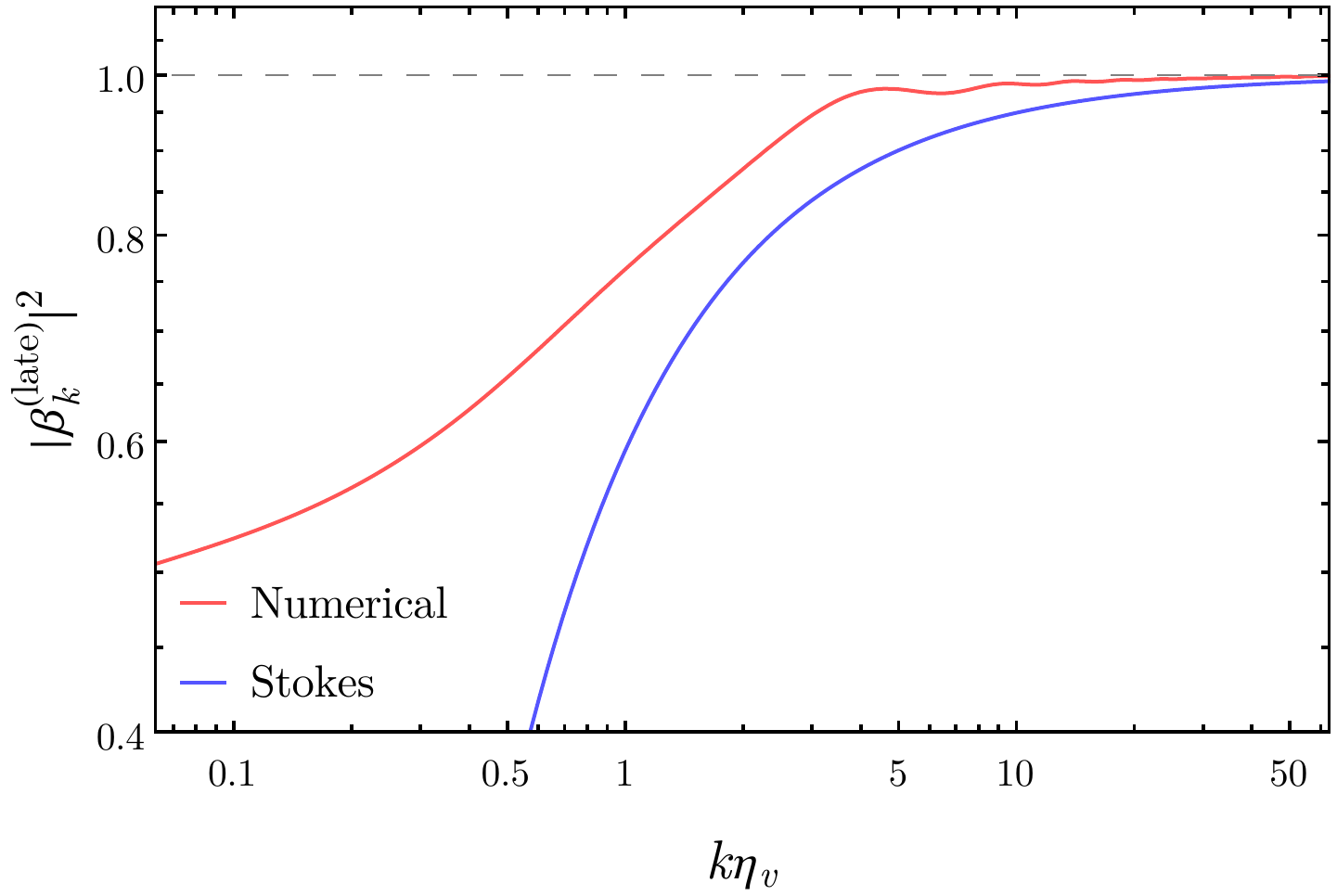}
	\includegraphics[width=.49\textwidth]{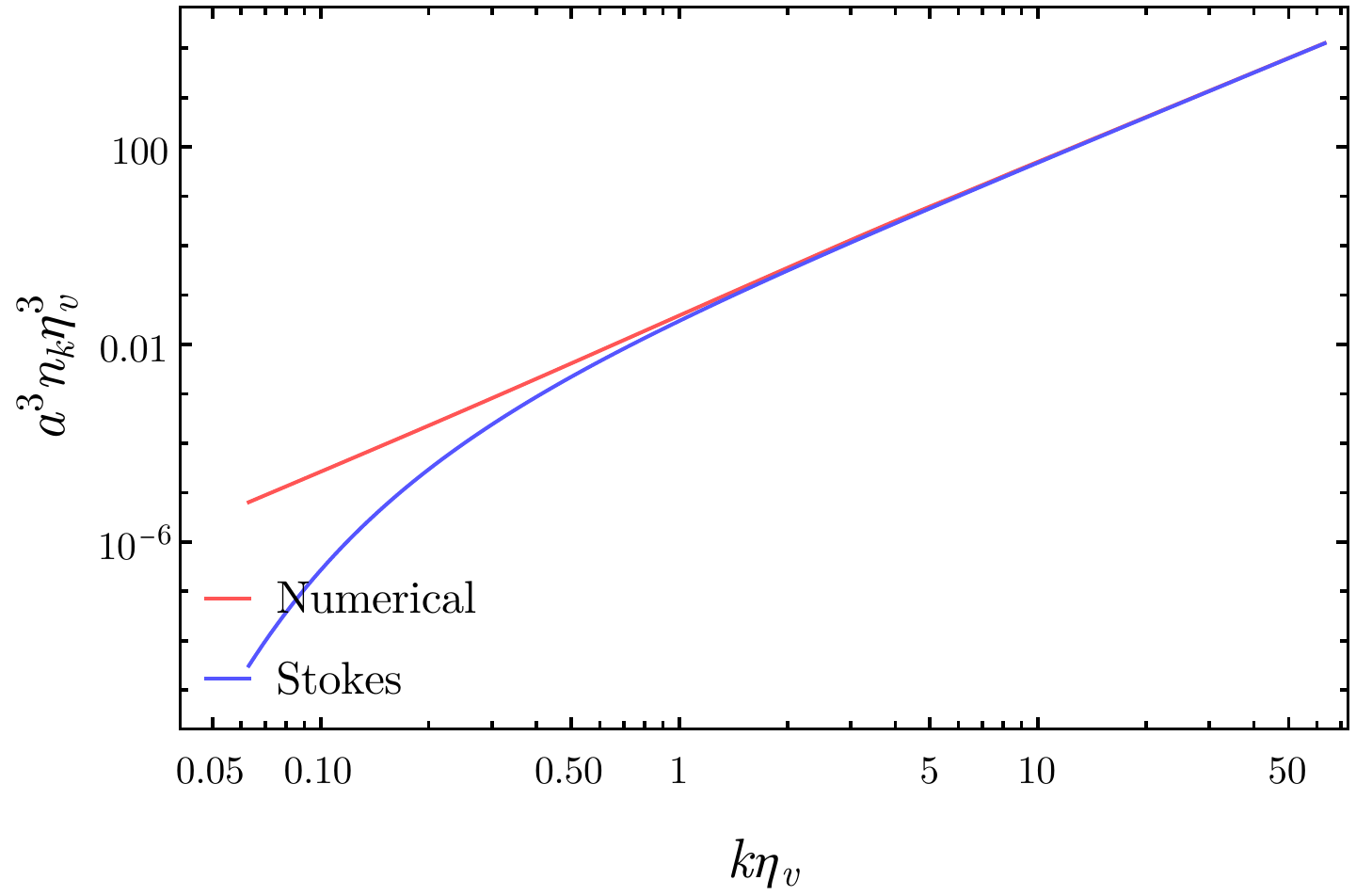}
	\caption{\label{fig:rs_spectrum}A comparison of the particle production obtained by solving the mode equations numerically (red curve) and by evaluating the Stokes phenomenon (blue curve). The horizontal axis is comoving wavenumber $k$ in units of $\eta_v^{-1}$.  The left panel shows the squared Bogoliubov coefficent $|\beta_k^{(\mathrm{late})}|^2$, and the right panel shows the comoving number density (per d.o.f) $a^3 n_k \propto k^3 |\beta_k^{(\mathrm{late})}|^2$ in units of $\eta_v^{-3}$.  The numerical and analytical calculations agree well at high $k$ where they exhibit $|\beta_k^{(\mathrm{late})}|^2 \approx 1$, confirming the phenomenon of `catastrophic' particle production observed in earlier work. 
	}	
\end{figure}

Fig.~\ref{fig:rs_spectrum} presents the results of the two approaches described above for calculating $|\beta_k|^2$.  
Direct numerical integration of the mode equation is used to generate the red curve, whereas the blue curve is a plot of Eq.~\eqref{eq:3/2_beta_k_stokes}.  
The two approaches agree well at large wavenumber $k \eta_v \gtrsim 1$.  
For $k \eta_v \gtrsim 5$ we find $|\beta_k^\mathrm{(late)}|^2 \approx 1$, corresponding to maximal enhancement of the negative frequency mode under the constraint of Fermi statistics, $|\alpha_k|^2 + |\beta_k|^2 = 1$.  
If the total particle number were calculated by integrating $n_k$, the result would be divergent (signaling that backreaction cannot be ignored). 
This behavior occurs when the spin-3/2 Rarita-Schwinger field experiences a vanishing sound speed; it has been called ``catastrophic particle production''~\cite{Kolb:2021xfn}.
One advantage that the analytical Stokes analysis has over the numerical approach is that it yields an analytic expression $\abs{\beta_k}^2$ that is valid at high-$k$, where numerical solutions of the equation is hard to obtain due to its rapidly oscillating structure. 
Moreover, with the analytic expression, we can immediately tell how fast $\abs{\beta_k}^2$ approach $1$ as $k\to \infty$, which is not obvious from numerical solutions.

\subsection{Implications for catastrophic particle production}\label{sec:catastrophic}

The previous subsection provides a concrete toy model cosmology in which the sound speed $c_s(\eta)$ \eqref{eq:cs_def} vanishes momentarily, leading to catastrophic production of the helicity-$1/2$ modes of spin-$3/2$ particles. 
More generally, the phenomenon of catastropic particle production is expected to occur whenever the sound speed vanishes \cite{Hasegawa:2017hgd,Kolb:2021xfn}.  
In this subsection, we present a more general argument that vanishing sound speed implies $\abs{\beta_k}^2 \sim \exp(-D/k)$ for some constant $D>0$ without reference to a specific spacetime $a(\eta)$.  

The argument develops from the following intuition.  
The evolution of the helicity-$1/2$ modes is primarily described by the time-dependent effective frequency $\omega_k^2 = c_s^2 k^2 + a^2 m^2$. 
If $c_s(\eta_v) = 0$ at some time $\eta_v$, then around $\eta_v$ we have $\omega_k^2(\eta) = C (\eta - \eta_v)^2 k^2 + a(\eta)^2 m^2 + \order{\eta - \eta_v}^3$, where $C$ is some constant independent of $k$.  
In this sense, the evolution of the helicity-$1/2$ modes is ``universal'' in $k$ at $\eta_v$.  
If backreaction is neglected, one concludes that arbitrarily large $k$ modes can be populated, and the number density of gravitationally produced particles \eqref{eq:number_density} is divergent.  

In order to calculate $|\beta_k|^2$ via the Stokes phenomenon analysis, it is first necessary to find the turning points, which are roots of $\omega_k(\eta) = \sqrt{c_s^2(\eta) k^2 + a(\eta)^2 m^2}$ [see below Eq.~\eqref{eq:rs_EWKB_soln}].  
Suppose the sound speed $c_s(\eta)$ vanishes at some real time $\eta_v$.  
The turning points are solutions of the equation $\omega_k^2(\eta) = 0$, which can be cast into the form:
\ba{
    -\frac{m^2}{k^2} = \frac{c_s^2}{a^2} = \eval{\frac{1}{2}\dv[2]{(c_s^2/a^2)}{\eta}}_{\eta_v} (\eta - \eta_v)^2 + \order{(\eta-\eta_v)^3}
    \;,
}
where we have Taylor expanded $c_s^2 / a^2$ around $\eta_v$.
Since $c_s(\eta)$ satisfies Eq.~\eqref{eq:cs_def}, the linear term in the Taylor expansion vanishes, and $c_s^2 / a^2 \geq 0$ implies that the coefficient for the quadratic term is positive. 
The solutions of this equation are the turning points
\ba{\label{eq:etac_from_etav}
    \eta_c, \eta_c^\ast = \eta_v \pm i \frac{B}{k} + \order{k^{-2}} ,\qq{where} 
    B\equiv \sqrt{2} m \left(\eval{\dv[2]{(c_s^2/a^2)}{\eta}}_{\eta_v}\right)^{-1/2} \; .
}
The turning points $\eta_c$ and $\eta_c^*$ approach $\eta_v$ as $k \to \infty$. 
Thus in general, for every real time $\eta_v$ at which the sound speed vanishes, $c_s(\eta_v) = 0$, the frequency $\omega_k(\eta)$ has a conjugate pair of turning points given by Eq.~\eqref{eq:etac_from_etav}. 
One can easily verify that these turning points are connected by a Stokes line.  
Crossing this Stokes line induces catastrophic particle production.

For simplicity we assume that the sound speed $c_s(\eta)$ vanishes only once at $\eta_v$, implying that the Stokes structure contains only two turning points~\eqref{eq:etac_from_etav}, connected by a single Stokes line, which divides the real time axis into two Stokes regions.  
It is straightforward to generalize to a more complex time dependence. 
We seek to compute the phase integral $\Psi_k$ in the limit of large wavenumber $k$. 
Since $\Psi_k = i \int_{\eta_c}^{\eta_c^\ast} \! \dd{\eta} \, \omega_k(\eta)$ is an integral of $\omega_k(\eta)$ between the turning points~\eqref{eq:Psi_def}, and since both turning points $\eta_c, \eta_c^\ast$ approach $\eta_v$ at large $k$, we can use the Taylor series approximation to $\omega_k(\eta)$ at $\eta_v$: 
\ba{
    \omega_k(\eta) 
    & = \sum_{n=0}^\infty \frac{1}{n!} \eval{\dv[n]{\omega_k}{\eta}}_{\eta_v} (\eta-\eta_v)^n 
    \;,
}
where the radius of convergence is $|\eta-\eta_v| < |\eta_c-\eta_v|$.  
The derivatives of $\omega_k(\eta)$ to leading order at large $k$ are given by 
\ba{\label{eq:omega2_Taylor}
    \eval{\dv[n]{\omega_k}{\eta}}_{\eta_v} = (-1)^{1+n/2} \, (n-3)!! \, (n-1)!! \, a(\eta_v) m \left( \frac{k}{B} \right)^n + \order{(k/B)^{n+2}}
    \;,
}
for even $n$.  
The phase integral $\Psi_k$ is evaluated by integrating term-by-term in the Taylor series: 
\bes{
    \Psi_k 
    & = i \int_{\eta_c}^{\eta_c^\ast} \! \dd{\eta} \, \omega_k(\eta) \\ 
    & = i \int_{\eta_c}^{\eta_c^\ast} \! \dd{\eta} \, \sum_{n=0}^\infty \frac{1}{n!} \eval{\dv[n]{\omega_k}{\eta}}_{\eta_v} (\eta-\eta_v)^n \\ 
    & = i \sum_{n=0}^\infty \frac{1}{(n+1)!} \eval{\dv[n]{\omega_k}{\eta}}_{\eta_v} \Bigl[ (\eta_c^\ast-\eta_v)^{n+1} - (\eta_c-\eta_v)^{n+1} \Bigr] \\ 
    & = i \sum_{n=0}^\infty \frac{1}{(n+1)!} \eval{\dv[n]{\omega_k}{\eta}}_{\eta_v} \Bigl[ (-i B/k)^{n+1} - (i B/k)^{n+1} + \order{(B/k)^{n+2}} \Bigr] \\ 
    & = 2 \sum_{n \ \mathrm{even}} \frac{1}{(n+1)!} \eval{\dv[n]{\omega_k}{\eta}}_{\eta_v} i^n (B/k)^{n+1} + \order{(B/k)^{n+2}} \\ 
    & = \frac{\pi}{2} \frac{a(\eta_v) m B}{k} + \order{k^{-2}}
    \;.
}
Note that $\Psi_k \to 0$ as $k \to \infty$.  
For an alternative derivation strategy, see also the approach in App.~\ref{app:spin_3/2_phase_integral}, which avoids calculating the full Taylor series \eqref{eq:omega2_Taylor}. 

Finally evaluating the Bogoliubov coefficient~\eqref{eq:formula} gives 
\ba{
    |\beta_k|^2 \approx \mathrm{exp}\biggl[ - \frac{\sqrt{2} \pi a(\eta_v) m^2}{k} \biggl( \eval{\dv[2]{(c_s^2/a^2)}{\eta}}_{\eta_v} \biggr)^{-1/2} \biggr] 
    \;.
}
This result allows the spectrum of gravitationally produced particles~\eqref{eq:number_density} to be evaluated for an arbitrary cosmology $a(\eta)$ and effective sound speed $c_s(\eta)$, provided that $c_s$ has one real root at $\eta_v$.  
As $k \to \infty$ the exponent decreases in magnitude and $|\beta_k|^2 \to 1$, implying unsuppressed particle production in the high wavenumber modes.  
We conclude that catastrophic production of spin-3/2 particles occurs whenever the sound speed $c_s$ momentarily vanishes.

\section{Conclusion}\label{sec:conclusion}

We have studied the phenomenon of gravitational particle production as applied to fermionic fields in expanding FRW spacetimes.  
In this work, we are particularly interested in analytic analysis, since a direct numerical solution of the mode equations is less informative and often computationally intensive.  
Our main result is an analytic expression for the connection matrix relating positive and negative frequency mode functions at asymptotically early and late times, and we also provide toy models to demonstrate potential applications.  

The key elements of our work are summarized as follows.  
We write down the time-dependent mode equations for spin-1/2 and 3/2 fields in an arbitrary FRW spacetime, which are available from the literature.
By employing the exact WKB method formalism, we construct formal, exact solutions to these equations.  
Using these solutions, we define the exact connection matrix $T$ that describes the mixing of positive and negative frequency modes at asymptotically early and late times.  
To obtain a tractable expression for $T$, we apply well-known theorems from WKB analysis, which lets us map the general fermion mode equations to the familiar Weber equation.  
Using identities of the standard solutions of the Weber equation, we derive an expression for $T$, which is one of the main results of our work.  
In the appendix we present the formal expression for $T$~\eqref{eq:connection_matrix_full} that is valid to all orders in the perturbation theory expansion parameter $\hbar$, receiving only non-perturbatively small corrections $\propto e^{-C/\hbar}$, and in the main text we show the leading $\hbar$ dependence~\eqref{eq:con_mat}.  
The expression allows the connection matrix $T$ to be evaluated once the FRW spacetime (time-dependent scale factor) and particle physics model parameters (mass and spin) have been specified.  
The off-diagonal entries in this matrix control the spectrum and abundance of gravitationally-produced particles.  

To illustrate the applications of our analytic results, we study two toy models.  
The first is a free spin-1/2 Dirac spinor field in an FRW spacetime with scale factor $a = a_0 (1+\tanh \eta/\tau)/2$ that evolves from $0$ to $a_0$, and the non-adiabatic mode evolution results from the field's effectively time-dependent mass $am$.  
The second model is a free spin-3/2 Rarita-Schwinger spinor field in an FRW spacetime with $a \sim \eta^{1/2}$, corresponding to a kination-dominated cosmological phase, and particle production results from the field's effectively time-dependent sound speed.  
For both models, we adapt our general analytic results to evaluate the spectrum of gravitationally produced particles, and we compare with a direct numerical integration of the mode equations.  
From this comparison we conclude that the two approaches agree extremely well.  
The asymptotic behavior at high comoving wavenumber $k$ is captured analytically by the Stokes phenomenon calculation, in a regime where the numerical calculation becomes challenging due to the rapidly oscillating integrand.  
Moreover, the Stokes approximation also works well around the peak of the spectrum, and therefore provides a good estimate of the total abundance of gravitationally produced particles, which is the integral of the spectrum.  

In addition, our study of the spin-3/2 Rarita-Schwinger field confirms the phenomenon of `catastropic' particle production resulting from a momentarily vanishing sound speed, which has been pointed out by earlier work, where it was identified through semi-analytic arguments and direct numerical integration of the mode equations.  
Here we provide also an analtyical understanding through the Stokes phenomenon.  

\section*{Acknowledgement}
The authors would like to thank Mustafa Amin, Daniel J. H. Chung, Mudit Jain, Edward Kolb, and Hong-Yi Zhang for helpful discussions.  
S.H.~is supported by JSPS KAKENHI, Grant-in-Aid for JSPS Fellows 20J10176 and the Advanced Leading Graduate Course for Photon Science (ALPS).  
S.H.~would also like to express his gratitude to the staff of Rice University who helped facilitate his visit to the campus during this severe situation.
A.J.L.~and S.L.~are supported in part by the National Science Foundation under Award No.~2114024. 

\appendix

\section{Diagonalize the two-level Schr\"odinger equation}\label{app:diagonalize}

The mode equation~\eqref{eq:mode_eqn}
\ba{
    i \begin{pmatrix} u_A^\prime \\ u_B^\prime \end{pmatrix} = \Omega(\eta) \begin{pmatrix} u_A \\ u_B \end{pmatrix}
    \qquad \text{with} \qquad 
    \Omega(\eta) = \begin{pmatrix} M(\eta) & \kappa(\eta) e^{i \zeta(\eta)} \\ \kappa(\eta) e^{-i \zeta(\eta)} & - M(\eta) \end{pmatrix}
    \;,
}
has the same form as the Schr\"odinger equation for a two level system with Hamiltonian $\Omega(\eta)$.  
In this appendix we demonstrate the unitary transformation that instantaneously diagonalizes the Hermitian matrix $\Omega(\eta)$, and we identify the instantaneous positive and negative frequency modes.  
Let the complex mode functions $\alpha(\eta)$ and $\beta(\eta)$ be defined by the unitary transformation 
\ba{
     & \mqty(\alpha(\eta) \\ \beta(\eta))
     = U(\eta) \mqty(u_A(\eta) \\ u_B(\eta)) 
}
where the matrix $U(\eta)$ is special ($\mathrm{det} \, U = 1$) and unitary ($U^\dagger U = U U^\dagger = I$).  
A general $\mathrm{SU}(2)$ matrix can be written as 
\ba{
     U(\eta) = 
     \mqty(\cos \theta(\eta) e^{-i \phi(\eta)} &
     \sin \theta(\eta) e^{-i \varphi(\eta)} \\
     -\sin \theta(\eta) e^{i \varphi(\eta)} &
     \cos \theta(\eta) e^{i \phi(\eta)}) 
}
where $\theta(\eta)$ is a real angle and where $\phi(\eta)$ and $\varphi(\eta)$ are real phases.  
In terms of the new mode functions, the mode equation becomes 
\ba{
    i \begin{pmatrix} \alpha^\prime \\ \beta^\prime \end{pmatrix} = \Bigl( U \Omega U^{-1} + i U^\prime U^{-1} \Bigr) \begin{pmatrix} \alpha \\ \beta \end{pmatrix} 
    \;,
}
where
\bsa{}{
    U \Omega U^{-1} & = \mqty( a & b \\ b^\ast & -a ) \\ 
    a & = M \, \cos 2\theta + \kappa \, \sin 2\theta \, \cos(\zeta-\phi+\varphi) \nonumber \\
    b & = -M \, \sin 2\theta \, e^{-i(\phi+\varphi)} + \kappa \, \cos^2 \theta \, e^{i(\zeta-2\phi)} - \kappa \, \sin^2 \theta \, e^{-i(\zeta+2\varphi)} \nonumber \\ 
    U^\prime U^{-1} & = \mqty( c & d \\ -d^\ast & -c ) \\ 
    c & = -i \bigl( \cos^2 \theta \, \phi^\prime + \sin^2 \theta \, \varphi^\prime \bigr) \nonumber \\ 
    d & = \Bigl( \theta^\prime + \frac{i}{2} \sin 2\theta \, \bigl( \phi^\prime - \varphi^\prime) \Bigr) \, e^{-i(\phi+\varphi)} \nonumber
    \;.
}
The angle and phases can be chosen such that $U \Omega U^{-1}$ is real and diagonal while $U \Omega U^{-1} + i U^\prime U^{-1}$ has vanishing diagonal entries.
Taking $\phi-\varphi = \zeta$ ensures that all the terms in $b$ have the same phase, and we have 
\bsa{}{
    a & = M \, \cos 2\theta + \kappa \, \sin 2\theta \\
    b & = - \bigl( M \, \sin 2\theta - \kappa \, \cos 2\theta \bigr) \, e^{-i(\phi+\varphi)} \\ 
    c & = -\frac{i}{2} \Bigl( \cos 2\theta \, \zeta^\prime + (\phi + \varphi)^\prime \Bigr) \\ 
    d & = \Bigl( \theta^\prime + \frac{i}{2} \sin 2\theta \, \zeta^\prime \Bigr) \, e^{-i(\phi+\varphi)} 
    \;.
}
Taking $\tan 2\theta = \kappa/M$ ensures that $b = 0$.  
Note that $\kappa = |K| \geq 0$, and we assume $M \geq 0$, which is the relevant regime for gravitational particle production [see \eqref{eq:choices}], implying $0 \leq \theta < \pi/4$ for the rotation angle. 
It also follows that $\cos 2\theta = M / \omega$, $\sin 2\theta = \kappa / \omega$, $\cos \theta = \sqrt{(1 + M /\omega)/2}$, $\sin \theta = \sqrt{(1 - M /\omega)/2}$, and $\theta^\prime = - (\kappa M^\prime - M \kappa^\prime)/(2 \omega^2)$ where $\omega \equiv \sqrt{\kappa^2 + M^2}$.  
Then we have 
\bsa{}{
    a & = \omega \,  \\
    b & = 0 \\ 
    c & = -\frac{i}{2} \Bigl( \frac{M}{\omega} \, \zeta^\prime + (\phi + \varphi)^\prime \Bigr) \\ 
    d & = - \frac{\kappa M^\prime - M \kappa^\prime - i \kappa \omega \zeta^\prime}{2 \omega^2} \, e^{-i(\phi+\varphi)} 
    \;.
}
These choices have led to $U \Omega U^{-1} = \mathrm{diag}(\omega,-\omega)$.  
To ensure that $U \Omega U^{-1} + i U^\prime U^{-1}$ has vanishing diagonal entries, we also need $a + i c = 0$.  
This is accomplished by taking 
\ba{
    \Bigl( \frac{\phi + \varphi}{2} \Bigr)^\prime = - \omega - \frac{M}{2\omega} \, \zeta^\prime 
    \;.
}
So in summary, the angle and phases are chosen such that 
\bsa{}{
    \tan 2\theta(\eta) & = \frac{\kappa(\eta)}{M(\eta)} \\ 
     \phi(\eta) & = \frac{1}{2} \zeta(\eta) - \Phi(\eta) - \delta(\eta) \\ 
     \varphi(\eta) & = - \frac{1}{2} \zeta(\eta) - \Phi(\eta) - \delta(\eta) \\ 
     \delta(\eta) & \equiv \int_{\eta_a}^\eta \! \dd{\eta^\prime} \, \frac{M(\eta^\prime)}{2\omega(\eta^\prime)} \, \zeta^\prime(\eta^\prime) \\ 
     \Phi(\eta) & \equiv \int_{\eta_a}^\eta \! \dd{\eta^\prime} \, \omega(\eta^\prime) 
    \;, 
}
where the lower limit of integration $\eta_a$ is arbitrary.  
The decomposition of $\phi+\varphi$ into $\Phi$ and $\delta$ is also arbitrary.\footnote{In Eq.~(3.35) of Ref.~\cite{Kolb:2021xfn} the constraint that $U \Omega U^{-1} + i U^\prime U^{-1}$ should have vanishing diagonal entries was not imposed, and the extra degree of freedom was used to set $\delta(\eta) = 0$.}  
With these choices, the unitary matrix is written as 
\begin{align}
    U = \frac{1}{\sqrt{2}} 
    \mqty(\sqrt{1+\frac{M}{\omega}} e^{-i\zeta/2} e^{i\delta} e^{i\Phi} & 
    \sqrt{1-\frac{M}{\omega}} e^{i\zeta/2} e^{i\delta} e^{i\Phi} \\ 
    -\sqrt{1-\frac{M}{\omega}} e^{-i\zeta/2} e^{-i\delta} e^{-i\Phi} & \sqrt{1+\frac{M}{\omega}} e^{i\zeta/2} e^{-i\delta} e^{-i\Phi}) 
    \;,
\end{align}
and it follows that 
\ba{
    U \Omega U^{-1} 
    = \begin{pmatrix} \omega(\eta) & 0 \\ 0 & -\omega(\eta) \end{pmatrix} 
    \qquad \text{and} \qquad 
    U^\prime U^{-1} 
    = \begin{pmatrix} i \omega(\eta) & -\mu(\eta) \, e^{2i\Phi(\eta)} \\  \mu^\ast(\eta) \, e^{-2i\Phi(\eta)} & -i \omega(\eta) \end{pmatrix}
    \;,
}
where $\mu = (\kappa M^\prime - M \kappa^\prime - i \kappa \omega \zeta^\prime) e^{2i\delta} / 2\omega^2$.  
Writing out the matrix products lets the new mode equations be written as 
\bsa{}{
    \alpha^\prime(\eta)
    & = - \mu(\eta) \, e^{2i \Phi(\eta)} \, \beta(\eta) \\ 
    \beta^\prime(\eta)
    & = \mu^\ast(\eta) \, e^{- 2i \Phi(\eta)} \, \alpha(\eta) 
    \;,
}
which is the same as Eq.~\eqref{eq:dalpha_dbeta} in the main text.  
Inverting the unitary transformation gives 
\bes{
    \mqty( u_A \\ u_B ) 
    & = U^{-1} \mqty( \alpha \\ \beta ) 
    = U^\dagger \mqty( \alpha \\ \beta ) \\ 
    & = \frac{1}{\sqrt{2}} 
    \mqty(\sqrt{1+\frac{M}{\omega}} e^{i\zeta/2} e^{-i\delta} e^{-i\Phi} & 
    - \sqrt{1-\frac{M}{\omega}} e^{i\zeta/2} e^{i\delta} e^{i\Phi} \\ 
    \sqrt{1-\frac{M}{\omega}} e^{-i\zeta/2} e^{-i\delta} e^{-i\Phi} & \sqrt{1+\frac{M}{\omega}} e^{-i\zeta/2} e^{i\delta} e^{i\Phi})  \mqty( \alpha \\ \beta ) \\ 
    & = \frac{1}{\sqrt{2}} 
    e^{-i\delta} e^{-i\Phi} \mqty( e^{i\zeta/2} \sqrt{1+\frac{M}{\omega}} \\ e^{-i\zeta/2} \sqrt{1-\frac{M}{\omega}} ) \alpha
    + \frac{1}{\sqrt{2}} e^{i\delta} e^{i\Phi} \mqty( - e^{i\zeta/2} \sqrt{1-\frac{M}{\omega}} \\ e^{-i\zeta/2} \sqrt{1+\frac{M}{\omega}} ) \beta
    \;,
}
which is the same as Eq.~\eqref{eq:uA_uB_solution} in the main text.

\section{WKB series for spinor field mode functions}\label{app:wkb_solution}

In this appendix, we construct a WKB series~\eqref{eq:wkb_solution} that solves the spinor field's mode equation~\eqref{eq:mode_eqn_with_hbar} order-by-order in powers of $\hbar$, and present the result in a standard normalized form \eqref{eq:normalized_EWKB_soln_for_u}.
To this end, we will first review the usual setting in which we construct the WKB series, show how the spinor field's mode equation can be reduced to the usual setting, give the recurrence relations that define the WKB series, and solve for the WKB series up to the leading order explicitly.

Recall that the WKB method is usually employed to solve second-order differential equations of the form 
\ba{
    \label{eqn:EWKB_setting}
    \bigg[\hbar^2 \dv[2]{x} + Q(x, \hbar)\bigg]y(x, \hbar) = 0 
    \;, \quad \text{where} \qquad 
    Q(x, \hbar) = \sum_{n=0}^{\infty} \hbar^n Q_n(x)
}
is a (formal) power series in $\hbar$. 
If solutions $y(x,\hbar)$ take the form of the WKB ansatz,
\begin{align}
    y(x,\hbar) = y_a \, e^{\frac{i}{\hbar} \int_{x_a}^x \! \dd{x^\prime} \, S(x^\prime, \hbar)}
    \quad \text{with} \quad  
    S(x, \hbar) = \sum_{n=0}^{\infty} \hbar^n S_n(x)
    \;,
\end{align}
then we have a first-order, nonlinear differential equation for $S(x,\hbar)$
\begin{align}
    i \hbar \dv{S}{x} - S(x, \hbar)^2 + Q(x, \hbar) = 0
    \;,
\end{align}
called a Riccati equation.  
Since the $S_n(x)$'s are assumed to be independent of $\hbar$, the $S(x,\hbar)$ equation must be solved order-by-order in powers of $\hbar$.  
This observation leads to these recurrence relations among the functions $S_n(x)$:
\ba{
& (S_0)^2 = Q_0 \nonumber \\ 
& S_n = \frac{1}{2 S_0} \bigg[i \dv{S_{n-1}}{x} + Q_n - \sum_{k=1}^{n-1} S_k S_{n-k} \bigg] \qq{for} n \geq 1
\;.
\label{eq:recurrence_relation}
}
The WKB series is constructed by solving these recurrence relations and plugging $S(x,\hbar)$ back into the WKB ansatz.

We are interested in solving the spinor field's mode equation~\eqref{eq:mode_eqn_with_hbar}:
\ba{\label{eq:u_eqn_1st_order}
    \hbar \begin{pmatrix} u_A^\prime \\ u_B^\prime \end{pmatrix} = -i \Omega(\eta) \begin{pmatrix} u_A \\ u_B \end{pmatrix}
    \qquad \text{where} \qquad 
    \Omega(\eta) = \begin{pmatrix} M(\eta) & K(\eta) \\ K^\ast(\eta) & -M(\eta) \end{pmatrix}
    \;,
}
where $M(\eta) = M(\eta)^\ast \geq 0$ is real and non-negative, and we can write $K(\eta) = \kappa(\eta) e^{i\zeta(\eta)}$ with real $\kappa$ and $\zeta$. To construct the WKB series, we assume that the mode equation has solutions of the form: 
\ba{\label{eq:WKB_ansatz_app}
    \mqty(u_A(\eta,\hbar) \\ u_B(\eta,\hbar)) = 
    \mqty(u_{A,a} \, e^{\frac{i}{\hbar} \int_{\eta_a}^{\eta} \dd{\eta'} \, S_A(\eta', \hbar)}  \\ 
    u_{B,a} \, e^{\frac{i}{\hbar} \int_{\eta_a}^{\eta} \dd{\eta'} \, S_B(\eta', \hbar)}) \qq{with}
    S_J(\eta, \hbar) = \sum_{n=0}^{\infty} \hbar^n S_{Jn}(\eta) 
}
for $J = A$ or $B$.  
The c-number coefficients are denoted by $u_{A,a}$ and $u_{B,a}$, and $\eta_a \in \mathbb{R}$ is an arbitrary anchor time.  

The mode equation takes the form of a Schr\"odinger equation for a two-level system with time-dependent Hamiltonian.  
It can be recast as a pair of oscillator equations, taking the form of Eq.~\eqref{eqn:EWKB_setting}, which is the usual setting for WKB analysis. 
We first differentiate with respect to $\eta$, which leads to a pair of second-order differential equations:
\bsa{eq:u_eqn_2nd_order}{
  0 &= \hbar^2 u_A'' + (\omega^2 + i \hbar M') u_A + i \hbar K' u_B \\
  0 &= \hbar^2 u_B'' + (\omega^2 - i \hbar M') u_B + i \hbar {K^*}' u_A
  \;,
}
where $\omega^2 \equiv |K|^2 + M^2$. 
Assuming that $K(\eta)$ doesn't vanish, we use the two first-order equations to decouple $u_A$ and $u_B$ in the second-order equations, which gives 
\bsa{}{
  0 & = \hbar^2 u_A'' - \hbar^2 \frac{K'}{K} u_A' + \Bigl( \omega^2 + i \hbar M' - i \hbar \frac{K'}{K} M \Bigr) u_A \\ 
  0 & = \hbar^2 u_B'' - \hbar^2 \frac{K^{\ast\prime}}{K^*} u_B' + \Bigl( \omega^2 - i \hbar M' + i \hbar \frac{K^{\ast\prime}}{K^*} M \Bigr) u_B 
  \;,
}
The $u_A^\prime$ and $u_B^\prime$ terms are removed with a change of variables, 
\bsa{eq:u_to_v}{
    v_A(\eta,\hbar) & = \exp(-\frac{1}{2} \int_{\eta_a}^\eta \! \dd{\eta'} \frac{K'}{K}) u_A(\eta,\hbar) 
    = \sqrt{\frac{K(\eta_a)}{K(\eta)}} \, u_A(\eta,\hbar) \\ 
    v_B(\eta,\hbar) & = \exp(-\frac{1}{2} \int_{\eta_a}^\eta \! \dd{\eta'} \frac{{K^*}'}{K^*}) u_B(\eta,\hbar) 
    = \sqrt{\frac{K^\ast(\eta_a)}{K^\ast(\eta)}} \, u_B(\eta,\hbar) 
    \;,
}
where the new mode functions, $v_A(\eta)$ and $v_B(\eta)$, must solve 
\bsa{eq:EWKBeq}{
    & \hbar^2 v_A'' + \biggl[ \omega^2 + \hbar \biggl( i M' - i \frac{K^\prime}{K} M \biggr) + \hbar^2 \biggl( \frac{K''}{2K} - \frac{3 K^{\prime 2}}{4 K^2} \biggr) \biggr] v_A = 0 \\
    & \hbar^2 v_B'' + \biggl[ \omega^2 + \hbar \biggl( -i M' + i \frac{K^{\ast\prime}}{K^\ast} M \biggr) + \hbar^2 \biggl( \frac{K^{\ast\prime\prime}}{2K^\ast} - \frac{3 K^{\ast \prime 2}}{4 K^{\ast2}} \biggr) \biggr] v_B = 0
    \;.
}
This sequence of manipulations has expressed the spinor field's mode equations in the familiar form of the WKB problem~\eqref{eqn:EWKB_setting}. 
The equations~\eqref{eq:EWKBeq} are written collectively as 
\ba{\label{eq:vJeqn}
    \biggl[ \hbar^2 \dv[2]{\eta} + Q_J(\eta,\hbar) \biggr] v_J(\eta,\hbar) = 0 
    \;, \quad \text{where} \qquad 
    Q_J(\eta, \hbar) = \sum_{n=0}^{\infty} \hbar^n Q_{Jn}(\eta)
}
for $J=A$ or $B$.
The power series coefficients are 
\ba{\label{eq:QJn_series}
\begin{array}{l}
    Q_{A0} = \omega^2 \equiv |K|^2 + M^2 \\ 
    Q_{A1} = i M^\prime - i K^\prime M / K \\ 
    Q_{A2} = K''/2K - 3 K^{\prime2}/4K^2 \\ 
    Q_{An} = 0 \quad \text{for $n \geq 3$} \\ 
\end{array}
\qquad \text{and} \qquad 
\begin{array}{l}
    Q_{B0} = \omega^2 \equiv |K|^2 + M^2 \\ 
    Q_{B1} = - i M^\prime + i K^{\ast\prime} M / K^\ast \\ 
    Q_{B2} = K^{\ast\prime\prime}/2K^\ast - 3 K^{\ast\prime2}/4K^{\ast2} \\ 
    Q_{Bn} = 0 \quad \text{for $n \geq 3$} \\ 
\end{array} 
    \;,
}
such that $Q_J$ includes terms up to $\order{\hbar^2}$, and $Q_{Bn} = Q_{An}^\ast$.   

We seek solutions of Eq.~\eqref{eq:vJeqn} in the form of:
\ba{
    v_J(\eta,\hbar) = v_{J,a} \, e^{\frac{i}{\hbar} \int_{\eta_a}^{\eta} \dd{\eta'} \, T_J(\eta', \hbar)} 
    \qquad \text{with} \qquad 
    T_J(\eta, \hbar) = \sum_{n=0}^{\infty} \hbar^n T_{Jn}(\eta) 
    \;.
    \label{eq:WKB_ansatz_app_v}
}
Equating Eqs.~\eqref{eq:WKB_ansatz_app}~and~\eqref{eq:u_to_v} gives
\bsa{}{
    v_A(\eta,\hbar) 
    & = u_{A,a} \, e^{\frac{i}{\hbar} \int_{\eta_a}^{\eta} \dd{\eta'} \, \left(S_A + i \hbar \frac{K'}{2 K} \right)} \\
    v_B(\eta,\hbar) 
    & = u_{B,a} \, e^{\frac{i}{\hbar} \int_{\eta_a}^{\eta} \dd{\eta'} \, \left(S_B + i \hbar \frac{{K^*}'}{2 K^*} \right)}
    \;,
}
and comparing with Eq.~\eqref{eq:WKB_ansatz_app_v} implies 
\bsa{eq:S_T_relation}{
    T_A(\eta,\hbar) & = S_A(\eta,\hbar) + i \hbar \frac{K'(\eta)}{2 K(\eta)} \\ 
    T_B(\eta,\hbar) & = S_B(\eta,\hbar) + i \hbar \frac{K^{\ast\prime}(\eta)}{2 K^\ast(\eta)} 
    \;.
}
In this way the WKB series for $u_J(\eta,\hbar)$ is related to the WKB series for $v_J(\eta,\hbar)$.  

We shall use the recurrence relations~\eqref{eq:recurrence_relation} to solve for $T_J$, and then Eq.~\eqref{eq:S_T_relation} determines $S_J$.
At order $n=0$, the recurrence relations provide two solutions to $T_{J0}$:
\ba{
    T_{A0} = \pm \omega \ , \quad T_{B0} = \pm \omega 
    \;.
}
We will see below that two of the four sign combinations are spurious solutions, and $T_{A0} = T_{B0} = \pm \omega$ is required, which correspond to the negative/positive frequency modes.  
At order $n=1$, we then have:
\bsa{}{
    & T_{A1} = \frac{1}{2(\pm\omega)}\Big[i (\pm\omega') + i M' - i M \frac{K'}{K} \Big] = i\frac{\omega'}{2\omega} \pm i\frac{M' \kappa - \kappa' M}{2 \omega \kappa} \pm \frac{M}{2 \omega} \zeta' \\
    & T_{B1} = \frac{1}{2(\pm\omega)}\Big[i (\pm\omega') - i M' + i M \frac{{K^*}'}{K^*} \Big] = i\frac{\omega'}{2\omega} \mp i\frac{M' \kappa - \kappa' M}{2 \omega \kappa} \pm \frac{M}{2 \omega} \zeta'
}
and
\bsa{eq:SJ1}{
    & S_{A1} = T_{A1} - i \frac{K'}{2K} = i\frac{\omega M' - M \omega'}{2 \omega (- M \pm \omega)} + \frac{\zeta'}{2} \pm \frac{\zeta' M}{2\omega} \\
    & S_{B1} = T_{B1} - i \frac{{K^*}'}{2K^*} = i\frac{-\omega M' + M \omega'}{2 \omega ( M \pm \omega)} - \frac{\zeta'}{2} \pm \frac{\zeta' M}{2\omega} .
}
Here we've used $\omega^2 = \kappa^2 + M^2$ to eliminate $\kappa$ and $\kappa^\prime$. 

For $(u_A,u_B)$ to be a WKB solution of the mode equation~\eqref{eq:mode_eqn_with_hbar}, we expect that $u_A$ and $u_B$ are either both negative or both positive frequency modes; in other words, $T_{A0} = T_{B0} = \pm \omega$. Here we show that this is indeed the case.
Plugging the WKB ansatz \eqref{eq:WKB_ansatz_app} back into the mode equation \eqref{eq:u_eqn_1st_order}, we get:
\ba{\label{eq:S_eqn_matrix}
    0 &= \mqty(S_A + M & K \\ K^* & S_B - M) \mqty(u_A\\ u_B) .
}
In this form $(u_A,u_B)$ is understood as an eigenvector of the matrix with eigenvalue equal to zero. 
Thus for this linear equation to have non-zero solution $(u_A, u_B)$, the determinant of the matrix must be zero. 
At order $n=0$, this condition implies:
\ba{
    (-S_{A0} + S_{B0}) M + S_{A0} S_{B0} = (-T_{A0} + T_{B0}) M + T_{A0} T_{B0} = \omega^2 .
}
Since $T_{A0} = \pm \omega$ and $T_{B0} = \pm \omega$, the only sign choices that satisfy the equation above for nonzero $K$ are $T_{A0} = T_{B0} = \pm \omega$. 
Thus the functions 
\ba{\label{eq:SBJpm}
    u_{J\pm}(\eta,\hbar) = u_{J\pm,a} \, e^{\frac{i}{\hbar} \int_{\eta_a}^\eta \! \dd{\eta'} S_{J\pm}(\eta^\prime,\hbar)}
    \quad \text{with} \quad
    S_{A0\pm}(\eta) = S_{B0\pm}(\eta) = \mp \omega(\eta) 
}
for $J = A$ or $B$, are called the positive/negative frequency solutions.  

It's worth noting that the $n=0$ and $n=1$ terms in the WKB series are related to the instantaneous positive and negative frequency mode functions \eqref{eq:uA_uB_solution} in the following way:
\ba{\label{eq:utilde}
    \exp(\frac{i}{\hbar} \int_{\eta_a}^{\eta} \! \dd{\eta^\prime} \, \sum_{n=0}^1 \hbar^n S_{Jn\pm}(\eta^\prime)) = \frac{\tilde{u}_{J\pm}(\eta,\hbar)}{\tilde{u}_{J\pm}(\eta_a,\hbar)} \;.
}
This demonstrates how the instantaneous mode functions are precisely the leading order WKB solutions.

The coefficients $u_{A,a}$ and $u_{B,a}$ appearing in the WKB series~\eqref{eq:WKB_ansatz_app} can be calculated using Eq.~\eqref{eq:S_eqn_matrix} and a normalization condition.  
Evaluating Eq.~\eqref{eq:S_eqn_matrix} at $\eta = \eta_a$ gives 
\bes{
    \frac{u_{A\pm,a}}{u_{B\pm,a}} 
    = - \frac{K(\eta_a)}{S_{A\pm}(\eta_a,\hbar) + M(\eta_a)} 
    = - \frac{S_{B\pm}(\eta_a,\hbar) - M(\eta_a)}{K^*(\eta_a)} 
    = \pm \eval{ \sqrt{ \frac{S_{B\pm} - M}{S_{A\pm} + M}} \, e^{i \zeta} }_{\eta_a}
    \;,
}
where $K/K^\ast = e^{2 i \zeta}$.  
The sign before the square root is chosen such that the expression is valid in the $\hbar \to 0$ limit, wherein $S_{J\pm} \to S_{J0\pm} = \mp \omega$. 
If we further impose the normalization condition $\abs{u_{A,a}}^2 + \abs{u_{B,a}}^2 = 1$, then the coefficients can be solved up to a common phase:
\bsa{eq:uAa_and_uBa}{
    u_{A\pm,a} & = C \eval{\frac{\pm 1}{\sqrt{2}} e^{i\zeta/2} \sqrt{1 - \frac{M}{S_{B\pm}}} \frac{1}{\sqrt{\mp S_{A\pm}}} }_{\eta_a} \\ 
    u_{B\pm,a} & = C \eval{\frac{1}{\sqrt{2}} e^{-i\zeta/2} \sqrt{1 + \frac{M}{S_{A\pm}}} \frac{1}{\sqrt{\mp S_{B\pm}}} }_{\eta_a} 
    \;,
}
where
\ba{
    C 
    = \eval{\sqrt{ \frac{2 \abs{S_{A+} S_{B+}}}{\abs{S_{B+} - M} + \abs{S_{A+} + M}} } }_{\eta_a}
    = \eval{\sqrt{ \frac{2 \abs{S_{A-} S_{B-}}}{\abs{S_{B-} - M} + \abs{S_{A-} + M}} } }_{\eta_a}
    \;.
}
Here, $C > 0$ is a normalization constant; the two expressions above for $C$ are equal due to Eq.~\eqref{eq:SA_SB_relation}, which we will show later. In the $\hbar \to 0$ limit, $S_{J\pm} \to \mp \omega$, and $C \to \sqrt{\omega}$.

Finally, combining Eqs.~\eqref{eq:WKB_ansatz_app}~and~\eqref{eq:uAa_and_uBa} gives an expression for the full WKB series: 
\bsa{eq:normalized_EWKB_soln_for_u}{
    u_{A\pm}(\eta,\hbar) 
    & = 
    \mathcal{N}_{A,\pm}(\hbar) \  
    \tilde{u}_{A,\pm}(\eta,\hbar) \, 
    \exp[\frac{i}{\hbar} \int_{\eta_a}^{\eta} \! \dd{\eta'} \sum_{n=2}^{\infty} \hbar^n S_{An\pm}(\eta')] 
    \\
    u_{B\pm}(\eta,\hbar) 
    & = 
    \mathcal{N}_{B,\pm}(\hbar) \  
    \tilde{u}_{B,\pm}(\eta,\hbar) \, 
    \exp[\frac{i}{\hbar} \int_{\eta_a}^{\eta} \! \dd{\eta'} \sum_{n=2}^{\infty} \hbar^n S_{Bn\pm}(\eta')] 
    \;,
}
where
\bsa{}{
    \mathcal{N}_{A,\pm}(\hbar) & \equiv 
    \eval{ \sqrt{ \frac{2 \abs{S_{A\pm} S_{B\pm}}}{\abs{S_{B\pm} - M} + \abs{S_{A\pm} + M}} \frac{1 - \frac{M}{S_{B\pm}}}{1 \pm \frac{M}{\omega}} \frac{\mp 1}{S_{A\pm}} } \  }_{\eta_a} \\ 
    \mathcal{N}_{B,\pm}(\hbar) & \equiv 
    \eval{ \sqrt{ \frac{2 \abs{S_{A\pm} S_{B\pm}}}{\abs{S_{B\pm} - M} + \abs{S_{A\pm} + M}} \frac{1 + \frac{M}{S_{A\pm}}}{1 \mp \frac{M}{\omega}} \frac{\mp 1}{S_{B\pm}} } \  }_{\eta_a}
    \;,
}
and where 
\bsa{eq:utilde_def}{
    \tilde{u}_{A,\pm}(\eta,\hbar) & \equiv 
    \pm \frac{1}{\sqrt{2}} \, e^{\mp i (\Phi/\hbar \, + \, \delta)} \, e^{i\zeta/2} \, \sqrt{1 \pm \frac{M}{\omega}} \\ 
    \tilde{u}_{B,\pm}(\eta,\hbar) & \equiv 
    \frac{1}{\sqrt{2}} \, e^{\mp i (\Phi/\hbar \, + \, \delta)} \, e^{-i\zeta/2} \, \sqrt{1 \mp \frac{M}{\omega}} 
    \;.
}
The first factors, $\mathcal{N}_{A,\pm}$ and $\mathcal{N}_{B,\pm}$ are evaluated at the anchor time $\eta_a$ and carry no time dependence; they are normalized to go to $1$ as $\hbar \to 0$.  
The second factors, $\tilde{u}_{A,\pm}$ and $\tilde{u}_{B,\pm}$ are the instantaneous positive/negative frequency mode functions for arbitrary $\hbar$; compare with Eq.~\eqref{eq:uA_uB_solution}.  
The phase integrals, 
\ba{
    \Phi(\eta) \equiv \int_{\eta_a}^{\eta} \! \dd{\eta^\prime} \, \omega(\eta') 
    \qquad \text{and} \qquad 
    \delta(\eta) \equiv \int_{\eta_a}^\eta \! \dd{\eta^\prime} \, \frac{M}{2 \omega} \zeta' 
    \;,
}
vanish at the anchor time $\eta_a$.  
The third factor starts at $\order{\hbar^1}$; the $n=0$ and $n=1$ terms are included in the second factor, since $\tilde{u}_{J,\pm}$ is given by Eq.~\eqref{eq:utilde}.  

The normalized solutions at the anchor time form a special unitary basis: 
\begin{subequations}
\ba{\label{eq:u_J_const_normalization}
    \mqty(\psi_{+,a} & \ \psi_{-,a})
    \equiv \mqty(u_{A+,a} & \ u_{A-,a} \\ u_{B+,a} & \ u_{B-,a})
    \in \mathrm{SU}(2) 
    \;,
}
or equivalently
\bes{
    & \psi_{\pm,a}^\dagger \psi_{\pm,a} = 1 
    \ , \quad 
    \psi_{\pm,a}^\dagger \psi_{\mp,a} = 0 
    \ , \quad 
    \psi_{+,a} \psi_{+,a}^\dagger + \psi_{+,a} \psi_{+,a}^\dagger = I 
    \ , \quad 
    \\ & \quad \text{and} \quad
    \det \mqty(\psi_{+,a} & \ \psi_{-,a}) = u_{A+,a} \, u_{B-,a} - u_{A-,a} \, u_{B+,a} = 1 
    \;.
}
\end{subequations}
One can verify these relations using the explicit expressions for $u_{A\pm,a}$ and $u_{B\pm,a}$ from Eq.~\eqref{eq:uAa_and_uBa} along with the relations 
\ba{
    S_{A-}^\ast(\eta,\hbar) = - S_{B+}(\eta,\hbar) 
    \qquad \text{and} \qquad S_{B-}^\ast(\eta,\hbar) = - S_{A+}(\eta,\hbar)
    \;.
    \label{eq:SA_SB_relation}
}
These relations derive from the following argument.  
Since $\Omega(\eta)$ is a Hermitian and traceless matrix, it obeys $\sigma \Omega^\ast \sigma = \Omega$ with $\sigma = ((0,1),(-1,0))$.  
So if $\psi = (u_A, u_B)$ is a solution of the mode equation~\eqref{eq:u_eqn_1st_order}, then $\psi = (u_B^\ast, -u_A^\ast)$ is also a solution.  
This implies that the positive and negative frequency solutions are related by $u_{B\mp}^\ast(\eta,\hbar) / u_{B\mp,a}^\ast = u_{A\pm}(\eta,\hbar) / u_{A\pm,a}$.  
Writing $u_{A\pm}$ and $u_{B\pm}$ in terms of the WKB series~\eqref{eq:SBJpm} implies the relations $S_{A-}^\ast(\eta,\hbar) = - S_{B+}(\eta,\hbar)$ and $S_{B-}^\ast(\eta,\hbar) = - S_{A+}(\eta,\hbar)$. 
This argument follows from the mode equation's invariance under charge conjugation. 

\section{Connection matrix for Landau-Zener problem}
\label{app:landau_zener}

In Sec.~\ref{sec:discussion}, we have defined the connection matrix $T$ as the matrix relating the expansion coefficients for different EWKB basis solutions $\psi_\pm^{(1)}$ and $\psi_\pm^{(2)}$ of the mode equation \eqref{eq:mode_eqn_with_hbar}. 
If $\psi_\pm^{(1)}$ and $\psi_\pm^{(2)}$ have anchor times in different Stokes regions, then $T$ is non-diagonal, corresponding to non-zero particle production. 
For pedagogical reasons, it is instructive to calculate $T$ for a concrete example of Stokes line crossing. 
One of such example is the Landau-Zener problem~\cite{196563, Zener:1932ws}, which can be solved exactly. 
We shall see in later appendices that a general mode equation \eqref{eq:mode_eqn_with_hbar} resembles the Landau-Zener problem as far as Stokes line crossing is concerned, so understanding the Landau-Zener problem is useful for understanding the general problem as well. 
Therefore, in this appendix, we consider the Landau-Zener problem, present two pairs of basis solutions with normalization \eqref{eq:unitary_normalization}, and derive the connection matrix between them explicitly.

The Landau-Zener problem corresponds to the Schr\"odinger equation for a two-state quantum system near the point of avoided crossing.  
The problem is specified by 
\ba{
  i \hbar \dv{\eta} \psi(\eta,\hbar) = \mqty(-\frac{v}{2}(\eta - \eta_*) & \Delta \, e^{i\zeta} \\ \Delta \, e^{-i\zeta} & \frac{v}{2}(\eta - \eta_*) ) \psi(\eta,\hbar) 
  \quad \text{with} \quad 
  \psi(\eta,\hbar) \equiv \mqty(u_A(\eta,\hbar) \\ u_B(\eta,\hbar)) 
  \;,
}
where $v > 0$, $\Delta > 0$, $\zeta \in \mathbb{R}$, and $\eta_* \in \mathbb{R}$ are constants. 
The matrix has eigenvalues $\pm \omega(\eta)$ with $\omega(\eta) = \sqrt{ v^2 (\eta-\eta_\ast)^2/4 + \Delta^2}$, and for $\Delta \neq 0$ the level-crossing at $\eta=\eta_\ast$ is avoided. 
Changing variables with 
\ba{\label{eq:lz_appendix_w_to_u}
    w_A(x,\hbar) = e^{-i\zeta/2} \, u_A(\eta,\hbar) 
    \ ,\quad 
    w_B(x,\hbar) = e^{i\zeta/2} \, u_B(\eta,\hbar) 
    \ ,\quad
    x = \sqrt{v}(\eta - \eta_*) 
    \;,
}
lets the equation be recast into 
\ba{\label{eq:lz_w_eqn}
    i \hbar \dv{x} \mqty(w_A(x,\hbar) \\ w_B(x,\hbar)) = \mqty(-\frac{x}{2} & \sqrt{\epsilon} \\ \sqrt{\epsilon} & \frac{x}{2}) \mqty(w_A(x,\hbar) \\ w_B(x,\hbar)) 
  \;,
}
where $\epsilon = \Delta^2 / v$.  
By taking an additional $x$ derivative, one can further obtain decoupled 2nd-order equations:
\bes{\label{eq:lz_weber}
    & \hbar^2 \dv[2]{w_A}{x} + \left(\epsilon - \frac{1}{2}i \hbar + \frac{x^2}{4} \right) w_A = 0 \\
    & \hbar^2 \dv[2]{w_B}{x} + \left(\epsilon + \frac{1}{2}i \hbar + \frac{x^2}{4} \right) w_B = 0 \;.
}
One can see that these are exactly the Weber equation \eqref{eq:Webereq} with $E = E_A$ and $E = E_B$, where $E_A \equiv \epsilon - i\hbar/2$ and $E_B \equiv \epsilon + i\hbar/2$. As discussed in Sec.~\ref{sec:discussion}, the solutions for $w_A$ and $w_B$ are given by the parabolic cylinder functions \eqref{eq:Weber_soln}, and the asymptotic forms of those functions are given by Eq.~\eqref{eq:ParaCyn_asym}. Furthermore, one can find the Stokes line for both $w_A$ and $w_B$ equation via definition \eqref{eq:Stokes1st}, and see that it is the straight line from $2i\sqrt{\epsilon}$ to $-2i\sqrt{\epsilon}$. The Stokes line crossing happens at $x = 0$, where the Stokes line intersects the real axis.

The general solution of the 1st-order mode equation \eqref{eq:lz_w_eqn} can be written in two different ways, in terms of the parabolic cylinder functions:
\bes{
  \mqty(w_A \\ w_B) =&\ 
  \lambda_+^{(1)} \mqty( D_n(-iz) \\ e^{-\frac14 i \pi} \sqrt{\epsilon/\hbar} D_{n-1}(-iz)  ) + 
  \lambda_-^{(1)} \mqty( D_{-n-1}(z) \\ e^{\frac34 i\pi} \frac{1}{\sqrt{\epsilon/\hbar}} D_{-n}(z) ) \\
  =&\ 
  \lambda_+^{(2)} \mqty( D_{-n-1}(-z) \\ e^{-\frac14 i \pi} \frac{1}{\sqrt{\epsilon/\hbar}} D_{-n}(-z) ) + 
  \lambda_-^{(2)} \mqty( D_{n}(iz) \\ e^{\frac34 i\pi} \sqrt{\epsilon/\hbar} D_{n-1}(iz) )  \;,
  \label{eq:lz_solution}
}
where $n \equiv i \epsilon / \hbar $ and $z \equiv e^{-\frac34 i \pi} x / \sqrt{\hbar} $.
In the first line, $\lambda_+^{(1)}$ and $\lambda_-^{(1)}$ are arbitrary constants, and the column vectors multiplied to them are exactly the positive/negative frequency solutions in the $x<0$ Stokes region; similarly, $\lambda_+^{(2)}$ and $\lambda_-^{(2)}$ are associated with the positive/negative frequency solutions in the $x>0$ Stokes region. One can also treat $w_A$ and $w_B$ as solutions to the (2nd-order) Weber equation \eqref{eq:lz_weber}, then the respective entries of the column vectors are the positive/negative frequency solutions to the Weber equation, as was discussed in \eqref{eq:w_to_ParaCyn}. 

From the solution in Eq.~\eqref{eq:lz_solution}, we can read off the positive and negative frequency mode functions:
\bsa{eq:lz_psi_full}{
    \psi_+^{(1)} &= e^{-\frac{\pi \epsilon}{4 \hbar}} \mqty( e^{i\zeta/2} & 0 \\ 0 & e^{-i\zeta/2}) \mqty( D_n(-iz) \\ e^{-\frac14 i \pi} \sqrt{\epsilon/\hbar} D_{n-1}(-iz) ) \\
    \psi_-^{(1)} &= \sqrt{\frac{\epsilon}{\hbar}} e^{-\frac{\pi \epsilon}{4 \hbar}} e^{-\frac34 i \pi} \mqty( e^{i\zeta/2} & 0 \\ 0 & e^{-i\zeta/2}) \mqty( D_{-n-1}(z) \\ e^{\frac34 i\pi} \frac{1}{\sqrt{\epsilon/\hbar}} D_{-n}(z) ) \\
    \psi_+^{(2)} &= \sqrt{\frac{\epsilon}{\hbar}} e^{-\frac{\pi \epsilon}{4 \hbar}} \mqty( e^{i\zeta/2} & 0 \\ 0 & e^{-i\zeta/2}) \mqty( D_{-n-1}(-z) \\ e^{-\frac14 i \pi} \frac{1}{\sqrt{\epsilon/\hbar}} D_{-n}(-z) ) \\
    \psi_-^{(2)} &= e^{-\frac{\pi \epsilon}{4 \hbar}} e^{-\frac34 i \pi} \mqty( e^{i\zeta/2} & 0 \\ 0 & e^{-i\zeta/2}) \mqty( D_{n}(iz) \\ e^{\frac34 i\pi} \sqrt{\epsilon/\hbar} D_{n-1}(iz) ) \;.
}
The factors of $e^{\pm i \zeta/2}$ arise from the transformation of $w$'s into $u$'s~\eqref{eq:lz_appendix_w_to_u}.  
The coefficients are fixed by imposing 
\ba{\label{eq:lz_app_SU2}
  \mqty(\psi_+^{(1)} & \psi_-^{(1)}) \in \mathrm{SU}(2)
  \qquad \text{and} \qquad 
  \mqty(\psi_+^{(2)} & \psi_-^{(2)}) \in \mathrm{SU}(2) 
}
at $z=x=0$, similar to the normalization conditions in Eq.~\eqref{eq:unitary_normalization}.  
The coefficients are only determined up to a phase since the transformation $\psi_\pm \mapsto e^{\pm i \theta} \psi_\pm$ maintains the normalization condition~\eqref{eq:lz_app_SU2}.  
This phase degree of freedom does not affect the unitarity of the connection matrix $T$ or observable quantities like $\abs{\beta}^2 = \abs{T_{12}}^2$.

The connection matrix $T$ is calculated from Eq.~\eqref{eq:connection_matrix_formula} with $x=0$, which evaluates to 
\bes{
    T 
    &= \mqty(
    \frac{e^{-\frac{3\pi\epsilon}{2\hbar}} (e^{\frac{2\pi\epsilon}{\hbar}}-1) \Gamma(1 + i\frac{\epsilon}{\hbar})}{\sqrt{2\pi} \sqrt{\epsilon/\hbar}} &
    e^{\frac14 i \pi} e^{-\frac{\pi\epsilon}{\hbar}} \\
    e^{\frac34 i \pi} e^{-\frac{\pi\epsilon}{\hbar}} &
    \frac{e^{-\frac{3\pi\epsilon}{2\hbar}} (e^{\frac{2\pi\epsilon}{\hbar}}-1) \Gamma(1 - i\frac{\epsilon}{\hbar})}{\sqrt{2\pi} \sqrt{\epsilon/\hbar}} 
    ) 
    \;.
}
The gamma function can be evaluated with help of the identity $\abs{\Gamma(1 + ib)}^2 = \pi b / \sinh(\pi b)$ for $b \in \mathbb{R}$.
Finally, the connection matrix in the Landau-Zener model, relating solutions in the two Stokes regions, is found to be~\cite{Enomoto:2020xlf} 
\ba{
    T = \mqty(
    \sqrt{1 - e^{-2\Psi/\hbar}} \, e^{i \theta_1} &
    e^{-\Psi/\hbar} \, e^{i \theta_2} \\
    - e^{-\Psi/\hbar} \, e^{-i \theta_2} &
    \sqrt{1 - e^{-\Psi/\hbar}} \, e^{-i \theta_1}
    ) \;,
}
where 
\ba{
    \Psi 
    = \pi \epsilon 
    = \pi \Delta^2 / v 
    \ , \quad 
    \theta_1 = \mathrm{arg} \, \Gamma(1+i\epsilon/\hbar) 
    \ , \quad \text{and} \quad 
    \theta_2 = \pi/4 
    \;.
}
Notice that $\Psi$ is equivalent to the phase integral from Eq.~\eqref{eq:Psi_def}:
\ba{
    \Psi 
    & = i \int_{\eta_c}^{\eta_c^\ast} \! \dd{\eta} \, \omega(\eta) 
    = i \int_{2i\sqrt{\epsilon}}^{-2i\sqrt{\epsilon}} \! \dd{x} \, \sqrt{\epsilon + \frac{x^2}{4}} 
    = \pi \epsilon 
    \;.
}
The phases, $\theta_1$ and $\theta_2$, don't affect observables such as $|\beta_k|^2$.  
Notice that $T \in \mathrm{SU}(2)$ is manifestly true. 
Also note that the expression for $T$ is exact (to all orders in $\hbar$).  

\section{Derivation of the connection matrix} \label{app:connection_matrix}
As was discussed in Sec.~\ref{sec:discussion}, the connection matrix between EWKB basis solutions is the primary object of study in gravitational particle production. 
In this appendix, we derive the connection matrix based on Theorem~2.4 of \cite{Aoki:1991} and Theorem~3.1 of \cite{10.2977/prims/1195166424}.

\subsection{Introduction of the theorem}
Before going into the technical details, we first address some generalities on the significance of Theorem~2.4 in \cite{Aoki:1991}.
In gravitational particle production, both fermionic and bosonic mode equations can be cast into the form:
\ba{
    \left(\hbar^2\dv[2]{\eta} + Q(\eta, \hbar) \right){v}(\eta, \hbar) = 0, \qq{where} Q(\eta, \hbar) = \sum_{n=0}^{\infty} \hbar^n Q_n(\eta)
    \;.
    \label{eq:tilde_psi_eqn}
}
For example, Eq.~\eqref{eq:EWKBeq} gives the $Q(\eta, \hbar)$ for fermionic mode equations and $Q(\eta,\hbar) = k^2 + a(\eta)^2 m^2$ for conformally coupled scalars \cite{Chung:1998zb, Birrell:1982ix}. 
If we draw the Stokes lines by looking at $Q_0(\eta)$, we will find that it is common for Stokes lines to appear in the form of Merged pair of simple Turning Points (MTP), wherein a turning point of $Q_0$ is connected to another turning point via a Stokes line; see Fig.~\ref{fig:Stokes_concept}. 
In fact, Ref.~\cite{Taya:2020dco} has shown that as long as $Q_0(\eta)$ is real for real $\eta$, a turning point $\eta_c$ will always be connected to $\eta_c^*$ via a Stokes segment, and the Stokes graph is symmetric with respect to the real axis. 

Theorem~2.4 of \cite{Aoki:1991} and Theorem~3.1 of \cite{10.2977/prims/1195166424} demonstrate that any Eq.~\eqref{eq:tilde_psi_eqn} with a MTP Stokes structure can be locally reduced to the Weber equation:
\ba{
    \left(\hbar^2\dv[2]{x} + 
    E(\hbar) + \frac{x^2}{4} \right)w(x, \hbar) = 0, 
    \qq{where} E = \sum_{n=0}^{\infty} \hbar^n E_n
    \;.
    \label{eq:psi_eqn}
}
More specifically, there exist appropriately normalized WKB solutions $\hat{v}_\pm(\eta, \hbar)$ of Eq.~\eqref{eq:tilde_psi_eqn}, and appropriately normalized WKB solutions $\hat{w}_\pm(x, \hbar)$ of Eq.~\eqref{eq:psi_eqn}, such that:
\bes{
    & \hat{v}_{\pm}(\eta, \hbar) = 
    \left( \pdv{x(\eta, \hbar)}{\eta} \right)^{-1/2} 
    \hat{w}_{\pm}(x(\eta, \hbar), \hbar) 
    \qq{where}  x(\eta, \hbar) \equiv \sum_{n=0}^{\infty} \hbar^n x_n(\eta) \;.
    \label{eq:v_w_relation}
}
Here, $x(\eta, \hbar)$ is a $\hbar$-dependent holomorphic coordinate that can be solved order-by-order in $\hbar$ solely from $Q(\eta, \hbar)$. Since any EWKB basis solution ${v}_\pm$ is a multiple of $\hat{v}_\pm$, one can use the above result to turn a pair of EWKB basis solutions ${v}_\pm^{(1)}$ and ${v}_\pm^{(2)}$ of Eq.~\eqref{eq:tilde_psi_eqn} into a pair of EWKB basis solutions $w_\pm^{(1)}$ and $w_\pm^{(2)}$ of Eq.~\eqref{eq:psi_eqn}, and infer the connection matrix between ${v}_\pm^{(1)}$ and ${v}_\pm^{(2)}$ from that between ${w}_\pm^{(1)}$ and ${w}_\pm^{(2)}$, which can be derived from exact solutions of the Weber equation \eqref{eq:psi_eqn}.

To clarify what ``appropriately normalized'' means for $\hat{v}_\pm(\eta, \hbar)$ and $\hat{w}_\pm(x, \hbar)$, we first write down a generic WKB ansatz: 
\ba{
    \hat{v}_\pm(\eta, \hbar) = \exp( \frac{i}{\hbar} \int^\eta \dd{\eta'} T_\pm)
    \;,
}
then $T_\pm(\eta,\hbar)$ satisfies a Riccati equation:
\ba{
    i \hbar \dv{T_\pm}{\eta} - T_\pm(\eta, \hbar)^2 + Q(\eta, \hbar) = 0
    \;.
}
By subtracting the $T_+$ and $T_-$ equations, one obtain:
\bes{
    & 2 i \hbar \dv{\eta} T_\mathrm{odd} = 4 T_\mathrm{odd} T_\mathrm{even} \qq{or}
    T_\mathrm{even} = \frac{i \hbar}{2} \dv{\eta}\log(T_\mathrm{odd}) 
    \;, \\
    & \mathrm{where} \quad T_\mathrm{odd} \equiv \frac12 (T_+ - T_-),\quad T_\mathrm{even} \equiv \frac12 (T_+ + T_-)
    \;.
}
Now let $\eta_c$ be one of the two turning points of $Q_0(\eta)$. By eliminating $T_\mathrm{even}$ from the WKB ansatz and setting the lower limit of integration to $\eta_c$, one can define a WKB ansatz with a fixed normalization:
\ba{
    \hat{v}_\pm(\eta, \hbar) = \frac{1}{\sqrt{T_\mathrm{odd}}} \exp(\pm \frac{i}{\hbar} \int_{\eta_c}^\eta \dd{\eta'} T_\mathrm{odd} ) \;.
    \label{eq:normalization_at_turning_point}
}
Upon Borel resummation, this ansatz gives the ``appropriately normalized'' EWKB solutions $\hat{v}_\pm(\eta, \hbar)$. We shall call such solutions to be ``normalized at turning point $\eta_c$.'' Similarly, one can define EWKB solutions 
\ba{
    \hat{w}_\pm(x, \hbar) = \frac{1}{\sqrt{R_\mathrm{odd}}} \exp(\pm \frac{i}{\hbar} \int_{x_c}^x \dd{x'} R_\mathrm{odd} ) \;,
    \label{eq:normalization_at_turning_point_what}
}
normalized at turning point $x_c = 2 i \sqrt{E_0}$. The relation \eqref{eq:v_w_relation} given by the theorem is valid with respect to the normalization given above. 

\begin{figure}[tb]
	\centering
	\includegraphics[width=0.5\textwidth]{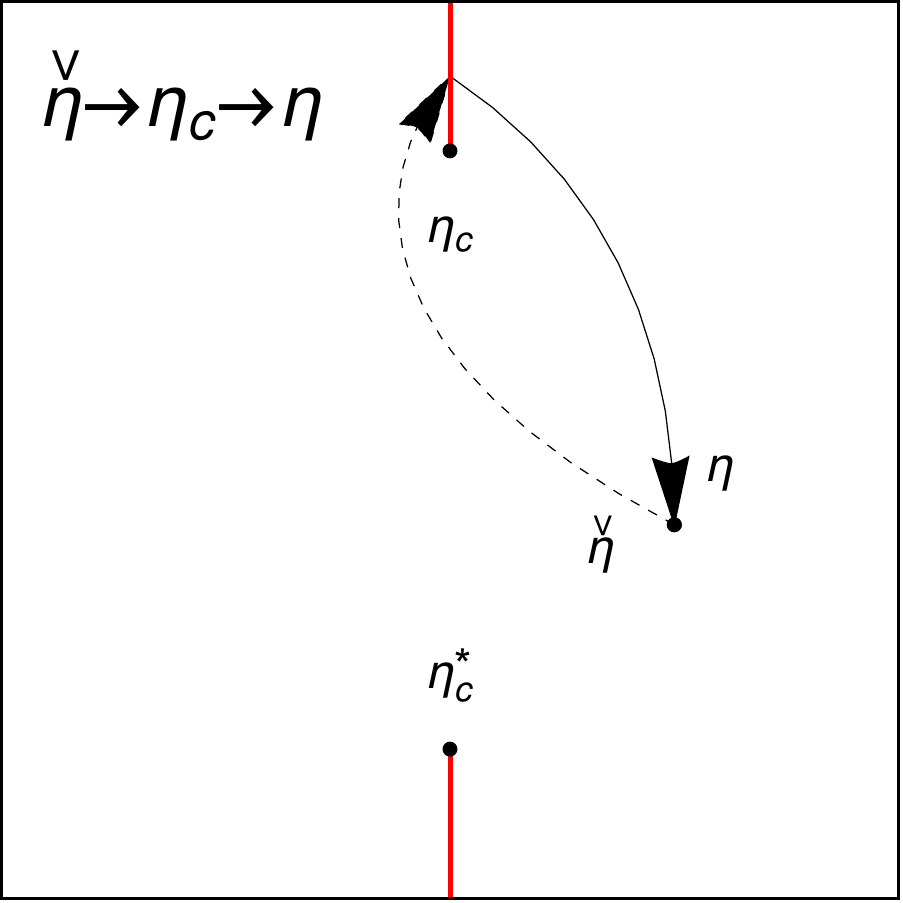}
	\caption{\label{fig:etac_to_eta}
	The integration contour $\check{\eta} \to \eta_c \to \eta$ used in Eq.~\eqref{eq:etac_to_eta}. The function $\omega(\eta)$ is double-valued, with branch cuts (red lines) emanating from the turning points $\eta_c$ and $\eta_c^*$. $\eta$ is on the ``positive'' sheet (on which $\omega(\eta) > 0$ for real $\eta$), and $\check{\eta}$ is the point corresponding to $\eta$ on the ``negative'' sheet. The integration contour starts from $\check{\eta}$ on the ``negative'' sheet, encircles $\eta_c$, crosses the branch cut, and ends at $\eta$ on the ``positive'' sheet. The part of the contour on the ``positive'' sheet is shown as a solid curve, and the part on the ``negative'' sheet is shown as a dashed curve. 
	}
\end{figure}

Before continuing, we address a potential problem in evaluating the integral $\int_{\eta_c}^\eta \dd{\eta'} T_\mathrm{odd}$. 
We write $T_\mathrm{odd}(\eta,\hbar) = \sum_{n=0}^\infty \hbar^n T_{\mathrm{odd},n}(\eta)$ and $R_\mathrm{odd}(\eta,\hbar) = \sum_{n=0}^\infty \hbar^n R_{\mathrm{odd},n}(\eta)$.  
Explicit computation will show that $T_{\mathrm{odd},n}$ has the form of $f(\eta) Q_0(\eta)^{-k/2}$, where $n \geq 1$, $k$ is some positive odd integer, and $f(\eta)$ is some analytic function. Such a function diverges near $\eta_c$, so its integral could be ill-defined. To sidestep this problem, one can interpret the integral to be:
\ba{\label{eq:etac_to_eta}
    \int_{\eta_c}^\eta \dd{\eta'} T_\mathrm{odd}
    \equiv \frac12 \int_{\check{\eta}}^{\eta} \dd{\eta'} T_\mathrm{odd} 
    \;,
}
where $\check{\eta}$ is the point corresponding to $\eta$ on the ``second sheet'' of the Riemann surface of $\sqrt{Q_0}$. The integration follows a contour going from $\check{\eta}$ to $\eta$, encircling $\eta_c$ and passing a branch cut in the process; see Fig.~\ref{fig:etac_to_eta} for an illustration. Since $T_\mathrm{odd}$ is of square-root type singularity at $\eta_c$, $T_\mathrm{odd}(\eta) = -T_\mathrm{odd}(\check{\eta})$, hence the integral from $\check{\eta}$ to $\eta$ is twice the integral from $\eta_c$ to $\eta$, which is accounted for by the $1/2$ factor. In subsequent discussions, we shall always interpret an integral of $T_\mathrm{odd}$ starting from a turning point in the above manner. See Lemma~2.3 of \cite{Aoki:1991} for more details.

We now move on to state an adapted version of Theorem~2.4 of \cite{Aoki:1991}.\footnote{For consistency of notation, we changed the variables names in Theorem~2.4 of \cite{Aoki:1991} via $\tilde x \to \eta$, $Q \mapsto -Q$, $\tilde{\psi} \to v$, $\psi \to w$, $\eta \to \hbar^{-1}$. We also make the change of variable $x \mapsto x' \equiv -ix$. Also note that our definition for $T_\pm$ and $R_\pm$ differ by a factor of $i/\hbar$ from that in \cite{Aoki:1991}.}
\newtheorem{theorem}{Theorem}
\begin{theorem}
We may construct $x(\eta, \hbar)$ and $E(\hbar)$ order-by-order in $\hbar$, such that the following properties hold:
\begin{enumerate}
    \item[i.] For solutions $T_\pm$ of the Riccati equation
    \ba{
        i \hbar \dv{T}{\eta} - T(\eta, \hbar)^2 + Q(\eta, \hbar) = 0
    }
    and solutions $R_\pm$ of the Riccati equation
    \ba{
        i \hbar \dv{R}{\eta} - R(\eta, \hbar)^2 + \left(E(\hbar) + \frac{x^2}{4} \right) = 0 \;,
    }
    it follows that, for $T_\mathrm{odd} \equiv (T_+ - T_-) / 2$ and $R_\mathrm{odd} \equiv (R_+ - R_-) / 2$,
    \ba{
        T_\mathrm{odd}(\eta, \hbar) = \pdv{x(\eta, \hbar)}{\eta} R_\mathrm{odd}(x(\eta, \hbar), \hbar)
    }
    holds if the branches of $T_{\mathrm{odd},0}$ and $R_{\mathrm{odd},0}$ are chosen so that
    \ba{
        \arg{T_{\mathrm{odd},0}}(\eta) = \arg{\left(\dv{x_0(\eta)}{\eta} R_{\mathrm{odd},0}(x_0(\eta))\right)}
    }
    may hold.
    \item[ii.] For a WKB solution $\hat{v}(\eta, \hbar)$ of the MTP equation
    \ba{
        \left( \hbar^2\dv[2]{\eta} + Q(\eta, \hbar) \right) \hat{v}_\pm = 0
    }
    that is normalized as in \eqref{eq:normalization_at_turning_point}, the WKB solution $\hat{w}(x,\hbar)$ of the Weber equation
    \ba{
        \left( \hbar^2\dv[2]{\eta} + \left(E(\hbar) + \frac{x^2}{4} \right) \right) \hat{w}_\pm = 0
    }
    satisfies the following relation:
    \ba{
        \hat{v}_{\pm}(\eta, \hbar) = 
        \left( \pdv{x(\eta, \hbar)}{\eta} \right)^{-1/2} 
        \hat{w}_{\pm}(x(\eta, \hbar), \hbar) \;.
    }
\end{enumerate}
\end{theorem}
Finally, we note that Theorem~2.4 of \cite{Aoki:1991} only deals with the case $Q(\eta, \hbar) = Q_0(\eta)$; Theorem~3.1 of \cite{10.2977/prims/1195166424} extends the result to the general $Q(\eta, \hbar)$ with higher orders in $\hbar$. In our presentation of the theorem above, we have combined Theorem~2.4 of \cite{Aoki:1991} with its extension.

\subsection{The coordinate $x(\eta,\hbar)$}
We introduced the holomorphic coordinate $x(\eta, \hbar)$ in \eqref{eq:v_w_relation}, and mentioned that it can be constructed order-by-order in $\hbar$ using $Q(\eta, \hbar)$ as an input. In this subsection, we will discuss some properties of $x_0(\eta)$ in detail and briefly talk about the construction of $x_n(\eta, \hbar)$ for $n \geq 1$.

Let $p_0$ and $p_1$ be the two turning points of $Q_0(\eta)$, such that they are connected by a Stokes line $\Gamma$. As in \eqref{eq:Psi_def}, we define $E_0$ by:
\ba{
    E_0 = \frac{i}{\pi} \int_{p_0}^{p_1} \dd{\eta'} \sqrt{Q_0(\eta')} \;.
}
We shall assume $E_0 > 0$. On a neighborhood of the Stokes line $\Gamma$, the holomorphic coordinate $x_0(\eta)$ satisfies:
\bes{
    & x_0(p_0) = 2i \sqrt{E_0},\quad x_0(p_1) = -2i \sqrt{E_0}, \\
    & Q_0(\eta) = \left(\dv{x_0}{\eta}\right)^2 \left(E_0 + \frac{x_0^2}{4}\right) \;.
}
Notice that $2i\sqrt{E_0}$ and $-2i\sqrt{E_0}$ are exactly the two turning points for the Weber equation \eqref{eq:psi_eqn}. In other words, $x_0(\eta)$ maps turning points for the $v(\eta, \hbar)$ equation \eqref{eq:tilde_psi_eqn} to turning points for the $w(x, \hbar)$ equation \eqref{eq:psi_eqn}. The second line above is an ODE for $x_0(\eta)$. With the initial condition $x_0(p_0) = 2i\sqrt{E_0}$, one can solve for $x_0(\eta)$ in a neighborhood of the Stokes line $\Gamma$.
Furthermore, for arbitrary $\eta$, we have:
\ba{
    i \int_{p_0}^{\eta} \dd{\eta'} \sqrt{Q_0(\eta')}
    = i \int_{x_0(p_0)}^{x_0(\eta)} \dd{x} \left( \dv{x_0}{\eta} \right)^{-1} \sqrt{Q_0(\eta)}
    = i \int_{2i\sqrt{E_0}}^{x_0(\eta)} \dd{x} \sqrt{E_0 + \frac{x^2}{4}} \;.
    \label{eq:x_eta_stokes}
}
If $\eta$ is on the Stokes line $\Gamma$, then from \eqref{eq:Stokes1st} we know that the LHS of the above equality is a non-negative real number. On the other hand, the Stokes line for the Weber equation \eqref{eq:psi_eqn} is defined by requiring the RHS of above to be a non-negative real number. In other words, the Stokes line for Eq.~\eqref{eq:tilde_psi_eqn} in $\eta$ coordinate is mapped to the Stokes line for the Weber equation \eqref{eq:psi_eqn} in $x$ coordinate. One can check that the Stokes line in $x_0$ coordinate is simply the straight line from $2i\sqrt{E_0}$ to $-2i\sqrt{E_0}$.

If $Q_0(\eta)$ is real on the real axis, then one can further show that $x_0(\eta_*) = 0$, where $\eta_*$ is the crossing point, the point where Stokes line $\Gamma$ intersects the real axis. 
To prove this fact, we parametrize the Stokes line $\Gamma$ by $i \int_{p_0}^{\eta} \dd{\eta'} \sqrt{Q_0}$. Namely, we choose parametrization $\gamma_\eta(t)$ such that:
\ba{
    \gamma_\eta \left(i \int_{p_0}^{\eta} \dd{\eta'} \sqrt{Q_0(\eta')} \right) = \eta,\quad \gamma_\eta: [0, \pi E_0] \to \Gamma \;.
}
Similarly, we can define a parametrization $\gamma_{x_0}: [0, \pi E_0] \to x_0(\Gamma)$ of the Stokes line in $x_0$ coordinate. Equality \eqref{eq:x_eta_stokes} then implies that $x_0(\gamma_\eta(t)) = \gamma_{x_0}(t)$. Now, by the Schwarz reflection principle, we know that $i \int_{p_0}^{\eta_*} \dd{\eta'} \sqrt{Q_0}$ should be half of $i \int_{p_0}^{p_1} \dd{\eta'} \sqrt{Q_0} = \pi E_0$, so $\gamma_\eta(\pi E_0 / 2) = \eta_*$, which implies $x_0(\eta_*) = \gamma_{x_0}(\pi E_0 / 2) = 0$.

Finally, we sketch the construction of the $x_n$'s. As was shown in Eq.~(3.9) of \cite{10.2977/prims/1195166424}, each $x_n(\eta)$ satisfies an ODE that depends only on $Q(\eta, \hbar)$ and $x_{n^\prime}$'s with $n^\prime < n$. With appropriate initial conditions, one can solve for the $x_n$'s recursively, order-by-order in $\hbar$. For instance, the solution for $x_1$ is given in integral form:
\ba{
    x_1(\eta) =&\ \frac{1}{\sqrt{E_0 + \frac{x_0(\eta)^2}{4}}} \int_{2 i \sqrt{E_0}}^{x_0(\eta)} \dd{y} \frac{Q_1(x_0^{-1}(y)) (\dv{x_0}{\eta})^{-2} - E_1}{2 \sqrt{E_0 + \frac{y^2}{4}}} 
    \;,
}
where $x_0^{-1}(y)$ is the inverse function of $x_0(\eta)$ evaluated at $y$. See Sec.~3 of \cite{10.2977/prims/1195166424} and App.~B of \cite{Aoki:1991} for more details.

\subsection{Calculating $E_J(\hbar)$}
\begin{figure}[tb]
	\centering
	\includegraphics[width=0.35\textwidth]{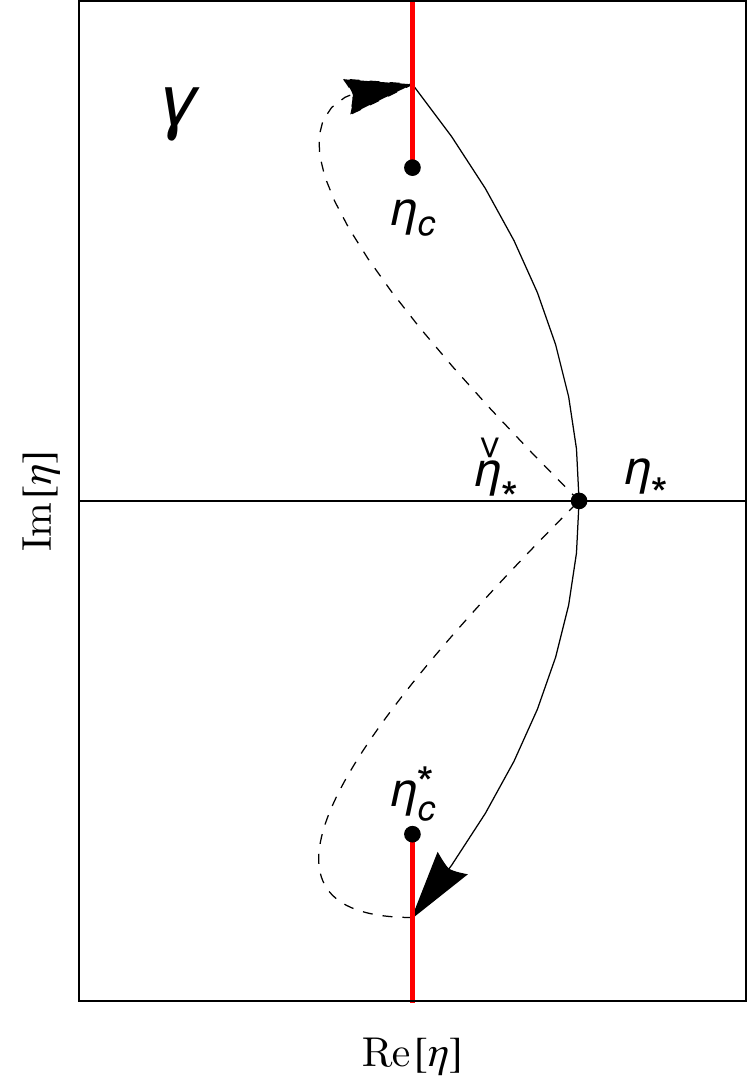}\ \ \ \
	\includegraphics[width=0.5\textwidth]{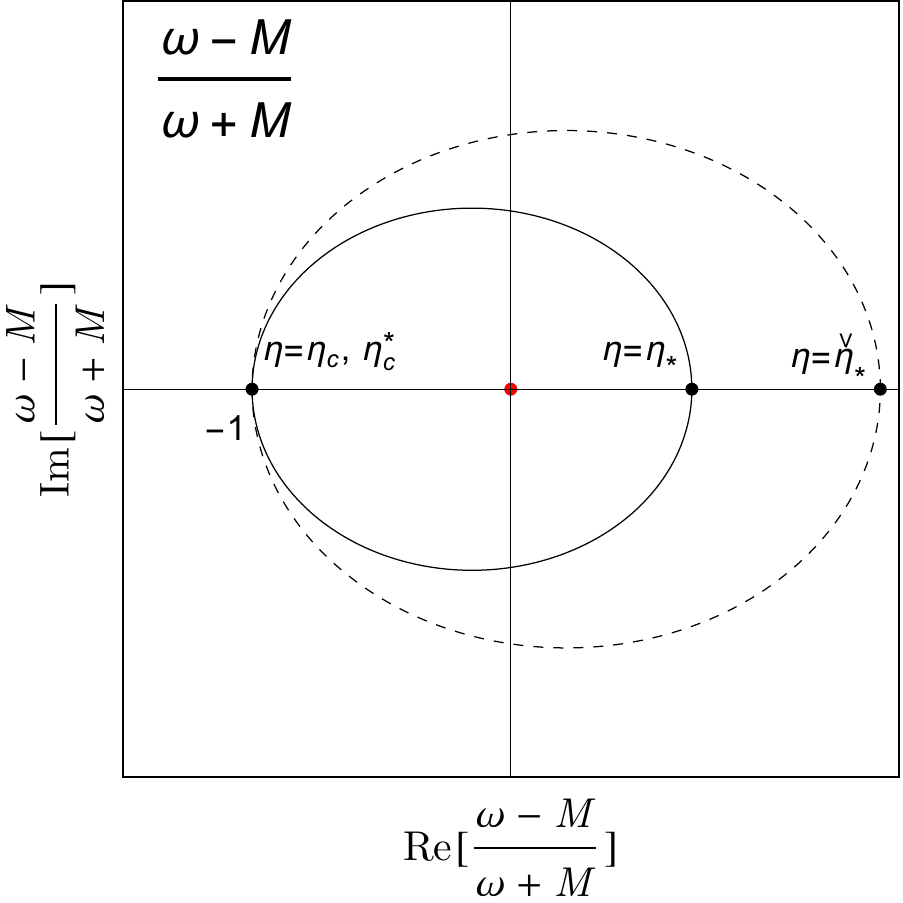}
	\caption{\label{fig:gamma}
	The left panel shows the loop contour $\gamma$. The function $\omega(\eta)$ is double-valued, with branch cuts (red lines) emanating from the turning points $\eta_c$ and $\eta_c^*$. $\eta_*$ is the crossing point on the ``positive'' sheet (on which $\omega(\eta) > 0$ for real $\eta$), and $\check{\eta}_*$ is the point corresponding to $\eta_*$ on the ``negative'' sheet. $\gamma$ starts from $\eta_c$, passes $\eta_*$ on the ``positive'' sheet, reaches $\eta_c^*$, crosses the branch cut, passes $\check{\eta}_*$ on the ``negative'' sheet, and returns back to $\eta_c$. The part of $\gamma$ on the ``positive'' sheet is shown in solid curve, and the part of $\gamma$ on the ``negative'' sheet is shown in dashed curve. 
	The right panel shows the value of $(\omega-M)/(\omega+M)$ as $\eta$ varies over the loop $\gamma$. At $\eta=\eta_c$ or $\eta=\eta_c^*$, $\omega = 0$, so the value $(\omega-M)/(\omega+M) = -1$; at $\eta=\eta_*$, the value is $(\omega(\eta_*)-M(\eta_*))/(\omega(\eta_*)+M(\eta_*))$; at $\eta=\check{\eta}_*$, $\omega(\check{\eta}_*) = -\omega(\eta_*)$, so the value is $(-\omega(\eta_*)-M(\eta_*))/(-\omega(\eta_*)+M(\eta_*))$. As $\eta$ varies along the path $\eta_c \to \eta_* \to \eta_c^* \to \check{\eta}_* \to \eta_c$, the value of $(\omega-M)/(\omega+M)$ forms two loops around the origin (red point). The orientation of the two loops depends on the function $(\omega-M)/(\omega+M)$.
	}
\end{figure}

We now apply the theorem on the $v_J$ equation \eqref{eq:EWKBeq} and derive $E(\hbar)$, which specifies the corresponding Weber equation \eqref{eq:psi_eqn}. The $v_J$ equation \eqref{eq:EWKBeq} is in the form of \eqref{eq:tilde_psi_eqn}, where $Q = Q_J$ ($J = A, B$) is given in \eqref{eq:QJn_series}. The WKB series for the $v_J$ equation is given in \eqref{eq:WKB_ansatz_app_v}, where $T_{J\pm}$ can be solved recursively order-by-order in $\hbar$. Theorem~2.5 of \cite{Aoki:1991} tells us:
\bes{
    & E_J
    = \frac{1}{2 \pi i} \oint_\gamma \! \dd{\eta} \, T_{J,\mathrm{odd}} 
    \qq{where} T_{J,\mathrm{odd}} \equiv \frac{1}{2}(T_{J+} - T_{J-}) , \\
    & E_J = \sum_{n=0}^{\infty} \hbar^n E_{Jn},\quad 
    T_{J,\mathrm{odd}} = \sum_{n=0}^{\infty} \hbar^n T_{J,\mathrm{odd},n}
    \; .
    \label{eq:Ej_formula}
}
Here, $\gamma$ is a closed loop that surrounds the two turning points $\eta_c$ and $\eta_c^*$ in a clockwise fashion, spanning both sheets on the Riemann surface for $\omega$; moreover, $\gamma$ is chosen such that $E_{J0} > 0$. See Fig.~\ref{fig:gamma} for an illustration of $\gamma$.

We now state the main result of this subsection:
\ba{
    & E_A = \epsilon \mp \frac{i\hbar}{2},\quad 
    E_B = \epsilon \pm \frac{i\hbar}{2}, \qq{where} 
    \pm = \mathrm{sgn}\left( \eval{\dv{\eta}(\frac{\omega-M}{\omega+M})}_{\eta_*} \right), \nonumber \\
    & \epsilon = \sum_{n=0}^{\infty} \hbar^n \epsilon_n,\quad \epsilon_n \in \mathbb{R},\quad \epsilon_0 = \frac{i}{\pi} \int_{\eta_c}^{\eta_c^*} \! \dd{\eta} \, \omega > 0
    \;.
    \label{eq:E_J_result}
}
The $E_J$'s here are similar to the $E_J$'s for the Landau-Zener model, introduced in App.~\ref{app:landau_zener}; the only difference is that $\epsilon$ is now a power series in $\hbar$. The Weber equations satisfied by $w_J$'s are the same for the Landau-Zener model and for the general problem here.

We first show that $E_A^* = E_B$. The identity $S_{A\pm}^* = -S_{B\mp}$ from Eq.~\eqref{eq:SA_SB_relation} implies $T_{A,\mathrm{odd}}^* = T_{B,\mathrm{odd}}$:
\bes{
    & T_{A\pm}^* 
    = S_{A\pm}^* - i\hbar \frac{{K^*}'}{2{K^*}} 
    = -S_{B\mp} - i\hbar \frac{{K^*}'}{2{K^*}}
    = -T_{B\mp} \\
    & T_{A,\mathrm{odd}}^* 
    = \frac{1}{2}(T_{A+}^* - T_{A-}^*) 
    = -\frac{1}{2}(T_{B-} - T_{B+})
    = T_{B,\mathrm{odd}}
    \;.
}
Note that here we are treating $T_{J,\mathrm{odd}}$ as functions in the real variable $\eta$. ($\eta$ should not take complex values here since complex conjugation is not analytic.) If we denote the real and imaginary parts of $T_{J,\mathrm{odd}}$ by real functions (in real variable $\eta$) $r(\eta, \hbar)$ and $\sigma(\eta, \hbar)$ then 
\ba{
    T_{A,\mathrm{odd}} = r + i \sigma,\quad 
    T_{B,\mathrm{odd}} = r - i \sigma \;,
    \label{eq:T_odd_r_mu}
}
then $E_A^* = E_B$ follows immediately from \eqref{eq:Ej_formula}:
\ba{
    E_A^* 
    = \left( \frac{1}{2 \pi i} \oint_\gamma \! \dd{\eta} \, (r + i \sigma) \right)^*
    = \frac{1}{2 \pi i} \oint_\gamma \! \dd{\eta} \, (r - i \sigma)
    = E_B \;.
}
Here, we have used the fact that $\oint \! \dd{\eta} \, r$ and $\oint \! \dd{\eta} \, \sigma$ are purely imaginary, a consequence of the Schwarz reflection principle. Note that we have extended $r(\eta,\hbar)$ and $\sigma(\eta,\hbar)$ to analytic functions in the complex variable $\eta$. 

To compute $E_J$, we proceed to evaluate the first few terms for $T_{J,\mathrm{odd},n}$:
\ba{
    T_{A,\mathrm{odd},0} =&\ -\omega \nonumber \\
    T_{A,\mathrm{odd},1} =&\ i\frac{M K' - K M'}{2 \omega K}
    = -\frac14 \csc(2\theta)\sin(4\theta) \zeta' + i \csc(2\theta) \theta' \nonumber \\
    T_{A,\mathrm{odd},2} =&\ \frac{
    2 M K' M' K
    + \omega^2 (3 K'^2-2 K'' K)
    - K^2 (M'^2+3 \omega'^2) 
    + 2 \omega \omega'' K^2
    - M^2 K'^2}{8 \omega^3 K^2} \nonumber \\
    =&\ \frac{
    4 \theta' \cot(2\theta) \omega' 
    + \omega \csc^2(2\theta) (-\zeta'^2 \sin^4(2\theta) + 2 \theta
   '^2 (\cos(4\theta) + 3) - 2 \sin(4\theta) \theta'')}{8 \omega^2} \nonumber \\
   &\ - i \dv{\eta}(\frac{\zeta'}{4\omega}) \nonumber \\
    T_{B,\mathrm{odd},0} =&\ -\omega \nonumber \\
    T_{B,\mathrm{odd},1} =&\ i\frac{{K^*} M' - M {K^*}'}{2 \omega {K^*}}
    = -\frac14 \csc(2\theta)\sin(4\theta) \zeta' - i \csc(2\theta) \theta' \nonumber \\
    T_{B,\mathrm{odd},2} =&\ \frac{
    2 K^* M M' {K^*}'
    + \omega^2 (3 {K^*}'^2-2 K^* {K^*}'') 
    - {K^*}^2 (M'^2+3 \omega'^2) 
    + 2 {K^*}^2 \omega \omega''
    - M^2 {K^*}'^2}{8 {K^*}^2 \omega ^3} \nonumber \\
    =&\ \frac{
    4 \theta' \cot(2\theta) \omega' 
    + \omega \csc^2(2\theta) (-\zeta'^2 \sin^4(2\theta) + 2 \theta
   '^2 (\cos(4\theta) + 3) - 2 \sin(4\theta) \theta'')}{8 \omega^2} \nonumber \\
   &\ + i \dv{\eta}(\frac{\zeta'}{4\omega})
    \;.
    \label{eq:T_A_odd_n}
}
Here, we have changed variables to $\omega$, $\theta$ and $\zeta$, where $\cos \theta = \sqrt{(1 + M /\omega)/2}$ and $\sin \theta = \sqrt{(1 - M /\omega)/2}$ as in App.~\ref{app:diagonalize}. For each term, we have separated the real part and the imaginary part; one can see that the identity $T_{A,\mathrm{odd}}^* = T_{B,\mathrm{odd}}$ holds order-by-order in $\hbar$. To evaluate $E_J$, one only needs to evaluate $E_A$. The first two terms for $E_{An}$ are given by:
\bes{
  E_{A0} =&\ \frac{1}{2 \pi i} \oint_\gamma \! \dd{\eta} \, T_{A,\mathrm{odd},0} 
  = \frac{1}{2 \pi i} \oint_\gamma \! \dd{\eta} \, (-\omega)
  = \frac{i}{\pi} \int_{\eta_c}^{\eta_c^*} \! \dd{\eta} \, \omega 
  = \frac{\Psi}{\pi} \\
  E_{A1} =&\ \frac{1}{2 \pi i} \oint_\gamma \! \dd{\eta} \, T_{A,\mathrm{odd},1} \\
  =&\ \frac{1}{2 \pi i} \oint_\gamma \! \dd{\eta} \, \frac{-1}{4} \csc(2\theta)\sin(4\theta) \zeta' 
  + \frac{1}{2 \pi i} \frac{i}{4} \oint_\gamma \dd(\log(\frac{\omega - M}{\omega + M})) \;.
}
$E_{A0}$ is given in terms of the phase integral $\Psi$, consistent with \eqref{eq:Psi_def}. The first term in $E_{A1}$ is a real geometric amplitude factor, which was derived by Berry in Ref.~\cite{10.2307/80000}. The second term in $E_{A1}$ is the integral of a total derivative; to find its value, we must study how $(\omega-M)/(\omega+M)$ changes as $\eta$ varies along $\gamma$. As shown in Fig.~\ref{fig:gamma}, as $\eta$ varies along the path $\eta_c \to \eta_* \to \eta_c^* \to \check{\eta}_* \to \eta_c$, the value of $(\omega-M)/(\omega+M)$ forms two loops around $0$, so $\log(\frac{\omega-M}{\omega+M})$ will pick up a factor of $\pm 4 \pi i$ along the path. To determine the sign, notice that:
\ba{
    \eval{\dd(\frac{\omega-M}{\omega+M})}_{\eta_*}
    = \eval{\dv{\eta}(\frac{\omega-M}{\omega+M}) \dd{\eta} }_{\eta_*} \;.
}
Since the loop $\gamma$ is clockwise, $\frac{\omega - M}{\omega + M}$ is also in the clockwise direction if $\dv{\eta}(\frac{\omega-M}{\omega+M}) > 0$ at $\eta = \eta_*$, and vice versa. We have thus evaluated the second term in $E_{A1}$:
\ba{
    \frac{1}{2 \pi i} \frac{i}{4} \oint_\gamma \dd(\log(\frac{\omega - M}{\omega + M})) = \mp \frac{i}{2} \qq{where}
    \pm = \mathrm{sgn}\left(\eval{\dv{\eta}(\frac{\omega-M}{\omega+M})}_{\eta_*} \right) \;.
}
The imaginary part for $E_{A2}$ is given by:
\ba{
    \Im[E_{A2}] = \frac{1}{2 \pi i} \oint_\gamma \dd(\frac{-\zeta'}{4\omega}) = 0 \;.
}
To show that the integral above is indeed $0$, we analytically continue $\zeta' / \omega$ along the loop $\gamma$ and confirm that its value doesn't change between the two endpoints. We know that $\omega$ returns to the original value along $\gamma$, and $\zeta'$ returns to its original value under the mild assumption that $\zeta'$ is analytic on a small neighborhood of the Stokes line $\Gamma$. Thus $\Im[E_{A2}] = 0$.
In fact, one can check that the imaginary part of $T_{A,\mathrm{odd},n}$ is a total derivative that evaluates to $0$ along the loop $\gamma$ for $n=2,3,4,5$, which suggests that $\Im[E_{An}] = 0$ for all $n\geq 2$. To show that this is indeed true, consider the analytic continuation of Eq.~\eqref{eq:S_eqn_matrix} along $\gamma$, starting from $\eta_*$:
\bes{
  & \eval{\left[ (S_{A-} + M) u_{A-,*} \exp(\frac{i}{\hbar} \oint_\gamma \! \dd{\eta} \, S_{A-}) + K u_{B-,*} \exp(\frac{i}{\hbar} \oint_\gamma \! \dd{\eta} \, S_{B-}) \right]}_{\eta_*} = 0 \;.
}
Writing $S_{J-}$ in terms of $T_{J,\mathrm{odd}}$ as in Eq.~\eqref{eq:normalization_at_turning_point} and using Eq.~\eqref{eq:T_odd_r_mu} to separate the real and imaginary parts, we have:
\bes{
  & \eval{\left[ (S_{A-} + M) u_{A-,*} \exp(- \frac{i}{\hbar} \oint_\gamma \! \dd{\eta} \, T_{A,\mathrm{odd}} ) + K u_{B-,*} \exp(- \frac{i}{\hbar} \oint_\gamma \! \dd{\eta} \, T_{B,\mathrm{odd}} ) \right]}_{\eta_*} = 0 \\
  & \eval{\left[ (S_{A-} + M) u_{A-,*} \exp(- \frac{i}{\hbar} \oint_\gamma \! \dd{\eta} \, i \sigma ) + K u_{B-,*} \exp(- \frac{i}{\hbar} \oint_\gamma \! \dd{\eta} \, (-i\sigma) ) \right]}_{\eta_*} = 0 \\
  & \exp(2 \frac{1}{\hbar} \oint_\gamma \! \dd{\eta} \, \sigma ) 
  = \frac{- K(\eta_*) }{S_{A-}(\eta_*) + M(\eta_*)} \frac{u_{B-,*}}{u_{A-,*}} 
  = 1 \\
  & \oint_\gamma \! \dd{\eta} \, \sigma = i \hbar n \pi \qq{for some} n \in \mathbb{Z}
  \;.
}
This is consistent with our calculation for $E_{A1}$. 
Moreover, since $\oint \! \dd{\eta} \, \sigma$ does not contain a term of order $\hbar^2$ or higher, $\Im[E_{An}] = 0$ for all $n\geq 2$. We conclude that $E_A = \epsilon \mp i\hbar/2$ and $E_B = \epsilon \pm i\hbar/2$.

\subsection{Deriving the connection matrix}
We now begin our derivation of the connection matrix. The overall strategy for the derivation will be as follows: we will first define EWKB basis solutions $u_{A\pm}^{(1)}$ and $u_{A\pm}^{(2)}$ in two different Stokes regions using the WKB series \eqref{eq:normalized_EWKB_soln_for_u}, and define the connection matrix $T$ between them as in \eqref{eq:connection_matrix}. We then use Eq.~\eqref{eq:u_to_v} to define $v_{\pm}^{(i)} \equiv v_{A\pm}^{(i)}$ from $u_{A\pm}^{(i)}$. $v_{\pm}^{(i)}$ are EWKB basis solutions of Eq.~\eqref{eq:tilde_psi_eqn}, hence they are multiples of $\hat{v}_{\pm}^{(i)}$ from Eq.~\eqref{eq:normalization_at_turning_point}, the EWKB basis solutions normalized at a turning point. On the other hand, the parabolic cylinder functions are multiples of $\hat{w}_{\pm}^{(i)}$ from Eq.~\eqref{eq:normalization_at_turning_point_what}, the EWKB basis solutions of the Weber equation \eqref{eq:psi_eqn} normalized at a turning point. Via theorem \eqref{eq:v_w_relation}, we can relate $\hat{v}_{\pm}^{(i)}$ with $\hat{w}_\pm^{(i)}$, and consequently relate $v_{\pm}^{(i)}$ with the parabolic cylinder functions. The known connection formulas between the parabolic cylinder functions can then be used to compute the connection matrix between $v_{\pm}^{(1)}$ and $v_{\pm}^{(2)}$, which in turn gives us the connection matrix $T$ between $u_{\pm}^{(1)}$ and $u_{\pm}^{(2)}$.

In Eq.~\eqref{eq:connection_matrix} of Sec.~\ref{sec:discussion}, we have defined  the connection matrix $T \in \mathrm{SU}(2)$ as the matrix that relates expansion coefficients between different normalized EWKB basis solutions. $T$ is not diagonal if the EWKB basis solutions are constructed in different Stokes regions, separated by one or more Stokes lines. Now suppose that there is a Stokes line from turning point $\eta_c$ to $\eta_c^*$ that crosses the real axis at $\eta_*$ as in Fig.~\ref{fig:terminology}. We construct EWKB basis solutions $u_{A\pm}^{(1)}$ and $u_{A\pm}^{(2)}$ at real $\eta_1$ and $\eta_2$, where $\eta_1 < \eta_* < \eta_2$, and $u_{A\pm}$ are Borel sums of the standard form WKB series \eqref{eq:normalized_EWKB_soln_for_u}.\footnote{The connection matrix $T$ relates $u_{A\pm}$ and $u_{B\pm}$ in different Stokes regions in the same way, so it is adequate for us to discuss only $u_{A\pm}$.}
For simplicity, we will also interpret $u_{A\pm}^{(1)}$ (and etc) as global solutions consistent with the EWKB basis solutions at $\eta_1$. The solution matching condition $\alpha^{(1)} u_{A+}^{(1)} + \beta^{(1)} u_{A-}^{(1)} = \alpha^{(2)} u_{A+}^{(2)} + \beta^{(2)} u_{A-}^{(2)}$ then implies:
\ba{
    \mqty(\alpha^{(2)}(\hbar) \\ \beta^{(2)}(\hbar)) = T(\hbar) \mqty(\alpha^{(1)}(\hbar) \\ \beta^{(1)}(\hbar)) \Rightarrow  \mqty(u_{A+}^{(1)}(\eta,\hbar) \\ u_{A-}^{(1)}(\eta,\hbar)) = T^T(\hbar) \mqty(u_{A+}^{(2)}(\eta,\hbar) \\ u_{A-}^{(2)}(\eta,\hbar)) 
    \;.
}
To be explicit, we define $a, b, c, d$ to be the entries of $T$:
\ba{
    T(\hbar) = \mqty(a(\hbar) & c(\hbar) \\ b(\hbar) & d(\hbar)) \; .
}
The vector equation above can then be written as $u_{A+}^{(1)} = a u_{A+}^{(2)} + b u_{A-}^{(2)}$ and $u_{A-}^{(1)} = c u_{A+}^{(2)} + d u_{A-}^{(2)}$, which correspond to the connection rules for $u_{A+}^{(1)}$ and $u_{A-}^{(1)}$. We shall determine $T$ by computing the numbers $a, b, c, d$. 

To apply the theorem, we change variables from $u_{A\pm}$ to $v_\pm \equiv v_{A\pm}$ as in \eqref{eq:u_to_v}: 
\ba{
    v_{\pm}^{(1)}(\eta, \hbar) = 
    \sqrt{\frac{K(\eta_1)}{K(\eta)}} \, u_{A\pm}^{(1)}(\eta, \hbar),\quad
    v_{\pm}^{(2)}(\eta, \hbar) = 
    \sqrt{\frac{K(\eta_2)}{K(\eta)}} \, u_{A\pm}^{(2)}(\eta, \hbar)
    \;,
}
where now the anchor times are chosen to be $\eta_a = \eta_1$ and $\eta_2$.  
The $v$'s satisfy Eq.~\eqref{eq:EWKBeq}, which is in the form of Eq.~\eqref{eq:tilde_psi_eqn} with
\ba{
    Q(\eta,\hbar) = \omega^2 + \hbar \biggl( i M' - i \frac{K^\prime}{K} M \biggr) + \hbar^2 \biggl( \frac{K''}{2K} - \frac{3 K^{\prime 2}}{4 K^2} \biggr) \;.
}
The connection rule for $v_\pm^{(i)}$ is then given by:
\ba{
    v_{+}^{(1)} = 
    \sqrt{\frac{K(\eta_1)}{K(\eta_2)}} \, (a v_{+}^{(2)} + b v_{-}^{(2)}) \;, \quad
    v_{-}^{(1)} = 
    \sqrt{\frac{K(\eta_1)}{K(\eta_2)}} \, (c v_{+}^{(2)} + d v_{-}^{(2)}) \;.
}
For simplicity, we choose $\eta_1$ and $\eta_2$ arbitrarily close to $\eta_*$, so the connection rule simplifies to:
\ba{
    v_{+}^{(1)} = a v_{+}^{(2)} + b v_{-}^{(2)},\quad v_{-}^{(1)} = c v_{+}^{(2)} + d v_{-}^{(2)} \;.
}
In other words, $T$ is also the connection matrix between EWKB basis solutions $v_\pm^{(1)}$ and $v_\pm^{(2)}$. If we do not choose $\eta_1$ and $\eta_2$ to be arbitrarily close to $\eta_*$, then connection matrix $T$ would be changed by a phase factor, as was shown in the discussion below Eq.~\eqref{eq:Tast_def}. 

Now, since the $v_\pm^{(i)}$'s are EWKB basis solutions of Eq.~\eqref{eq:tilde_psi_eqn}, they must be constant multiples of the EWKB basis solutions normalized at turning point $\eta_c$:
\ba{
    v_\pm^{(i)}(\eta, \hbar) = \Lambda_\pm^{(i)}(\hbar) \, \hat{v}_\pm^{(i)}(\eta, \hbar) \qq{for} i = 1, 2 \;,
}
where $\hat{v}_\pm^{(i)}$ are the Borel sums of the WKB ansatz \eqref{eq:normalization_at_turning_point}. Using relation \eqref{eq:v_w_relation} provided by the theorem, we can write:
\bes{
    & v_\pm^{(i)}(\eta, \hbar) 
    = \Lambda_\pm^{(i)}(\hbar) \, \hat{v}_\pm^{(i)}(\eta, \hbar) 
    = \left( \pdv{x(\eta, \hbar)}{\eta} \right)^{-1/2}  w_\pm^{(i)}(x(\eta, \hbar), \hbar) \\
    \qq{where} & w_\pm^{(i)}(x, \hbar) \equiv \Lambda_\pm^{(i)}(\hbar) \, \hat{w}_\pm^{(i)}(x, \hbar),\quad i = 1, 2
    \;.
}
Here, the $w_\pm^{(i)}$'s and $\hat{w}_\pm^{(i)}$'s are EWKB basis solutions of the Weber equation \eqref{eq:psi_eqn}, with the Borel resummation taken in the corresponding Stokes regions. 
As was shown in Sec.~\ref{sec:Weber}, the EWKB basis solutions $w_\pm^{(1)}$ and $w_\pm^{(2)}$ can be written as constant multiples of the parabolic cylinder functions: 
\bes{
    & w_+^{(1)}(x,\hbar) = \lambda_+^{(1)}(\hbar) \, D_{-\frac{1}{2}+i\xi}(-iz),\quad 
    w_-^{(1)}(x,\hbar) = \lambda_-^{(1)}(\hbar) \, D_{-\frac{1}{2}-i\xi}(z), \\
    & w_+^{(2)}(x,\hbar) = \lambda_+^{(2)}(\hbar) \, D_{-\frac{1}{2}-i\xi}(-z),\quad 
    w_-^{(2)}(x,\hbar) = \lambda_-^{(2)}(\hbar) \, D_{-\frac{1}{2}+i\xi}(iz) ,
    \label{eq:lambda_def}
}
where $\xi(\hbar) \equiv E(\hbar) / \hbar$ and $z(x, \hbar) \equiv e^{-3i\pi/4}x / \sqrt{\hbar}$.
Note that the constants $\lambda_\pm^{(i)}$ and $\Lambda_\pm^{(i)}$ are now fully specified, and can be computed explicitly in principle.

The ratios between the $\lambda_\pm^{(i)}$'s, along with the identities of parabolic cylinder functions, can be used to calculate the connection matrix $T$. To see that, note that the connection rule $v_{+}^{(1)} = a v_{+}^{(2)} + b v_{-}^{(2)}$ along with \eqref{eq:v_w_relation} gives us:
\bes{
    & v_+^{(1)}(\eta, \hbar) = a v_+^{(2)}(\eta, \hbar) + b v_-^{(2)}(\eta, \hbar) \\
    \Rightarrow\ & w_+^{(1)}(x(\eta, \hbar), \hbar) = a w_+^{(2)}(x(\eta, \hbar), \hbar) + b w_-^{(2)}(x(\eta, \hbar), \hbar) \\
    \Rightarrow\ & D_{-\frac{1}{2}+i\xi}(-iz) = 
    a \frac{\lambda_+^{(2)}}{\lambda_+^{(1)}} D_{-\frac{1}{2}-i\xi}(-z) 
    + b \frac{\lambda_-^{(2)}}{\lambda_+^{(1)}} D_{-\frac{1}{2}+i\xi}(iz) \;.
}
Here, we have shown that $w_+^{(1)}$ satisfies the same connection rule as $v_+^{(1)}$; moreover, the connection rule for $w_+^{(1)}$ can be written as a relation between the parabolic cylinder functions. We can perform a similar computation for $v_-^{(1)}$ and summarize the result in a formula for $T$:
\ba{
    & T = \mqty(a & c \\ b & d) = \mqty(
    A \frac{\lambda_+^{(1)}}{\lambda_+^{(2)}} &
    C \frac{\lambda_-^{(1)}}{\lambda_+^{(2)}} \\
    B \frac{\lambda_+^{(1)}}{\lambda_-^{(2)}} &
    D \frac{\lambda_-^{(1)}}{\lambda_-^{(2)}}
    ) \qq{where} \mqty(D_{-\frac{1}{2}+i\xi}(-iz) \\ D_{-\frac{1}{2}-i\xi}(z)) = \mqty(A & B \\ C & D) \mqty(D_{-\frac{1}{2}-i\xi}(-z) \\ D_{-\frac{1}{2}+i\xi}(iz)) ,\nonumber \\
    & A = \frac{\sqrt{2\pi}}{\Gamma(\frac{1}{2}-i\xi)} e^{\frac14 i \pi - \pi \xi},\quad
    B = -i e^{-\pi\xi},\quad
    C = i e^{-\pi\xi},\quad
    D = \frac{\sqrt{2\pi}}{\Gamma(\frac{1}{2}+i\xi)} e^{-\frac14 i \pi - \pi \xi} \; .
    \label{eq:ABCD_def}
}
The entries $A$, $B$, $C$, $D$ are extracted from known identities\footnote{See 6.1.18, 6.1.32 and 19.4.6 of Ref.~\cite{abramowitz+stegun}.} of the parabolic cylinder functions. The task of determining $T$ is thus reduced to computing $\xi = E / \hbar$ and the ratios between $\lambda_\pm^{(i)}$'s.

In order to obtain $\xi = E / \hbar$, we need $E = E_A$, whose derivation was detailed in previous subsections. Result \eqref{eq:E_J_result} states that $E_A = \epsilon \mp i \hbar / 2$, where the $\mp$ sign is determined by the specific form of $\omega$ and $M$. For brevity and for ease of comparison to the Landau-Zener problem in App.~\ref{app:landau_zener}, we will only do the subsequent calculations for the case $E_A = \epsilon - i \hbar / 2$.
The calculations for the case $E_A = \epsilon + i \hbar / 2$ can be done in a similar way. 
To summarize:
\ba{
    E = E_A = \epsilon - \frac{i \hbar}{2},\quad 
    E(\hbar) = \sum_{n=0}^{\infty} \hbar^n E_n,\quad 
    E_0 > 0,\quad 
    \xi = \frac{\epsilon}{\hbar} - \frac{i}{2} \;.
    \label{eq:E_assumption}
}
For small $\hbar$ we have $\epsilon = E_0 + \order{\hbar}$, and since $E_0 > 0$, we can treat $\epsilon$ as real and positive as well.  

We now proceed to compute the $\lambda_\pm^{(i)}$ ratios. We will first compute $\lambda_\pm^{(1)} / \lambda_\pm^{(2)}$ by taking $z = x = 0$:
\bes{
    & \frac{\lambda_+^{(1)}}{\lambda_+^{(2)}} 
    = \frac{w_+^{(1)}(0,\hbar) / D_{-\frac12 +i\xi}(0)}{w_+^{(2)}(0,\hbar) / D_{-\frac12 -i\xi}(0)} 
    = \frac{v_+^{(1)}(x^{-1}(0))}{v_+^{(2)}(x^{-1}(0))} \frac{D_{-\frac12 -i\xi}(0)}{D_{-\frac12 +i\xi}(0)} \\ 
    & \frac{\lambda_-^{(1)}}{\lambda_-^{(2)}} 
    = \frac{w_-^{(1)}(0,\hbar) / D_{-\frac12 -i\xi}(0)}{w_-^{(2)}(0,\hbar) / D_{-\frac12 +i\xi}(0)} 
    = \frac{v_-^{(1)}(x^{-1}(0))}{v_-^{(2)}(x^{-1}(0))} \frac{D_{-\frac12 +i\xi}(0)}{D_{-\frac12 -i\xi}(0)}
    \;,
}
where $x^{-1}$ denotes the inverse of the single variable function $\eta \mapsto x(\eta, \hbar)$. Since $v_+^{(1)}$ and $v_+^{(2)}$ are Borel sums of the same WKB ansatz \eqref{eq:normalized_EWKB_soln_for_u} in two different Stokes regions, their difference is the Laplace integral of $\mathcal{B} v_+$ over one or more Hankel contours; see Fig.~\ref{fig:Borel}. The contribution from such an integral is exponentially small\footnote{The precise meaning of ``exponentially small'' can be formalized via the so called trans-series expansion. See \cite{Dorigoni:2014hea, Ito:2018eon} for more details.}, in the sense that it is equal to $e^{-S / \hbar} f(\hbar)$ for some $S > 0$ and some analytic function $f$. As $\hbar \to 0^+$, all derivatives of $e^{-S / \hbar} f(\hbar)$ with respect to $\hbar$ vanish, so it cannot be treated perturbatively in $\hbar$. We claim that the difference between $v_+^{(1)}$ and $v_+^{(2)}$ is exponentially small and $v_+^{(1)} / v_+^{(2)} = 1 + \order{e^{-\pi E_0 / \hbar}}$. Similarly, $v_-^{(1)} / v_-^{(2)} = 1 + \order{e^{-\pi E_0 / \hbar}}$. On the other hand, the ratio between the parabolic cylinder functions is given by:
\bes{
    \frac{D_{-\frac12+i\xi}(0)}{D_{-\frac12-i\xi}(0)}
    &= \sqrt{\frac{2}{\pi}} \cosh(\frac{\pi \epsilon}{2 \hbar}) \Gamma(1 + i \frac{\epsilon}{\hbar}) 
    = \sqrt{\frac{2}{\pi}} \cosh(\frac{\pi \epsilon}{2 \hbar}) \sqrt{\frac{\pi \epsilon / \hbar}{\sinh(\pi \epsilon / \hbar)}} e^{i \phi_1} \\
    &= \sqrt{\frac{\epsilon}{\hbar}} e^{i \phi_1} \left( 1 + \order{e^{-\pi E_0 / \hbar}} \right) 
    \;,
}
where $\phi_1 = \arg \Gamma(1 + i \epsilon / \hbar)$. We conclude:
\bes{
    & \frac{\lambda_+^{(1)}}{\lambda_+^{(2)}} 
    = \sqrt{\frac{\hbar}{\epsilon}} e^{-i \phi_1} \left(1 + \order{e^{-\pi E_0 / \hbar}} \right)  \\
    & \frac{\lambda_-^{(1)}}{\lambda_-^{(2)}} 
    = \sqrt{\frac{\epsilon}{\hbar}} e^{i \phi_1} \left(1 + \order{e^{-\pi E_0 / \hbar}} \right)
    \;.
    \label{eq:lambda_ratio_1}
}
Now we compute $\lambda_\pm^{(1)} / \lambda_\mp^{(2)}$:
\bes{
    & \frac{\lambda_-^{(1)}}{\lambda_+^{(2)}} 
    = \frac{\Lambda_-^{(1)}}{\Lambda_+^{(2)}} \frac{\hat{w}_-^{(1)}(0,\hbar)}{\hat{w}_+^{(2)}(0,\hbar)} / \frac{D_{-\frac12-i\xi}(0)}{D_{-\frac12-i\xi}(0)}
    = \frac{v_-^{(1)}(\eta_*, \hbar) / \hat{v}_-^{(1)}(\eta_*, \hbar)}{v_+^{(2)}(\eta_*, \hbar) / \hat{v}_+^{(2)}(\eta_*, \hbar)}  \frac{\hat{w}_-^{(1)}(0,\hbar)}{\hat{w}_+^{(2)}(0,\hbar)} \\
    & \frac{\lambda_+^{(1)}}{\lambda_-^{(2)}} 
    = \frac{\Lambda_+^{(1)}}{\Lambda_-^{(2)}} \frac{\hat{w}_+^{(1)}(0,\hbar)}{\hat{w}_-^{(2)}(0,\hbar)} / \frac{D_{-\frac12+i\xi}(0)}{D_{-\frac12+i\xi}(0)}
    = \frac{v_+^{(1)}(\eta_*, \hbar) / \hat{v}_+^{(1)}(\eta_*, \hbar)}{v_-^{(2)}(\eta_*, \hbar) / \hat{v}_-^{(2)}(\eta_*, \hbar)}  \frac{\hat{w}_+^{(1)}(0,\hbar)}{\hat{w}_-^{(2)}(0,\hbar)}
    \;.
}
Note that we are evaluating the $v_\pm^{(i)}$'s at $\eta_*$, which is not necessarily the same as $x^{-1}(0)$; the fact that $\Lambda_\mp^{(1)} / \Lambda_\pm^{(2)}$ is a constant means we can evaluate the $v_\pm^{(i)}$'s at any $\eta$ we want. We will first evaluate $v_\mp^{(1)} / v_\pm^{(2)}$. The normalization factors \eqref{eq:uAa_and_uBa} imply, on the WKB ansatz level (before Borel resummation): 
\bes{
    \frac{v_\mp^{(1)}(\eta_*, \hbar)}{v_\pm^{(2)}(\eta_*, \hbar)}
    = \frac{u_{A\mp,*}}{u_{A\pm,*}}
    = - \eval{ \sqrt{\frac{1 - M / S_{B\mp}}{1 - M / S_{B\pm}}} \sqrt{\frac{- S_{A\pm}}{S_{A\mp}}} }_{\eta_*}
    \;.
}
Note that this expression is exact to all orders in $\hbar$. Now we evaluate $\hat{v}_\mp^{(1)} / \hat{v}_\pm^{(2)}$ using definition \eqref{eq:normalization_at_turning_point}:
\ba{
    \frac{\hat{v}_\mp^{(1)}(\eta_*, \hbar)}{\hat{v}_\pm^{(2)}(\eta_*, \hbar)}
    = \exp(\mp 2 \frac{i}{\hbar} \int_{\eta_c}^{\eta_*} \dd{\eta'} T_{A,\mathrm{odd}})
    = \exp(\mp \frac{i}{\hbar} \int_{\check{\eta}_*}^{\eta_*} \dd{\eta'} T_{A,\mathrm{odd}})
    \;,
}
where $\check{\eta}_*$ is $\eta_*$ on the ``second sheet'' of $\sqrt{Q_0}$. Using results from \eqref{eq:T_A_odd_n}, one can show that:
\bes{
    & \int_{\check{\eta}_*}^{\eta_*} \dd{\eta'} T_{A,\mathrm{odd},n} 
    = 
    \begin{cases}
    i \pi \epsilon_0 & n = 0 \\
    i \pi \epsilon_n + \phi_{2,n} + i \nu_n & n \geq 1
    \end{cases} \\
    &\int_{\check{\eta}_*}^{\eta_*} \dd{\eta'} T_{A,\mathrm{odd}} 
    = i \pi \epsilon + \phi_2 + i \nu \\
    & \nu_1 = \frac{1}{2} \log(\frac{\omega(\eta_*) - M(\eta_*)}{\omega(\eta_*) + M(\eta_*)}),\quad \nu_2 = - \frac{\zeta'(\eta_*)}{2 \omega(\eta_*)},\quad \hdots
}
where $\phi_2 = \sum_{n=1}^\infty \hbar^n \phi_{2,n}$ and $\nu = \sum_{n=1}^{\infty} \hbar^n \nu_n$ are purely real numbers. Note that although the contour $\check{\eta}_* \to \eta_c \to \eta_*$ is the ``upper half'' of the loop $\gamma$, as seen in Fig.~\ref{fig:gamma}, the integral above does not evaluate to half of $2\pi i E$, as one might expect from the expression for $E \equiv E_A$ in Eq.~\eqref{eq:Ej_formula}.
Instead, we see an extra phase $\phi_2$ and an extra factor $\nu$. The phase $\phi_2$ comes from the real contribution of the real part of $T_{A,\mathrm{odd}}$; see Eq.~\eqref{eq:T_odd_r_mu}. If the integration contour were the loop $\gamma$, then the Schwarz reflection principle guarantees that the ``lower half'' of $\gamma$ would contribute $-\phi_2$, cancelling the $\phi_2$ from the ``upper half'' we see here. The factor $\nu$ comes from the imaginary part of $T_{A,\mathrm{odd},n}$; it appears because we are taking a path different from $\gamma$ for analytic continuation. To evaluate the $\nu_n$'s explicitly, notice that the imaginary part of $T_{A,\mathrm{odd},n}$ for $n\geq 1$ are total derivatives, so the integral is simply the difference between the anti-derivative evaluated at $\eta_*$ and $\check{\eta}_*$. In conclusion:
\ba{
    \frac{\hat{v}_\mp^{(1)}(\eta_*, \hbar)}{\hat{v}_\pm^{(2)}(\eta_*, \hbar)}
    = e^{\pm \pi \epsilon / \hbar} e^{\mp i \phi_2 / \hbar} e^{\pm \nu / \hbar} 
    \left( 1 + \order{e^{-\pi E_0 / \hbar}} \right)
    \;,
}
where the exponentially small factor comes from Borel resummation.
Finally, we evaluate $\hat{w}_\mp^{(1)} / \hat{w}_\pm^{(2)}$. 
Using the results of App.~\ref{sub:Rodd_equation} and arguments similar to those provided above, one can also show that 
\bes{\label{eq:Rodd_equation}
    \int_{\check{0}}^{0} \! \dd{x} \, R_{A,\mathrm{odd},n} 
    &= 
    \begin{cases}
    i \pi E_0 & n = 0 \\
    i \pi E_n + \phi_{3,n} & n \geq 1
    \end{cases} \\
    \int_{\check{0}}^{0} \! \dd{x} \, R_{A,\mathrm{odd}} 
    & = i \pi \epsilon + \frac{\pi \hbar}{2} + \phi_3
    \;,
}
where $\phi_3 = \sum_{n=1}^\infty \hbar^n \phi_{3,n}$ is some purely real number. $\check{0}$ is the point on the ``second sheet'' corresponding to $0$, and the integration path is $\check{0} \to 2i\sqrt{E_0} \to 0$. Thus we have, upon Borel resummation:
\ba{
    \frac{\hat{w}_\mp^{(1)}(0,\hbar)}{\hat{w}_\pm^{(2)}(0,\hbar)}
    = \exp(\mp \frac{i}{\hbar} \int_{\check{0}}^{0} \! \dd{x} \, R_{A,\mathrm{odd}})
    = \mp i \, e^{\pm \pi \epsilon / \hbar} e^{\mp i \phi_3 / \hbar}
    \left(1 + \order{e^{-\pi E_0 / \hbar}} \right)
    \;.
}
Combining the above results, we have: 
\bes{
    & \frac{\lambda_\mp^{(1)}}{\lambda_\pm^{(2)}}
    = \pm i\, e^{\pm i (\phi_2 - \phi_3) / \hbar} 
    X^{\pm 1}
    \left( 1 + \order{e^{-\pi E_0 / \hbar}} \right) \\ 
    & \mathrm{where} \quad 
    X \equiv e^{-\nu / \hbar} \eval{ \sqrt{\frac{1 - M / S_{B-}}{1 - M / S_{B+}}} \sqrt{\frac{- S_{A+}}{S_{A-}}} }_{\eta_*}
    \;.
}
Using explicit expressions for $\nu_n$ and $S_{J\pm}$, one can evaluate $X^* X$ to order $\hbar$ and show that $\left| X \right|^2 = 1 + \order{\hbar^2}$, which suggests that $X$ is simply a phase factor. To show that this is indeed true, consider the analytic continuation of Eq.~\eqref{eq:S_eqn_matrix} from $\eta_*$ to $\check{\eta}_*$:
\bes{
    & (S_{A-}(\check{\eta}_*) + M(\check{\eta}_*)) u_{A-,*} \exp(\frac{i}{\hbar} \int_{\eta_*}^{\check{\eta}_*} \! \dd{\eta} \, S_{A-} ) + K(\check{\eta}_*) u_{B-,*} \exp(\frac{i}{\hbar} \int_{\eta_*}^{\check{\eta}_*} \! \dd{\eta} \, S_{B-} ) = 0 \\
    & \eval{ \left[ (S_{A+} + M) u_{A-,*} \exp(- \frac{i}{\hbar} \int_{\eta_*}^{\check{\eta}_*} \! \dd{\eta} \, T_{A,\mathrm{odd}} ) - K u_{B-,*} \exp(- \frac{i}{\hbar} \int_{\eta_*}^{\check{\eta}_*} \! \dd{\eta} \, T_{B,\mathrm{odd}} ) \right] }_{\eta_*} = 0 
    \;.
}
Note that $S_{A-}(\check{\eta}_*) = S_{A+}(\eta_*)$. The minus sign before $K$ comes from the analytic continuation of $1 / \sqrt{T_{J,\mathrm{odd}}}$ factors. Breaking $T_{J,\mathrm{odd}}$ into real and imaginary parts, we have:
\bes{
    & \eval{ \left[ (S_{A+} + M) u_{A-,*} e^{- \nu / \hbar} - K u_{B-,*} e^{+ \nu / \hbar} \right] }_{\eta_*} = 0 \\
    & e^{- 2 \nu / \hbar} 
    = \eval{\frac{K}{S_{A+} + M}}_{\eta_*} \frac{u_{B-,*}}{u_{A-,*}} 
    = -\eval{\frac{S_{A-} + M}{S_{A+} + M}}_{\eta_*}  \\
    & X = 
    \eval{\sqrt{\frac{(S_{A-} + M) (S_{B-} - M)}{(S_{A+} + M) (S_{B+} - M)}} }_{\eta_*}
    \eval{\sqrt{\frac{S_{A+} S_{B+}}{S_{A-} S_{B-}}} }_{\eta_*}
    \;.
}
One can show that both square roots above have modulus $1$ via the zero-determinant condition from \eqref{eq:S_eqn_matrix} and Eq.~\eqref{eq:SA_SB_relation}. Thus $X$ is a phase factor.
We define a phase $\phi_4$ by $X = e^{i\phi_4 / \hbar}$, and conclude:
\bes{
    & \frac{\lambda_\mp^{(1)}}{\lambda_\pm^{(2)}}
    = \pm i\, e^{\pm i (\phi_2 + \phi_4 - \phi_3) / \hbar} 
    \left( 1 + \order{e^{-\pi E_0 / \hbar}} \right) 
    \;.
    \label{eq:lambda_ratio_2}
}
We see that $\lambda_\mp^{(1)} / \lambda_\pm^{(2)}$ is a phase factor up to order of small exponentials.

We now collate all the $\lambda_\pm^{(i)}$ ratios \eqref{eq:lambda_ratio_1} and \eqref{eq:lambda_ratio_2}, and evaluate the four entries of $T$: 
\ba{
    a =&\ A \frac{\lambda_+^{(1)}}{\lambda_+^{(2)}}
    = \frac{\sqrt{2\pi}}{\Gamma(- i \frac{\epsilon}{\hbar})} e^{i\pi (1 + i \frac{\epsilon}{\hbar}) /2} \sqrt{\frac{\hbar}{\epsilon}} e^{-i \phi_1} \left(1 + \order{e^{-\pi E_0 / \hbar}} \right) \nonumber \\
    =&\ \sqrt{1 - e^{-2 \pi \epsilon / \hbar}} \left(1 + \order{e^{-\pi E_0 / \hbar}} \right) \nonumber \\
    b =&\ B \frac{\lambda_+^{(1)}}{\lambda_-^{(2)}}
    = -i e^{-\pi(\frac{\epsilon}{\hbar} - \frac{i}{2})} (- i e^{-i (\phi_2 + \phi_4 - \phi_3) / \hbar}) \left(1 + \order{e^{-\pi E_0 / \hbar}} \right) \nonumber \\
    =&\ - i e^{-i (\phi_2 + \phi_4 - \phi_3) / \hbar} e^{-\pi \epsilon / \hbar} \left(1 + \order{e^{-\pi E_0 / \hbar}} \right) \nonumber \\
    c =&\ C\frac{\lambda_-^{(1)}}{\lambda_+^{(2)}}
    = i e^{-\pi(\frac{\epsilon}{\hbar} - \frac{i}{2})} (+ i e^{+i (\phi_2 + \phi_4 - \phi_3) / \hbar}) \left(1 + \order{e^{-\pi E_0 / \hbar}} \right) \nonumber \\
    =&\ -i e^{i (\phi_2 + \phi_4 - \phi_3) / \hbar} e^{-\pi \epsilon / \hbar} \left(1 + \order{e^{-\pi E_0 / \hbar}} \right) \nonumber \\
    d =&\ D\frac{\lambda_-^{(1)}}{\lambda_-^{(2)}}
    = \frac{\sqrt{2\pi}}{\Gamma(1 + i\frac{\epsilon}{\hbar})} e^{i\pi (i \frac{\epsilon}{\hbar}) /2} \sqrt{\frac{\epsilon}{\hbar}} e^{i \phi_1} \left(1 + \order{e^{-\pi E_0 / \hbar}} \right) \nonumber \\
    =&\ \sqrt{1 - e^{-2 \pi \epsilon / \hbar}} \left(1 + \order{e^{-\pi E_0 / \hbar}} \right)
    \;,
}
where we have used Gamma function identities:
\ba{
    \Gamma(1 \pm i \frac{\epsilon}{\hbar}) = \sqrt{\frac{\pi \epsilon / \hbar}{\sinh(\pi \epsilon / \hbar)}} e^{\pm i \phi_1},\quad \Gamma(1 - i \frac{\epsilon}{\hbar}) = (- i \frac{\epsilon}{\hbar}) \Gamma(- i \frac{\epsilon}{\hbar})
    \;.
}
In conclusion, $T$ is given by: 
\bes{
    & T = \mqty(
    \sqrt{1 - e^{-2 \pi \epsilon / \hbar}} & 
    -i e^{i (\phi_2 + \phi_4 - \phi_3) / \hbar} e^{-\pi \epsilon / \hbar} \\
    - i e^{-i (\phi_2 + \phi_4 - \phi_3) / \hbar} e^{-\pi \epsilon / \hbar} &
    \sqrt{1 - e^{-2 \pi \epsilon / \hbar}} )
    \left(I + \order{e^{-\pi E_0 / \hbar}} \right) \;, \\
    & \mathrm{where}\quad 
    \epsilon = \frac{i}{\pi} \int_{\eta_c}^{\eta_c^*} \omega \dd{\eta} + \hbar \frac{1}{2 \pi i} \oint_\gamma \frac{-M}{2\omega} \zeta' \dd{\eta} + \order{\hbar^2} \;.
    \label{eq:connection_matrix_full}
}
One can see that, up to order small exponentials, $T$ is manifestly in $\mathrm{SU}(2)$. The master formula \eqref{eq:con_mat} is obtained by dropping terms of order $\hbar$ in $\epsilon$. The order $\hbar^1$ term in $\epsilon$ is a geometric amplitude factor, consistent with the one derived by Berry~\cite{10.2307/80000}.

\subsection{Properties of $R_\mathrm{odd}$}\label{sub:Rodd_equation}
In this subsection, we will construct a WKB series for a generic Weber equation \eqref{eq:psi_eqn} and discuss some of its properties. 
The Weber equation is given by:
\ba{
    \left(\hbar^2 \dv[2]{x} + P(x, \hbar) \right)w(x,\hbar) = 0 \qq{where} 
    P(x, \hbar) \equiv E(\hbar) + \frac{x^2}{4} \;.
}
As usual, we take the WKB ansatz for $w(x, \tau)$:
\ba{
    w_\pm(x, \hbar) = \exp(\frac{i}{\hbar} \int^x \dd{x'} R_\pm) \;,
}
then the function $R$ satisfies equation:
\ba{
    i \hbar \dv{R}{x} - R(x, \hbar)^2 + P(x, \hbar) = 0
    \;.
}
$R$ can be obtained recursively, order-by-order in $\hbar$. The first few terms for $R$ is given by:
\bsa{}{
    R_\pm =&\ \sum_{n=0}^{\infty} \hbar^n R_{n\pm} \\
    R_{0\pm} =&\ \mp \sqrt{E_0 + \frac{x^2}{4}} \\
    R_{1\pm} =&\ \frac{i x}{2 (x^2 + 4 E_0)} \mp \frac{E_1}{\sqrt{x^2 + 4 E_0}} \;.
}
Defining $R_\mathrm{odd} = (R_+ - R_-) / 2$ as usual, we have:
\bsa{}{
    R_{\mathrm{odd},0} =&\ - \sqrt{E_0 + \frac{x^2}{4}} \\
    R_{\mathrm{odd},1} =&\ - \frac{E_1}{\sqrt{x^2 + 4 E_0}} \\
    R_{\mathrm{odd},2} =&\ - \frac{E_2}{\sqrt{x^2 + 4 E_0}} + \frac{-3 x^2 + 8 E_0 + 4 x^2 E_1 + 16 E_0 E_1^2}{4 (4 E_0 + x^2)^{5/2}} \;.
}
By observing the above pattern, one may speculate that the coefficient for $E_n$ in $R_{\mathrm{odd},n}$ is always $-1 / \sqrt{x^2 + 4 E_0}$. This is indeed true, and it can be easily seen from the recurrence relation for $R_\pm$:
\bes{
    & R_{n\pm} = \frac{E_n}{\mp 2 \sqrt{E_0 + \frac{x^2}{4}}} + \frac{1}{\mp 2 \sqrt{E_0 + \frac{x^2}{4}}} \left(i \dv{R_{n-1,\pm}}{x} - \sum_{k=1}^{n-1} R_{k\pm} R_{n-k,\pm} \right) \\
    &\Rightarrow R_{\mathrm{odd},n} = -\frac{E_n}{\sqrt{x^2 + 4 E_0}} + \qq{(remainder)} \;.
}
Corollary~2.5 of \cite{Aoki:1991} implies:
\ba{
    \oint_\gamma \dd{x} R_{\mathrm{odd},n} = 2 \pi i E_n \;.
}
Also, an argument similar to the one we used to compute $E_{A1}$ in the previous section shows:
\ba{
    \oint_\gamma \dd{x} \frac{-E_n}{\sqrt{x^2 + 4 E_0}} 
    = \frac{E_n}{2} \oint_\gamma \dd( \log(\frac{\sqrt{x^2 + 4 E_0} - x}{\sqrt{x^2 + 4 E_0} + x}) ) = 2 \pi i E_n \;.
    \label{eq:R_odd_integral}
}
We see that the ``remainder'' term in $R_{\mathrm{odd},n}$ does not contribute to the contour integral. If we were to change the integration contour from $\gamma$ to $\check{0} \to x_c \to 0$, then the Schwarz reflection principle implies that the integral of the ``remainder'' term would become $\phi_{3,n}$, where $\phi_{3,n}$ is some real number. This fact was used to show \eqref{eq:Rodd_equation}.

\section{Analytic formula for the phase integral}\label{app:phase_integral}
\subsection{spin-1/2 particle production}\label{app:spin_1/2_phase_integral}
In this Appendix, we present the derivation of the phase integral \eqref{eq:dirac_phase_integral} given by:
\begin{align}
  \Psi_k \equiv i \int_{\eta_c}^{\eta_c^*} \! \dd{\eta} \, \omega_k  = i \int_{\eta_c}^{\eta_c^*} \! \dd{\eta} \, \sqrt{k^2 + \frac{a_0^2}{4}\left(1 + \tanh(\frac{\eta}{\tau})\right)^2 m^2} .
\end{align}
With \emph{Mathematica}, we can find an expression for $\Psi_k$:
\begin{align}
  \Psi_k=& \frac{\tau}{2} \Big[\pi a_0 m + \pi \sqrt{k^2 + a_0^2 m^2} + 2 i a_0 m \,\mathrm{arctanh}(\frac{a_0 m / 2}{\sqrt{k^2 + a_0^2 m^2/4}}) + 2 \pi a_0 m \left\lfloor \frac{\theta}{4\pi} \right\rfloor \nonumber \\
       & + 2 i k \,\mathrm{arctanh}(\frac{\sqrt{k^2 + a_0^2 m^2/4}}{k}) - 2 i \sqrt{k^2 + a_0^2 m^2} \,\mathrm{arctanh}(\frac{k^2 + a_0^2 m^2 / 2}{\sqrt{k^2 + a_0^2 m^2 / 4} \sqrt{k^2 + a_0^2 m^2}}) 
         \nonumber \\
       & + 2 \pi \sqrt{k^2 + a_0^2 m^2} \left(\left\lfloor \frac{\pi - 2 \arg(k - i a_0 m) + \theta}{4\pi} \right\rfloor - \left\lfloor \frac{3\pi - 2 \arg(k - i a_0 m) + \theta}{4\pi} \right\rfloor\right) \Big] ,
\end{align}
where
\bes{
  \theta =& \arg(i k + a_0 m) + \arg(\mathrm{arctanh}(1 + \frac{2 i k}{a_0 m})) \\
  &+ 2\pi \left\lfloor - \frac{\arg(i k + a_0 m) + \arg(\mathrm{arctanh}(1 + \frac{2 i k}{a_0 m}))}{2\pi} \right\rfloor .
}
Due to the continuity of the power spectrum, the values of the floor functions or their differences are constant in $k$. One can then obtain $\lfloor \theta/(4\pi) \rfloor = -1$ by evaluating it in the $k \to \infty$ limit. Plugging in the values for the floor functions, we now have:
\begin{align}
  \Psi_k=& \frac{\tau}{2} \Big[\pi \sqrt{k^2 + a_0^2 m^2} - \pi a_0 m + 2 i a_0 m \,\mathrm{arctanh}(\frac{a_0 m / 2}{\sqrt{k^2 + a_0^2 m^2/4}}) 
  \nonumber \\
       & + 2 i k \,\mathrm{arctanh}(\frac{\sqrt{k^2 + a_0^2 m^2/4}}{k}) - 2 i \sqrt{k^2 + a_0^2 m^2} \,\mathrm{arctanh}(\frac{k^2 + a_0^2 m^2 / 2}{\sqrt{k^2 + a_0^2 m^2 / 4} \sqrt{k^2 + a_0^2 m^2}}) \Big] ,
\end{align}
Since $\Psi_k > 0$, the imaginary part of the above expression must vanish, so we only need to keep the real contribution from each term.  The term $2ik\, \mathrm{arctanh}(\sqrt{k^2 + a_0^2 m^2 / 4} / k)$ gives a real contribution of $\pi k$, and all other terms involving $\mathrm{arctanh}$'s are purely imaginary. Summarizing, we have:
\ba{
\Psi_k = \frac{\pi \tau}{2} \Bigl( \sqrt{k^2 + a_0^2 m^2} + k - a_0 m \Bigr) .
}

\subsection{spin-3/2 particle production in kination-dominated universe}\label{app:spin_3/2_phase_integral}
In this Appendix, we present the derivation of the phase integral \eqref{eq:3/2_psi}.
In order to find approximate solutions for the turning points \(\eta_c\) at large wavenumber \(k\),
we want to solve \(\omega_k^2(\eta) = 0\).
We do this by first expanding $c_s^2 / a^2$ around $\eta_v$:
\begin{align}
  & \frac{c_s^2}{a^2} 
  = \eval{\frac{1}{2} \dv[2]{(c_s^2 / a^2)}{\eta}}_{\eta_v} (\eta - \eta_v)^2 + \order{(\eta-\eta_v)^3} 
  = 9m^2 (\eta-\eta_v)^2 + \order{(\eta-\eta_v)^3} .
\end{align}
Noticing that $\omega_k^2(\eta) = 0$ is equivalent to $c_s^2 / a^2 = - m^2 / k^2$, we have:
\bes{
    9m^2 (\eta-\eta_v)^2 + \order{(\eta-\eta_v)^3}  = - \frac{m^2}{k^2} 
    \qq{}\Rightarrow\qq{} \eta - \eta_v = \pm i \frac{1}{3k} + \order{k^{-2}}.
    \label{eq:rs_eta_c_expansion}
}
The solution in the upper half plane is thus \(\eta_c = \eta_v + i / (3k) + \order{k^{-2}}\).
The other solution is simply $\eta_c^*$.

We define $\Delta\eta \equiv \eta - \eta_v$ and restrict $\Delta\eta$ to the region $\abs{\Delta\eta} \leq \abs{\eta_c - \eta_v}$. Around \(\eta_v\), we have the following approximation for \(\omega_k\):
\bes{
  \omega_k(\eta_v + \Delta\eta) =&\ a(\eta_v) m \sqrt{\frac{c_s^2(\eta) k^2}{a(\eta_v)^2 m^2} + \frac{a^2(\eta)}{a^2(\eta_v)}} \\
  =&\ a(\eta_v) m \sqrt{1 + 9 k^2 \Delta\eta^2 + \order{k^{-1}}} \; ,
}
where the expression inside the square root is due to Taylor expansion to order $\Delta\eta^2$, and the $\order{\Delta\eta}$ term is collected into $\order{k^{-1}}$. This expression for $\omega_k$ lends us a way to evaluate the phase integral:
\bes{
    \Psi =&\ i \int_{\eta_c}^{\eta_c^*} \! \dd{\eta} \, \omega_k(\eta) \\
    =&\ i \int_{i / (3k)}^{- i / (3k)} \! \dd{\Delta \eta} \, \omega_k(\eta_v + \Delta \eta) + \order{k^{-2}} \\
    =&\ i \int_{i / (3k)}^{- i / (3k)} \! \dd{\Delta \eta} \, a(\eta_v) m \sqrt{1 + 9k^2 \Delta \eta^2 } + \order{k^{-2}} \; .
}
Here, the contour of integration has length $\order{k^{-1}}$, so to extract the $\order{k^{-1}}$ part of $\Psi$ we only keep the $\order{k^0}$ term in the integrand. The integral can then be explicitly integrated to give:
\ba{
    \Psi 
    = i \int_{\eta_c}^{\eta_c^*} \! \dd{\eta} \, \omega_k(\eta) 
    = \frac{\pi a(\eta_v) m}{6 k} + \order{k^{-2}} \; .
}

\bibliographystyle{JHEP}
\bibliography{main}

\end{document}